\documentclass[]{aa}

\usepackage{graphicx}
\usepackage{txfonts}
\usepackage{natbib}
\usepackage{graphicx}
\usepackage{color}
\usepackage{txfonts}
\usepackage{hyperref}
\usepackage{booktabs}
\usepackage{amsmath}
\usepackage{amssymb}
\usepackage{mathtools}
\usepackage{txfonts}
\usepackage{longtable}
\usepackage[normalem]{ulem}
\usepackage{subcaption}
\usepackage[flushleft]{threeparttable} 

\providecommand{\gmag}{\ensuremath{G}}

\newcommand\gaia{\textit{Gaia}\xspace}

\newcommand\gdrtwo{\gaia DR2\xspace}
\newcommand\gdrthree{\gaia DR3\xspace}

\def\deg{\ensuremath{^\circ}}
\def\arcmin{\ensuremath{^\prime}\xspace}
\def\arcsec{\ensuremath{^{\prime\prime}}\xspace}

\def\farcs{\ensuremath{.\!\!^{\prime\prime}}}

\makeatletter
\renewcommand*\aa@pageof{, page \thepage{} of \pageref*{LastPage}}
\makeatother

\setcitestyle{citesep={,}}

\newcommand{\orcit}[1]{\protect\href{https://orcid.org/#1}{\protect\includegraphics[width=8pt]{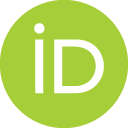}}}

\makeatletter
\renewcommand*\maketitle{%
  \thispagestyle{firstpage}
\begingroup
    \if@wideboxfn
    \setlength\bibindent{1.4\parindent}
    \else
    \setlength\bibindent{\parindent}
    \fi
    \renewcommand*\thefootnote{\@fnsymbol\c@footnote}%
    \renewcommand\@makefntext[1]{%
    \ifaa@longfn\hsize\textwidth\fi
    \noindent
    \hb@xt@\bibindent{\hss\@makefnmark\enspace}##1}
  \ifaa@twocolumn
  \begingroup
    \begin{aa@strip}
          \aa@maketitle
    \end{aa@strip}
    \@thanks            
  \endgroup
  \else
    \begingroup
      \let\thanks\footnote
      \aa@maketitle
    \endgroup
  \fi
\endgroup
  \setcounter{footnote}{0}%
}
\makeatother

\begin{document} 

\title{\gaia Data Release 3: the Solar System survey}
\subtitle{}

\author{
P.   ~Tanga                         \orcit{0000-0002-2718-997X}\inst{\ref{inst:0001}}
\and T.        ~Pauwels                       \inst{\ref{inst:0013}}
\and F.        ~Mignard                       \inst{\ref{inst:0001}}
\and K.        ~Muinonen \orcit{0000-0001-8058-2642}\inst{\ref{inst:0011}},\inst{\ref{inst:0012}}
\and A.        ~Cellino                       \orcit{0000-0002-6645-334X}\inst{\ref{inst:0008}}
\and P.        ~David                         \inst{\ref{inst:0002}}
\and D.        ~Hestroffer                    \orcit{0000-0003-0472-9459}\inst{\ref{inst:0002}}
\and F.        ~Spoto                         \orcit{0000-0001-7319-5847}\inst{\ref{inst:0021}}
\and J.        ~Berthier                      \orcit{0000-0003-1846-6485}\inst{\ref{inst:0002}}
\and J.        ~Guiraud                       \inst{\ref{inst:0015}}
\and W.        ~Roux                          \inst{\ref{inst:0015}}
\and B.        ~Carry               \orcit{0000-0001-5242-3089}\inst{\ref{inst:0001}}
\and M.        ~Delbo                         \inst{\ref{inst:0001}}
\and A.        ~Dell'Oro            \orcit{0000-0003-1561-9685}\inst{\ref{inst:0009}}
\and C.        ~Fouron                        \inst{\ref{inst:0018}}
\and L.        ~Galluccio                     \orcit{0000-0002-8541-0476}\inst{\ref{inst:0001}}
\and A.        ~Jonckheere                    \inst{\ref{inst:0013}}
\and S.A.      ~Klioner                       \orcit{0000-0003-4682-7831}\inst{\ref{inst:0026}}
\and Y.        ~Lefustec                      \inst{\ref{inst:0015}}
\and L.        ~Liberato   \orcit{0000-0003-3433-6269}\inst{\ref{inst:0001}},\inst{\ref{inst:0019}}
\and C.        ~Ord\'{e}novic        \orcit{0000-0003-0256-6596}\inst{\ref{inst:0001}}
\and I.        ~Oreshina-Slezak               \inst{\ref{inst:0001}}
\and A.        ~Penttil\"{ a}               \orcit{0000-0001-7403-1721}\inst{\ref{inst:0011}}
\and F.        ~Pailler                       \inst{\ref{inst:0015}}
\and Ch.       ~Panem                         \inst{\ref{inst:0015}}
\and J.-M.     ~Petit                 \orcit{0000-0003-0407-2266}\inst{\ref{inst:0014}}
\and J.        ~Portell               \orcit{0000-0002-8886-8925}\inst{\ref{inst:0017}}
\and E.        ~Poujoulet                        \inst{\ref{inst:0020}}
\and W.        ~Thuillot                      \inst{\ref{inst:0002}}
\and E.        ~Van~Hemelryck                 \inst{\ref{inst:0013}}
\and A.        ~Burlacu                        \inst{\ref{inst:0018}}
\and Y.        ~Lasne                          \inst{\ref{inst:0016}}
\and S.        ~Managau                        \inst{\ref{inst:0016}}
}
\institute{
     Universit\'{e} C\^{o}te d'Azur, Observatoire de la C\^{o}te d'Azur, CNRS, Laboratoire Lagrange, Bd de l'Observatoire, CS 34229, 06304 Nice Cedex 4, France\relax                                        \label{inst:0001}
\and Royal Observatory of Belgium, Ringlaan 3, 1180 Brussels, Belgium\relax                                                                                                                                  \label{inst:0013}
\and University of Helsinki, Department of Physics, P.O. Box 64, 00014 Helsinki, Finland\relax                                                                                                               \label{inst:0011}
\and Finnish Geospatial Research Institute FGI, Geodeetinrinne 2, 02430 Masala, Finland\relax                                                                                                                \label{inst:0012}
\and INAF - Osservatorio Astrofisico di Torino, via Osservatorio 20, 10025 Pino Torinese (TO), Italy\relax                                                                                                   \label{inst:0008}
\and IMCCE, Observatoire de Paris, Universit\'{e} PSL, CNRS,  Sorbonne Universit\'{e}, Univ. Lille, 77 av. Denfert-Rochereau, 75014 Paris, France\relax                                                      \label{inst:0002}
\and Minor Planet Center - Center for Astrophysics, Harvard \& Smithsonian, 60 Garden St., MS 15, Cambridge, MA, USA \label{inst:0021}
\and CNES Centre Spatial de Toulouse, 18 avenue Edouard Belin, 31401 Toulouse Cedex 9, France\relax  \label{inst:0015}
\and INAF - Osservatorio Astrofisico di Arcetri, Largo Enrico Fermi 5, 50125 Firenze, Italy\relax                                                                                                            \label{inst:0009}
\and Institut UTINAM UMR6213, CNRS, OSU THETA Franche-Comt\'{e} Bourgogne, Universit\'{e} Bourgogne Franche-Comt\'{e}, 25000 Besan\c{c}on, France\relax                                                      \label{inst:0014}
\and Thales Services for CNES Centre Spatial de Toulouse, 18 avenue Edouard Belin, 31401 Toulouse Cedex 9, France\relax  \label{inst:0016}
\and Institut de Ci\`{e}ncies del Cosmos (ICCUB), Universitat  de  Barcelona  (IEEC-UB), Mart\'{i} i  Franqu\`{e}s  1, 08028 Barcelona, Spain\relax \label{inst:0017}
\and Telespazio for CNES Centre Spatial de Toulouse, 18 avenue Edouard Belin, 31401 Toulouse Cedex 9, France \label{inst:0018}
\and Lohrmann Observatory, Technische Universit\"{ a}t Dresden, Mommsenstra{\ss}e 13, 01062 Dresden, Germany\relax  \label{inst:0026}
\and UNESP - Sao Paulo State University, Grupo de Dinamica Orbital e Planetologia, CEP 12516-410, Guaratingueta, SP, Brazil \label{inst:0019}
\and AKKA for CNES Centre Spatial de Toulouse, 18 avenue Edouard Belin, 31401 Toulouse Cedex 9, France\relax \label{inst:0020}
}

\date{Received / Accepted}

 
\abstract
   {The third data release by the Gaia mission of the European Space {Agency (DR3)} is the first release to provide the community with a large sample of observations for more than 150 thousand Solar System objects, including asteroids and natural planetary satellites. The release contains astrometry (over 23 million epochs) and photometry, along with average reflectance spectra of 60518 asteroids and osculating elements.}
   {We present an overview of the procedures that have been implemented over several years of development and tests to process Solar System data at the level of accuracy that Gaia can reach. We illustrate the data properties and potential with some practical examples.}
   {In order to allow the users of \gdrthree to best exploit the data, we explain the assumptions and approaches followed in the implementation of the data processing pipeline for Solar System processing, and their effects in terms of data filtering, optimisation, and performances. We then test the data quality by analysing post-fit residuals to adjusted orbits, the capacity of detecting subtle dynamical effects (wobbling due to satellites or shape and Yarkovsky acceleration), and to reproduce known properties of asteroid photometry (phase curves and rotational light curves).}
   {The DR3 astrometric accuracy is a clear improvement over the data published in DR2, which concerned a very limited sample of asteroids. The performance of the data reduction is met, and is illustrated by the capacity of detecting milliarcsecond-level wobbling of the asteroid photocentre that is due to satellite or shape effects and contributes to Yarkovsky effect measurements.}
   {The third data release can in terms of data completeness and accuracy be considered the first full-scale realisation of the Solar System survey by Gaia.}


\titlerunning{\gaia DR3: Solar System survey}
\authorrunning{Tanga et al.}
\maketitle
%
\section{Introduction}
\begin{figure*}[ht]
\centering
\includegraphics[width=1.0\textwidth]{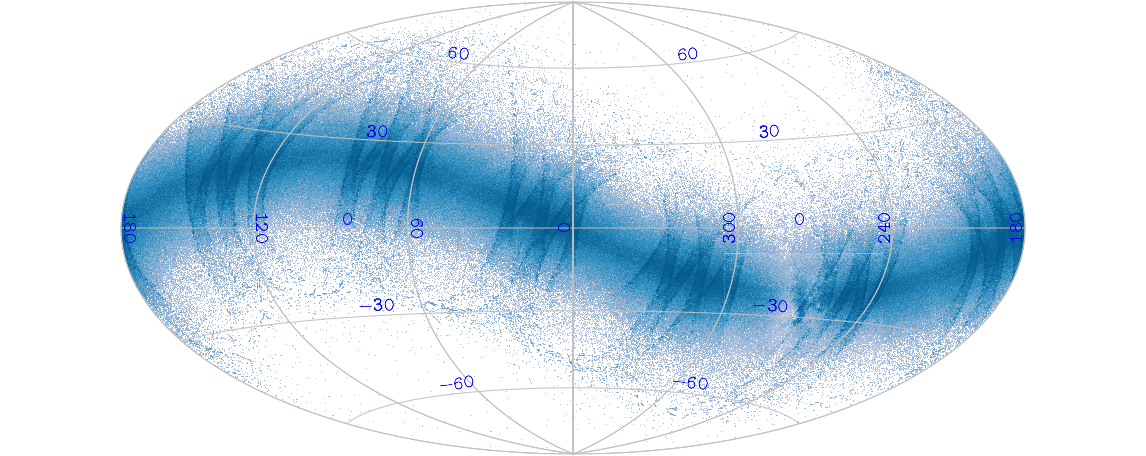}
\caption{Positions of the transits of sources published in \gdrthree in a full-sky Hammer-Aitoff projection in equatorial coordinates. The origin of RA and Dec is at the centre, north is up, and $\alpha$ increases from right to left. Darkness is related to the density of observations. Some density patterns related to the Gaia scanning law are clearly present. It is interesting to note the emergence of the highest stellar density regions as less dark areas close to the Galactic centre (around RA=270$\deg$ and Dec=-30$\deg$), where the efficiency of unambiguous identifications decreases.}
\label{F:positions}
\end{figure*}
The potential capacity of \gaia to provide an outstanding survey of Solar System objects (SSO) became clear already during the preparation studies \citep{hestrofferScienceSolarSystem1999, mignardObservationsSolarSystem2002}. The expected sample of $\sim$3$\times 10^5$ objects, including astrometric positions, photometry, and spectra, and the unprecedented accuracy, were soon considered unique properties in the landscape of large surveys, fostering new science achievements \citep{cellinoAsteroidScienceGaia2007b, mignardGaiaMissionExpected2007, tangaGaiaObservationsSolar2007}. With the accumulation of more accurate information about the mission performance, applications of the \gaia Solar System survey were identified in the improvement of our knowledge of dynamical properties of asteroids through the ultra-accurate astrometry \citep{tangaGaiaUnprecedentedObservatory2008, bancelinDynamicsAsteroidsNearEarth2012}, leading in particular to the determination of asteroid masses \citep{mouretAsteroidMassesImprovement2007}, the measurement of the Yarkovsky effect \citep{delboDetectionYarkovskyEffect2008, desmars15-yarko},
the discovery and characterisation of asteroid satellites \citep{pravecSmallBinaryAsteroids2012,Oszkiewicz13}, the improvement of the dynamical models of satellite orbits \citep{arlotAstrometryNaturalPlanetary2012}, a long-standing impact on { ground-based} observations of stellar occultations \citep{tangaAsteroidOccultationsToday2007}, and new tests of General Relativity \citep{hees18IAUs,hestrofferGaiaAsteroidsLocal2009}. Epoch brightness measurements and low-resolution reflectance { spectra} were identified as an unprecedented source of knowledge about asteroid physical properties: global shape properties and rotation parameters \citep{cellinoDerivationAsteroidPhysical2012}, compositions, and taxonomic classification \citep{delboAsteroidSpectroscopyGaia2012}.

While \gdrtwo provided a very limited and preliminary high-quality sample of astrometric and photometric data \citep{Spoto18}, \gdrthree for the first time reaches the level of quality, variety, and volume that was expected for the Solar System (Fig.~\ref{F:positions}).
The goal of this article is to illustrate the properties of the processing pipeline and the quality of the data that are obtained through the example of some significant applications. 

The implementation of the data processing for SSO was a long process that originated in preliminary studies that started at the end of the 1990s. These led to a functional analysis of the possible pipeline in 2006, several years before the launch of the satellite. Over time, the different processing modules have been developed, tested, qualified, and gradually entered into operations with each data release. Solar System objects benefit from the improvement of all aspects of the \gaia data processing, with an increase in data quality from one release to the next. At the same time, the structure of the pipeline increases in complexity. It treats new features and produces a more complete data set at its output.

While the fast daily processing that feeds asteroid alerts was illustrated elsewhere \citep{tangaDailyProcessingAsteroid2016,carryPotentialAsteroidDiscoveries2021}, we focus here on the procedures that have been implemented for \gdrthree, aiming to exploit the whole accuracy of the data. We also intend to show how \gaia data, which are peculiar in many aspects, should be used in practice. By doing so, we illustrate  their potential
for science with the example of some applications.
Conversely, processing and validation of asteroid spectra are not discussed here as they are extensively presented { by} \citet{DR3-DPACP-89}.

The article starts with a summary of the peculiarities of \gaia observations for the Solar System and of the general properties of the data present in \gdrthree (Section~\ref{S:general}). We then illustrate the data-processing pipeline (Sect.~\ref{S:SSO_processing}), starting with a description of the input data. We then describe the principles adopted to identify SSOs in the general data stream, to derive the astrometry, and to compute the orbits and the calibrated photometry. We provide information about the result of a match with a recent orbit catalogue for moving sources that are listed as not identified in \gdrthree (Sect.~\ref{S:unmatched_validation}). 
The quality of the astrometric and photometric data is then illustrated by several examples in Sect.~\ref{S:astrometry_performance} and \ref{S:photometry_performance}.

\section{Solar System data in \gdrthree}
\label{S:general}

\subsection{Summary of the general properties of Gaia observations}
\label{S:genprop}

\begin{figure*}
\centering
\includegraphics[trim= 80 0 90 0, clip, width=1.0\textwidth]{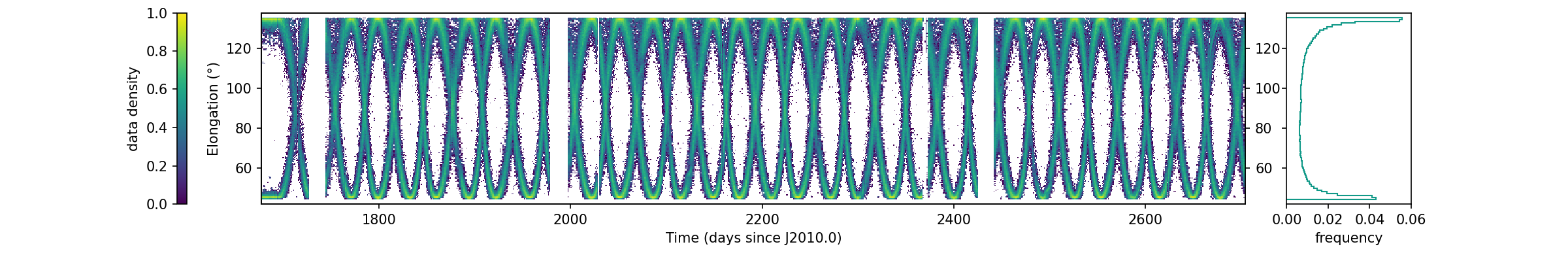}
\caption{
Distribution of the observations of Solar System objects in solar elongation as a function of time. Colour is related to data density. Two overlapping sinusoids appear, corresponding to the variation in the directions in which the scanning plane of $\gaia$ intersects the ecliptic due to the precession of the spin axis of the satellite. The peaks correspond to periods in which the ecliptic is crossed by the scan both at $\sim$45$\deg$ and in the opposite direction $\sim$135$\deg$. In these conditions, the scanning plane is perpendicular to the ecliptic. Scattered data correspond to detections of objects at high ecliptic inclination. The nodes of the sinusoids are around quadrature (90$\deg$ elongation), when the scanning plane cuts the ecliptic at 45$\deg$. An initial period without precession (enforcing the ecliptic pole scanning law) is visible. Three gaps appear, corresponding to technical operations preventing the normal collection of data. In the right panel, the cumulative distribution shows the preferential accumulation of observations at the extremes.
}  
\label{F:obs_time}
\end{figure*}
We recall some basic properties that strongly drive the processing, the results, and the use of the data with the relevant terminology. Exhaustive descriptions can be found in the  \href{https://gea.esac.esa.int/archive/documentation/GDR3/Catalogue\_consolidation/chap\_cu9val/}{online
  documentation} and in \citet{2016A&A...595A...1G}. General properties of Solar System observations were also provided with \gdrtwo in \citep{Spoto18}. 

The continuous rotation of the satellite results in the drift of all sources across the CCD matrix on the focal plane. Each focal plane passage (called ''transit'') can provide nine positions at most (in the Astrometric Field instrument, AF) and two low-dispersion spectra (in the Blue and Red Photometers, BP and RP). The AF is unfiltered  and produces the G-band photometry. At the beginning of each transit, sources are first detected by the Sky Mapper instrument (SM). While the SM is essential for the on-board assignment of pixel windows to track the transiting sources, the astrometry that it provides has lower quality and is not published.

The scanning motion of the \gaia telescopes combines the rotation of the satellite (period of six hours), the precession cycle of the rotation axis (68 days) on a Sun-centred cone, and the revolution around the Sun (one year). These three motions determine the typical timescales and locations when and where SSOs are observed. Single CCD measurements providing positions during a transit are spaced by 4.4\,s. G-band photometry, averaged over the AF, combines observations over $\sim$40\,s. 

In \gdrthree, the astrometry of an SSO is provided for each CCD. Hereafter, ''position'' refers to this single CCD measurement. G--band photometry is also derived from CCD-level measurements, but is provided as an average value over all CCDs that are available during a transit. This is implicitly assumed when the brightness of an object is mentioned. In summary, astrometric data of SSOs are provided at individual CCD levels, while photometric data are provided as averages over a transit (see Table~\ref{tab:DR3_summary_obs}).

The orientation of the scanning motion with respect to the Sun results in two avoidance cones with a semi-aperture of 45\deg, one centred on the Sun, and the other at solar opposition. As a consequence, observations are always obtained at solar elongations in the range between 45\deg and 135\deg. The distribution within this interval is not uniform and strongly favours the extreme values of elongation, where more time is spent by the scanning motion (Fig.~\ref{F:obs_time}).

The two telescopes on board \gaia sweep almost the same sky area 106\ minutes apart (over a single rotation of the satellite). Conventionally, they are referred to as \textup{\textit{preceding}} and \textup{\textit{following}} fields of view (FOVs), or FOV1 and FOV2. For a given asteroid, this scanning law usually results in short sequences of consecutive observations, separated by an absence of detection over several weeks or months.

By design, focal plane pixels are rectangular with an aspect ratio of three. Their short side is oriented in the scan direction (along scan, AL). With the exception of the brightest sources (G$<$13), the signal is also binned in the across-scan (AC) direction. Only the AL accuracy (at milliarcsecond level) is fully preserved, while only an approximate position is available for AC ($\text{with an accuracy of about one }$arcsecond). To reduce telemetry volumes, only a window (a limited surface of pixels around each source) is acquired by the image-processing system on board \gaia. As the window coordinates are computed at the beginning of the transit following the detection in SM, and then propagated across the focal plane by the predicted motion expected for a stellar source, moving objects (e.g. the SSOs) will drift with respect to the window centre, and their signal can be truncated at the edge of the window. 

The global astrometric solution of Gaia, provides absolute positions on the sky. This also applies to SSOs. The exact definition of the reference system and timescale is critical for the best exploitation of \gaia data (Sect.~\ref{sect:reference_frames}).

\subsection{Definition of the reference frames} \label{sect:reference_frames}
Because the astrometric positions resulting from the \gaia data are highly accurate, both the reference frame and the timescale must be clearly defined and well understood by the user. The individual observations at the CCD level and the orbital elements, or equivalently, the state vector of the observed objects, must be distinguished first. 

\subsubsection{Astrometric positions}
The positions given as Gaia-centric right ascension (RA) and declination (Dec) are derived from the local coordinates in the physical pixels and are transformed to astronomical coordinates as explained in Sect.~\ref{sect:coordinates}. The final positions are given in the Barycentric Celestial Reference System (BCRS) with the origin at \gaia, and this is achieved by ultimately referring the attitude of \gaia to the Gaia-CRF3  axes, which are aligned to ICRF3 \citep{EDR3-DPACP-133}. The directions provided in the form of RA, Dec are similar to the astrometric positions, meaning that they are corrected for the annual aberration, but not for the relativistic light deflection by the Solar System gravity field.  This last correction for sources at a finite distance requires knowledge of the distance, which is known for all the SSOs in \gdrthree, with the exception of a fraction of asteroids in the unmatched category. Taking the presence of light deflection into account is left to users computing orbits, as for the light travel time. For the relativistic framework adopted for \gaia, we refer to \citet{Klioner2003Practical,Klioner2004Gaia}.

\begin{figure*}[ht]
\centering
\includegraphics[width=0.75\textwidth]{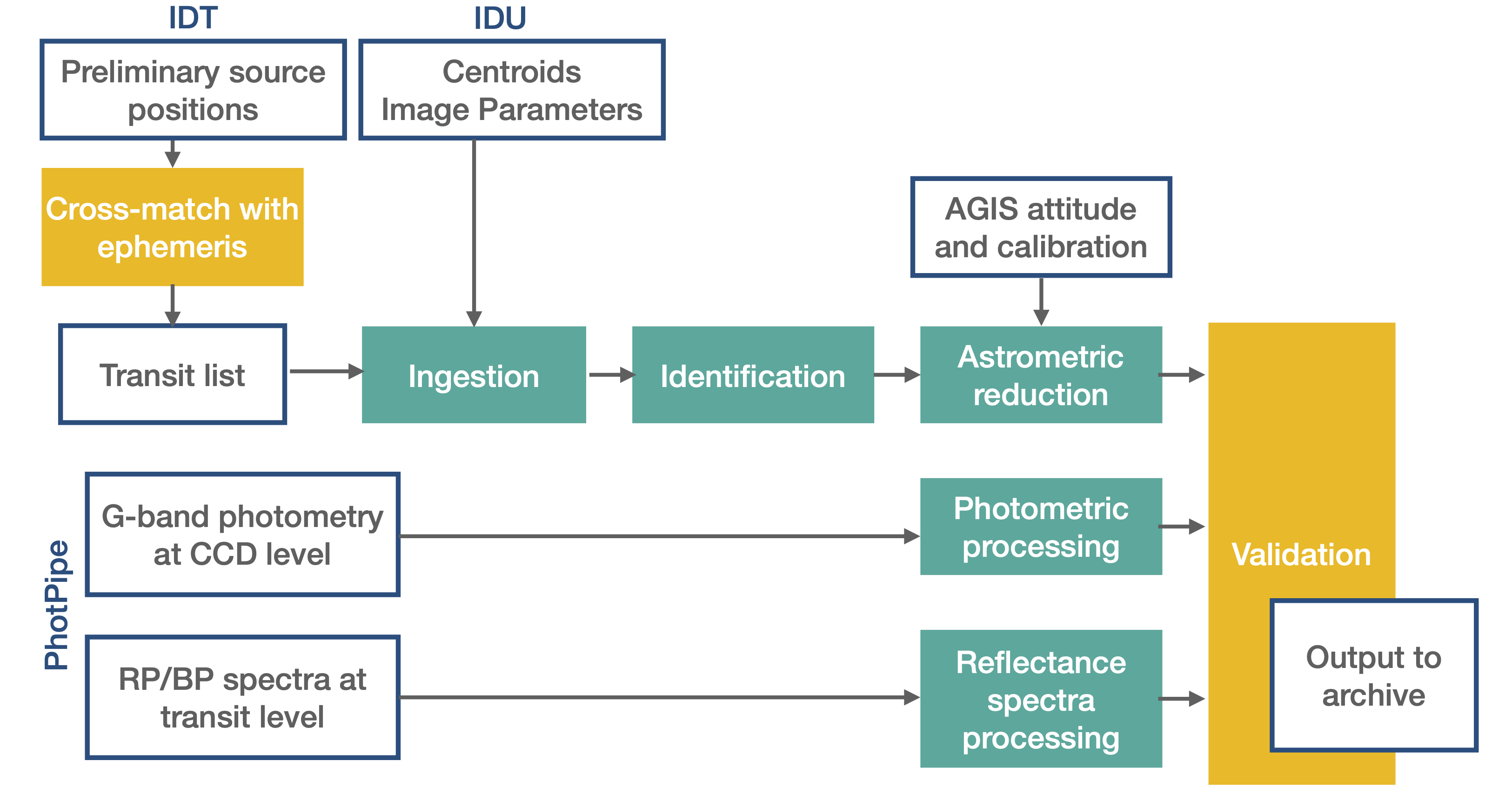}
\caption{General structure of the data processing. The turquoise (filled) boxes are modules developed specifically for Solar System objects, which run on the servers of one of the Data Processing Centers of \gaia (CNES in our case). White boxes represent input and output data. Yellow boxes are tasks that are developed, managed, and run by the Solar System scientific team that are essential to the preparation of the input data or to the validation of the output. Other modules providing input data, developed by other teams, are initial data treatment (IDT), intermediate data updating (IDU), the astrometric global iterative solution (AGIS), and the photometric pipeline (PhotPipe).
}  
\label{F:processing}
\end{figure*}
\subsubsection{Orbital elements}
Orbits are computed from a least-squares fitting of a dynamical model to the observed \textit{Gaia}-centric directions at transit times. This is detailed in Sect.~\ref{S:orbit_processing}. For each SSO, the unknowns are the six components of the state vector (position vector, velocity vector) at a certain epoch, nominally, the median of the transit times of the SSO, in order to minimize the correlations. The state vector is heliocentric and given in the ICRF axes. The Keplerian orbital elements are derived from the state vector with the ecliptic as reference plane. In this context, the ecliptic is defined by two rotations from the ICRF, the obliquity and the origin bias. This results in a rotation matrix. The values used throughout the \gaia data processing are those from \cite{2002A&A...387..700C}, with the origin bias of $\phi_\text{bias}= -55.42$\,mas defined as the right ascension of $\gamma_\text{ICRS}$ on the ICRF fundamental plane. In this { definition,} $\gamma_\text{ICRS}$ is the intersection of the dynamical ecliptic with the ICRS fundamental plane,  which is not the same as the J2000 celestial equator. The obliquity that is used corresponding to the inclination between the two planes at $\gamma_\text{ICRS}$ is $\epsilon_\text{ICRS} = 84381\farcs4110$. These values are not exactly identical to those adopted in the SOFA standards ($-52.928$\,mas and $84381\farcs412819$). The $\gamma_\text{ICRS}$ so defined is used as the origin of longitude in the ecliptic plane for the longitude of node in the orbital elements. It differs from  $\gamma_\text{J2000}$, the intersection of the mean ecliptic with the celestial equator at J2000 by an angle of $\sim\! 42$\,mas, with the sign convention such that this angle is the longitude of $\gamma_\text{J2000}$ referred to $\gamma_\text{ICRS}$. The \gaia -derived longitudes of node may therefore show a systematic when compared to other sources of orbital elements using the origin of longitude in the celestial equator of J2000.

Finally, the rotation matrix to transform a vector given in the ICRF to the same vector expressed in the ecliptic as defined above is 

\begin{equation*}
  \mathcal{R} =  \mathcal{R}(1,\epsilon_\text{ICRS})\,  \mathcal{R}(3,\phi_\text{bias}). 
\end{equation*}
This matrix has been applied to the orbit determination to express the state vector in the ecliptic frame before the heliocentric orbital elements were computed. The same matrix is used for the transformation of the covariance matrix, but in this case, the exact definitions are less critical.

\subsubsection{Timescale}
Very early in the data processing, it was decided to use the TCB as the astronomical timescale for all the \gaia computations. This was a logical choice after the BCRS and the associated relativistic metric were chosen as the framework for the astrometric modelling. The on-board time tagging is calibrated against the TCB on the ground to obtain the correspondence between the two scales and to provide a final timing of all observations and  \gaia events in TCB.  Because of the unique accuracy of  \gaia astrometry, the internal consistency was an essential requirement for the processing. This means that every  ephemeris used in the processing has TCB as an independent variable for the Solar system (major and minor planets, natural satellites), but also for the orbit of the spacecraft itself. This contrasts with the more common use of TDB for the public ephemeris and also as the timescale for the epoch of the orbital elements in
\texttt{astorb} \citep{2021DPS....5310104M} or at the Minor Planet Center. The transformation between the two scales is given by \citet{IEPA2021} and  \citet{klionerUnitsRelativisticTime2010} following IAU resolution 2006 B3\footnote{ \url{https://iau.org/static/resolutions/IAU2006_Resol3.pdf}},
\begin{align*}
    \textrm{TDB} &=  \textrm{TCB} \\ 
    & -L_\text{B} (JD_\text{TCB} -2\,443\,144.500\,3725)\times 86400~s \\
    & -6.55\times10^{-5}~s,
\end{align*}
where the time is expressed in seconds, and $L_\text{B} = 1.550\,519\,768\times 10^{-8}$ is a defining constant in the astronomical system of units. During the period covered by the \gdrthree, the difference $\textrm{TDB}\,-\, \textrm{TCB}$  is $\sim -19\,$s. 

As a purely indicative approximation, the UTC at the position of \gaia as derived from TCB is provided in the astrometry table \texttt{gaiadr3.sso\_observation} of \gdrthree. Nevertheless, UTC should not be used for an accurate exploitation of \gaia astrometry.  

\subsection{Population of Solar System objects}

\begin{figure*}
\centering
\includegraphics[width=0.95\textwidth]{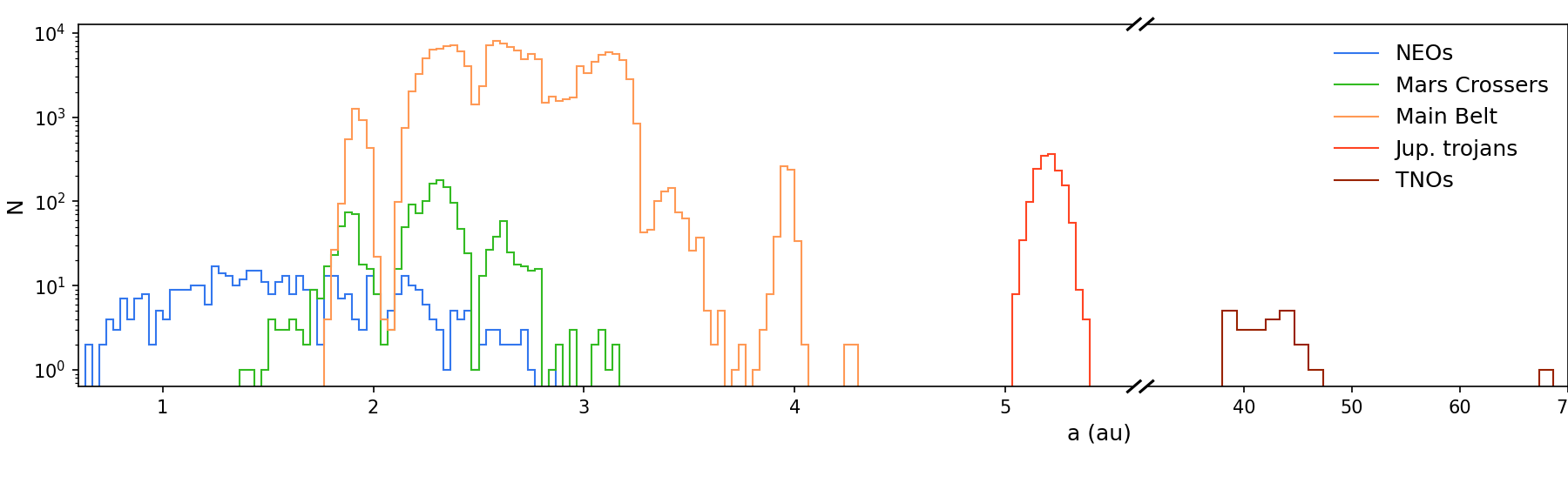}
\includegraphics[trim= 0 30 0 0, clip, width=0.85\textwidth]{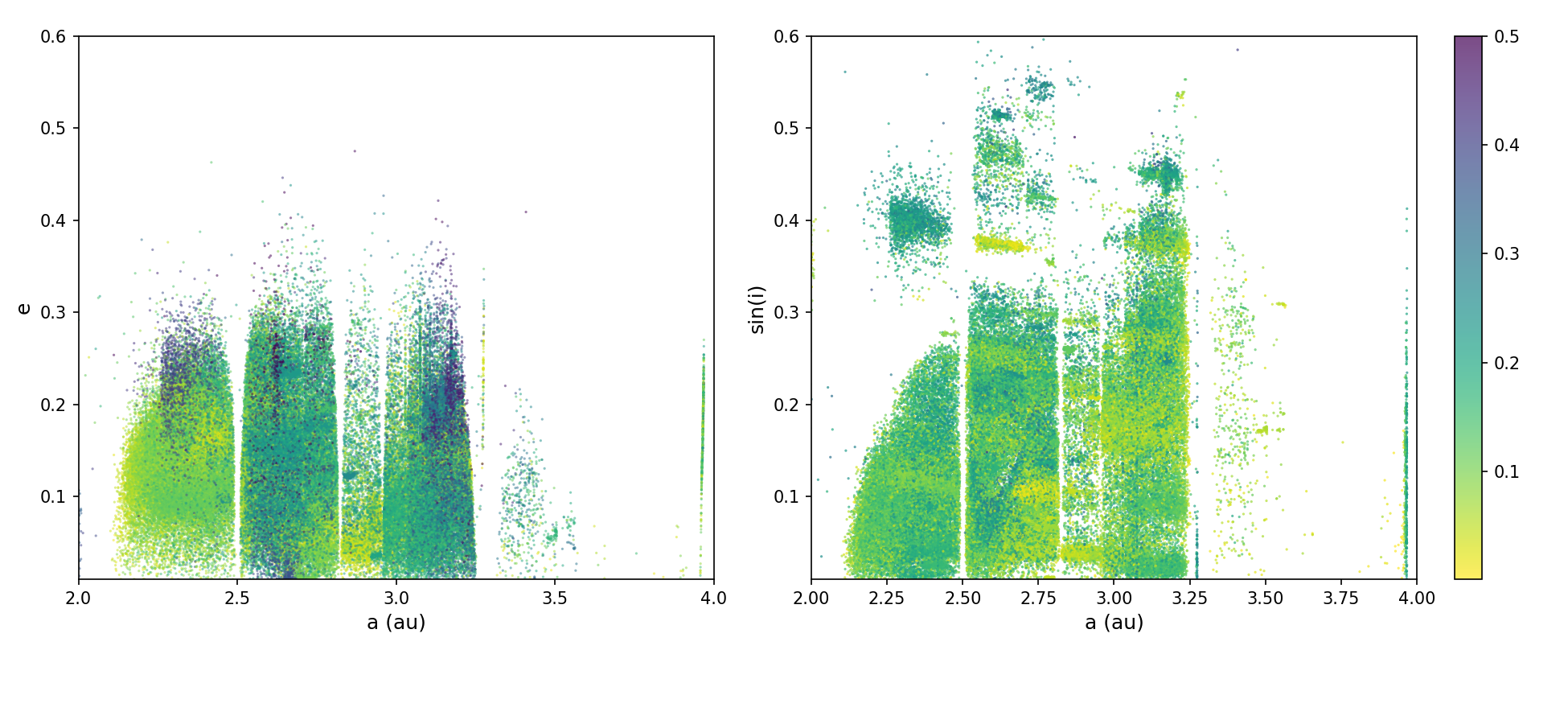}
\includegraphics[width=0.48\textwidth]{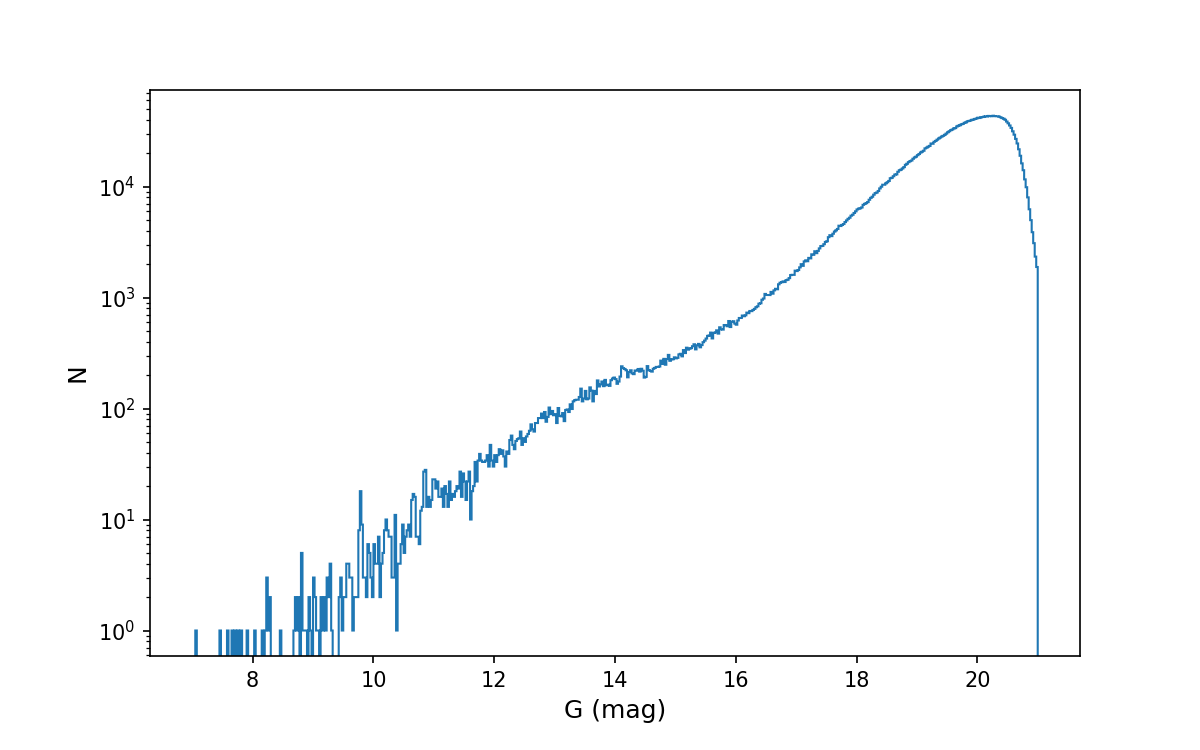}
\includegraphics[width=0.48\textwidth]{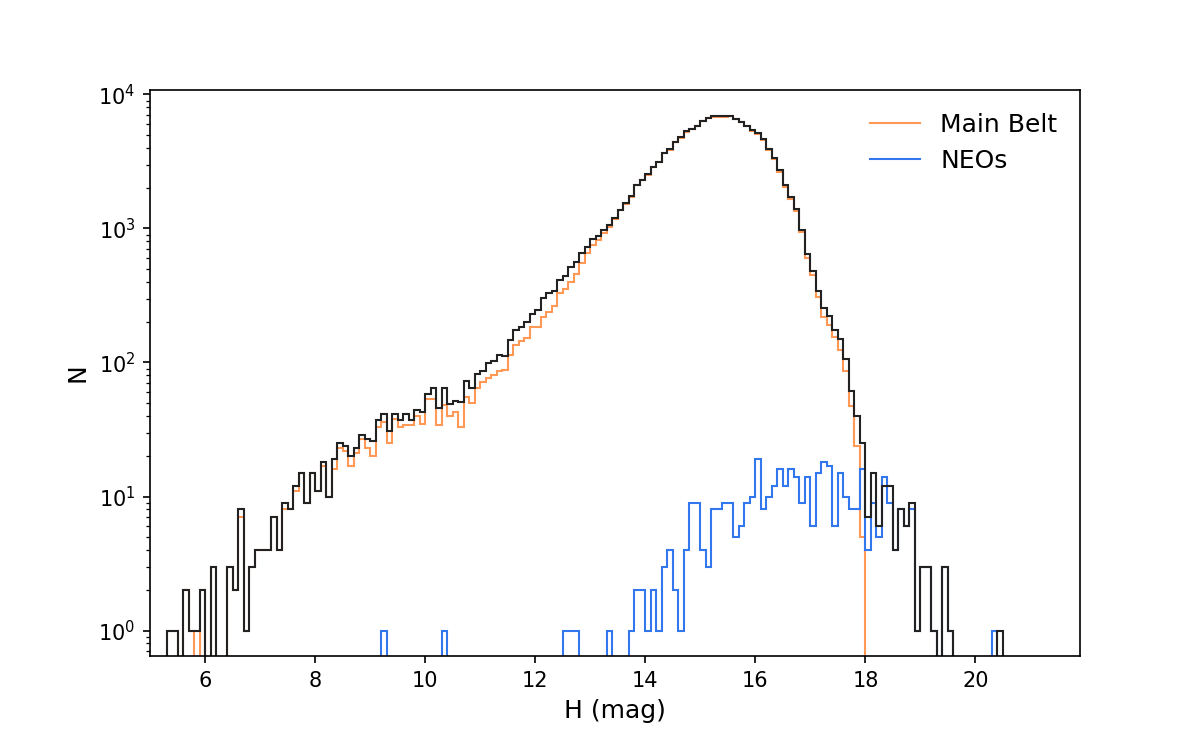}
\caption{
Visual portfolio of some data properties for asteroids in \gdrthree. In the top panel, we show the distribution on the semi-major axis of objects belonging to the main populations (eight Centaurs are excluded). The bin size is $1/30$~au in the left part and $1/§$~au for TNOs. Middle panel: Distribution of main-belt asteroids in the proper elements as provided by Astdys on the $a,e$ (left) and $a,sin(i)$ planes (right). Colours represent $sin(i)$ and $e$, respectively. The left bottom panel presents the distribution of the published G magnitude per transit. The cut imposed at $G=21$ is visible. The right bottom panel shows that the distribution of the H magnitudes of the objects, as provided by the Minor Planet Center, is strongly dominated by main-belt asteroids, with a contribution from NEOs for the faintest sources.}
\label{F:portfolio}
\end{figure*}

\begin{figure}
\centering
\includegraphics[trim= 32 0 50 0, clip,width=0.5\textwidth]{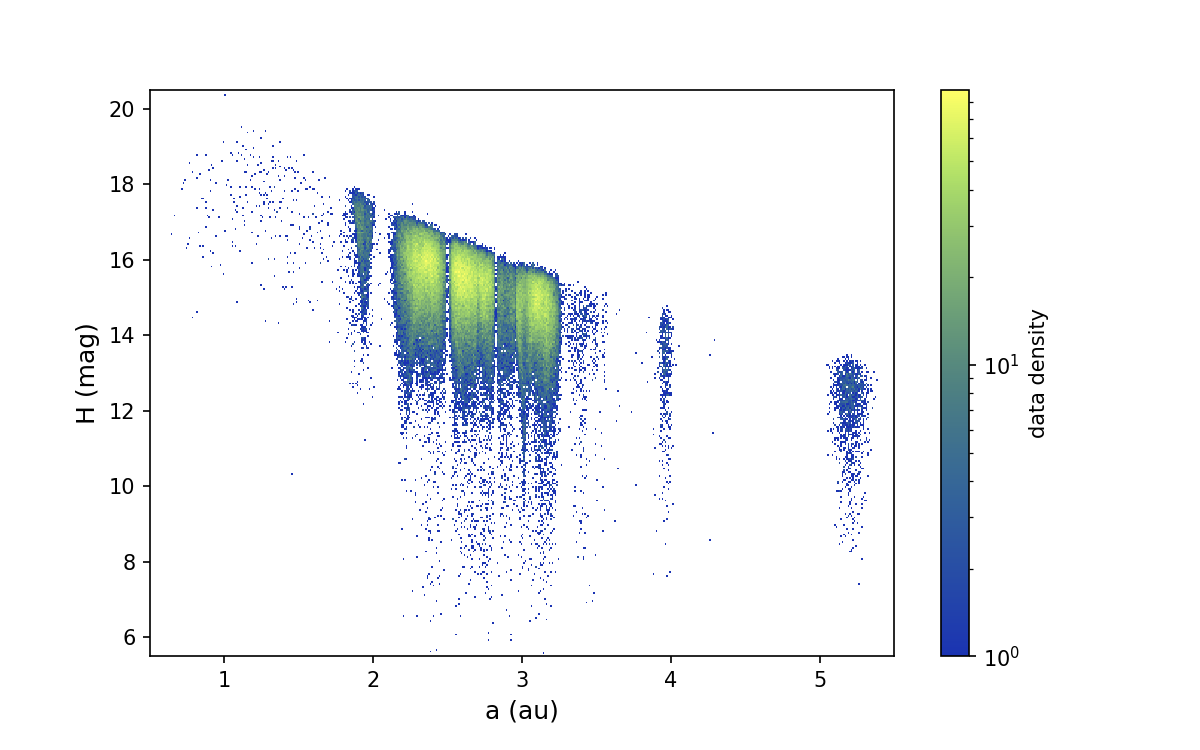}
\caption{
Distribution of H magnitudes for asteroids in the semi-major axis range from NEOs to Trojans. The trend followed by the detection limit is visible. In the mid-belt at 2.8~au, the limit G$\sim$21 corresponds to H$\sim$16.5. 
}  
\label{F:H_limit}
\end{figure}

{In comparison to} DR2, \gdrthree is more than ten times richer in terms of objects, a factor {of} $>$1.5 longer in time span, and provides a more complete data set. A summary of the published data is provided in Tables \ref{tab:DR3_summary_obs} and \ref{tab:DR3_SSO_type}. Some general properties are illustrated by Figs.~\ref{F:portfolio} and \ref{F:H_limit}. All published data concern small SSOs, with the following main categories (Table \ref{tab:DR3_SSO_type}): main-belt asteroids (MBAs), near-Earth objects (NEOs), outer Solar System populations, unmatched (unidentified) moving objects, and planetary satellites.

\renewcommand{\arraystretch}{1.5}
\begin{table}[h!]
    \centering
 \parbox{7.5cm}{\caption{SSO data summary in DR3 at CCD level, at transit level, or per object. See text for explanations.}
\begin{tabular}{lcc}
     \toprule
        Data type & CCD-level & transits  \\
     \midrule
        astrometry & 23~336~467 & 3~214~776 \\
        G-band photometry &  & 3~069~170\\
    \bottomrule
      & \multicolumn{2}{c}{number of objects}  \\
    \midrule
            reflectance spectra & \multicolumn{2}{c}{60~518} \\
        orbits &  \multicolumn{2}{c}{154~741} \\
    \bottomrule
\end{tabular}       \label{tab:DR3_summary_obs}}
\end{table}

\begin{table}[h!]
    \centering
 \parbox{7.5cm}{\caption{Object types in DR3.}  
    \begin{tabular}{lr}
    \toprule
       Object type & \multicolumn{1}{c}{number of objects}  \\
    \midrule
    Atira           &   1\\
    Aten            &   43\\
    Apollo          &   230\\
    Amor            &   173\\
    Mars Crossers            &   1550\\
    Inner Main Belt            &   3305\\
    Main Belt            &   144~975\\
    Outer Main Belt            &   4940\\
    Jupiter Trojans            &   1550\\
    Centaurs            &   8\\
    TNOs           &   24\\
    Others            &   2  \\
    \midrule
        Total asteroids & 156~801\\
        Unmatched moving objects &  1~320 \\
        Planetary satellites & 31\\
    \bottomrule
        Total  & 158~152\\
    \bottomrule
    \noalign{\smallskip}
    \end{tabular}
  \label{tab:DR3_SSO_type}}
\end{table}


\section{Data processing for the Solar System}
\label{S:SSO_processing}

All data products are the result of the data treatment pipeline developed by scientists of the \gaia Data Processing and Analysis Consortium (DPAC), implemented in the computing facilities of the French space agency (CNES). Strict qualification and validation protocols have been followed to ensure correct results and full consistency with the other subsystems of DPAC. 

The general structure of the data processing is illustrated in Fig.~\ref{F:processing}. We describe it below by focusing on the aspects that more strongly define the properties of SSO data found in \gdrthree to allow potential users in the scientific community to become familiar with the data and their exploitation. Orbits have not been computed by the core pipeline, but by an offline procedure (Sect.~\ref{S:orbit_processing}) with a dedicated validation.

The core of the input data comes from the intermediate data updating (IDU) and from the astrometric global iterative solution (AGIS), wich are two components of the general data processing of \gaia. Details about their implementation principles can be found in \citet{DR1-DPACP-7, EDR3-DPACP-73, torraGaiaEarlyData2020} and \citet{EDR3-DPACP-128}.

IDU in particular provides all the information required to reconstruct the position and brightness of a source in its window, starting from the image parameter determination (IPD; i.e. the determination of the centroid of the signal). The data exploited for each transit are
\begin{itemize}
\item the value of the magnitude determined on board by the video-processing unit (VPU), as a preliminary estimation.
\item The window class generated by the VPU. It is possible to reconstruct the window geometry for each strip based on this parameter.
\item The along-scan and across-scan window coordinates are related to the timing and position of the window in the 
  corresponding CCD (reconstructed by IDU).
\item The data also include the list of along-scan centroids $x_s$ (IDU),
\item the list of the fluxes $f_s$ (photo-electrons per second, by IDU),
\item the list of the across-scan centroids $y_s$ for bi-dimensional windows (IDU),
\item and a list of flags, generated by IDU, which describes the quality of the IPD output and the encountered issues, such as the formal errors, the goodness-of-fit (GoF) to the PSF/LSF model, the presence of secondary peaks in the window, or the background estimation.
\end{itemize}

Along with ($x_s$, $y_s$), the window reference system coordinates (WRS), the corresponding epochs are present, expressed in the internal on-board mission time line \citep[OBMT; for its definition, see][]{DR1-DPACP-8}. AGIS provides all the information required to calibrate the astrometry and to transform the WRS coordinates into BCRS. A particularly relevant input is a pre-computed list of transits that provides a first identification of observations associated with SSOs. We
illustrate this in the following section.

\subsection{Selection of  sources}
\label{S:input_list}

Two main reasons prevent an automated selection of the SSOs within their dedicated pipeline: first, the huge volume of data in the automated pipeline, of which only a small fraction must be selected and exploited; and second, the need to optimise the extraction of SSOs while avoiding contaminants as much as possible. 

The SSO selection must therefore be performed by a pre-processor and is then exploited at the ingestion of the pipeline to select only the corresponding IDU and AGIS data. We adopted two approaches. The first approach is the most relevant for processing the bulk of the SSO sample. It is devoted to objects that can be identified by a direct match to their predicted positions. The second approach retrieves sequences (bundles) of moving-object detections that are not matched to known SSOs.  

For the first selection, several criteria were applied to obtain a subset of objects that was about ten times larger than in the previous \gaia release (DR2).
\begin{itemize}
\item We aimed for between 100\,000 and 150\,000 objects in the input list that were representatives of all the broad categories of asteroids were sought, such as NEOs, MBAs, Jupiter Trojans, and {transneptunian objects} (TNOs). Transits of several planetary satellites were also included. Comets alone are not present in \gdrthree.
\item A transit was not selected if a star, another SSO, or a contaminant generated by a bright star was found too close to the object during its observation by \gaia.
\item Each selected SSO had to be detected on at least eight transits over the 34 months covered by the \gdrthree data.
\end{itemize}

Known SSOs were searched for by matching all the observed transits \citep[from the output of the initial data treatment, IDT; ][]{DR1-DPACP-7} to computed transits of {SSOs} over the \gdrthree time span. The computed positions as seen by \gaia were obtained by the available information on the position of \gaia in space (the satellite orbit), the scanning law, and a numerical integration of the SSO motion. This last procedure starts from the osculating elements and osculating epoch given in the \texttt{astorb}\footnote{\url{https://asteroid.lowell.edu/main/astorb/}} database \citep{bowellPublicDomainAsteroidOrbit1993}. As the selection had to be finalised well before the operations of the SSO pipeline, the \texttt{astorb} version we used is that of 13 December 2017.

The SSO cross--matching proceeds in two steps: first by the crossing time (required match within 0.1~s), and then by the sky coordinates within a window of 1\farcs5.
After the list was filtered for possible contaminants, the final input selection had 3~513~248 transits for 156~837 known asteroids. To meet the criteria mentioned above for \gdrthree, a search for numbered asteroids was sufficient. Unnumbered asteroids do not appear in this release, with the exception mentioned further below. 

The search for observations of planetary satellites observable with \gaia has proceeded like for the asteroids. Their passages were first predicted using the ephemeris of the satellites provided by V. Lainey from IMCCE. Due to the angular proximity of the planet, the number of contaminants is much higher than average, and the selection of a single match may become somewhat arbitrary at times. This is the case in particular for the Galilean satellites. Unlike for the asteroids, there has been no filtering for the number of detected transits.
Out of the 44 possible satellites in the appropriate magnitude range (2 for Mars, 18 for Jupiter, 14 for Saturn, 7 for Uranus, and 3 for Neptune), 42 have been matched to at least one IDT transit. The two missing objects are the Jovian satellites Amalthea and Thebe, whose signal is dominated by the noise due to their proximity to Jupiter.

The astrometry of the potentially unknown moving sources constitutes a smaller data set that appears in \gdrthree for the first time. As mentioned above, their selection is based on an independent search that heavily relies on the along-scan motion ($V_{\rm AL}$) that is reliably provided by IDT \citep[see Sect. 5.2 in ][]{DR1-DPACP-7} only after 1 December 2016. Therefore, for \gdrthree, there were altogether only six months of data that could be exploited for this search. 

The selection recovers the position of sources above a chosen AL motion threshold ($>1.5$~mas~s$^{-1}$) provided they can be paired with at least one other after an interval of time corresponding to transits in the preceding-following field of view (or following-preceding), consecutive or separated by a satellite rotation (6 hours; Sect.~\ref{S:genprop}). The pairing is validated only if the estimated $V_{\rm AL}$ based on the two positions and the two $V_{\rm AL}$ values of the potential pair are also compatible. In a second step, all the pairs are examined in succession to connect to other pairs that may belong to the same source and are thus appended to form  longer chains, with three or more observations in a sequence found in consecutive IDT runs. In chains of four or more transits, the number of constraints is so large that the link of the detections to a single source is certain.

The acceptance thresholds and, at the end of the process, the reliability of the detection, {depend} on the number of consecutive observations. While sources with only two or three transits dominate the sample, approximately 10\% of the sources have four or more transits, the largest number of transits is $42$. In the process, all the known asteroids (whose ephemerides are computed based on the \texttt{astorb} file mentioned above) are also found and discarded. 

The final input list of unidentified SSO for DR3 comprises 4522 transits of unmatched asteroids, corresponding to 1531 groups of chained transits. Unmatched sources do not necessarily correspond to new asteroids. This sample includes asteroids that were not in the orbital data at the end of 2017 because their orbit was not available at that time or because it was too poor to have a successful position match in the window of 1\farcs5. This is a vivid illustration that the population of asteroids is not sharply divided between known and unknown SSOs. These are just the two boundaries of a continuous spectrum of knowledge. A further exploration of this selected sample of unmatched objects, based on an updated version of the orbit data base, is provided in Sect.~\ref{S:unmatched_validation}.

\subsection{Identification of Solar System objects}
\label{S:SSO_identification}

The first step of the SSO pipeline is identifying each source entering the processing based on the average transit position provided by the initial data treatment \citep[Sect. 6.4 in][]{DR1-DPACP-7}. This is basically done by matching the approximate source position ($\sim$0\farcs1) to the ephemerides of {SSOs}.
This procedure essentially duplicates the one adopted to build the input list, but with the fundamental difference that most of the potential sources that are not SSOs have not been ingested in the pipeline. The new identification is thus performed on a rather clean data set, in which sources that are not SSOs, in overwhelming number at the input list selection, are now reduced to a minimum.  Another relevant difference {is as follows: at this level,} no attempt is made to {identify unmatched sources or to rebuild their bundles.} 

As time passes, the inventory of known SSOs grows. Since the time of the \gaia launch, hundreds of thousands of new asteroids have been discovered or their orbits were considerably improved, bringing the total number of known asteroids with an orbit to more than 1\,170\,000 as early as 2022. A database of precomputed ephemerides of all known SSOs is regularly updated for the need of the processing of \gaia data. Time-dependent \gaia-centric positions of SSOs are arranged by using a HEALPix spatial index \citep{2005ApJ...622..759G}, a grid resolution $N_{\rm side} = 2^{10}$, and a time resolution adapted to each object. They are stored into an Apache Cassandra database\footnote{\url{https://cassandra.apache.org/}}. During each processing cycle, packets of transits are cross-matched with known SSOs extracted from the database. For this purpose, the pairs \{HEALPix, Epoch\} of each transit are used to extract a sample of zero to a few dozen SSO candidates. Their \gaia-centric accurate positions at the transit epoch are then recomputed by means of two-body numerical integration perturbed by $n$ bodies, providing equatorial coordinates that can be directly compared to measured transit coordinates.

The first criterion of candidate selection relies on the accuracy of SSO orbits. For each target, the ephemeris uncertainty at the epoch of each transit is computed based on the 1$\sigma$ RMS ($\sigma_o$) of its orbit, again adopting the \texttt{astorb} database. A candidate is retained if its current ephemeris uncertainty (CEU) fulfils  the condition
\begin{equation} 
 \mathrm{CEU} = \sigma_o + (t-t_0)\,\dot{\sigma}_o < \epsilon, 
\end{equation} 
where $t$ is the observation epoch of transit, $t_0$ is the reference epoch of the orbital elements, $\dot{\sigma}_o$ is the rate of change of $\sigma_o$, and $\epsilon$ is a given threshold. The adopted value $\epsilon = 10$\arcsec\ leads to the rejection of all SSOs with uncertain orbits, which could lead to spurious identifications.

The second criterion takes into account the relative positions of SSO candidates compared to the recorded transit position, as described by \citet{2011A&A...527A.126P} in their cross-correlation algorithm. The SSO position is projected onto a 2D plane centred on the transit position, so that the relative coordinates of the SSO are $x = d$, $y = 0$, where $d$ is the angular distance between the two sources calculated by the haversine function,
\begin{equation}  
d = 2 \arcsin\sqrt{\sin^2\left(\frac{\delta_s - \delta_t}{2}\right) + \sin^2\left(\frac{\alpha_s - \alpha_t}{2}\right)\cos\delta_t\cos\delta_s}, 
\end{equation} 
where $\alpha_t, \delta_t$ and $\alpha_s, \delta_s$ are the equatorial coordinates of the transit and the SSO candidate, respectively. In this plane, an SSO candidate is retained if its coordinates satisfy the condition
\begin{equation}  
\frac{d}{\sigma_{x_c}\sqrt{1-(\rho_c\sigma_{x_c}\sigma_{y_c})^2}} \leq k 
,\end{equation} 
where $k = 3.43935$ is the 2D completeness value for a 3$\sigma$ criterion (e.g. 99.7\%), and where $\sigma_{x_c} = \sqrt{\sigma_{x_t}^2 + \sigma_{x_s}^2}$ and $\rho_c\sigma_{x_c}\sigma_{y_c} = \rho_t\sigma_{x_t}\sigma_{y_t} + \rho_s\sigma_{x_s}\sigma_{y_s}$ represent the uncertainties on the positions of the transit ($t$) and the SSO candidate ($s$) expressed by their covariance matrix, assuming Gaussian uncertainties \citep[see Appendix A of][]{2011A&A...527A.126P}. For SSOs, the positional uncertainty is taken as the current ephemeris uncertainty, for instance, $\sigma_{\alpha_s} = \sigma_{\delta_s} = \mathrm{CEU}$. For transits, the positional uncertainty is fixed to $0.5$\arcsec, providing a large margin over the formal uncertainty of IDT (about $0\farcs06$ in each coordinate).

A third criterion based on the difference in magnitude between the observed transit and SSO candidates might be used in principle to distinguish between different candidates. Nevertheless, the uncertainty on the predicted apparent magnitudes of many SSOs can reach values of about 1~mag or more because their albedos, light-scattering properties, and shapes are poorly known. No magnitude-based criterion was used in \gdrthree. A consequence of this choice is that a faint source very close to an asteroid (e.g., a possible satellite) might in principle be matched to the asteroid itself. Despite the fact that in \gdrthree close couples are filtered at the input list generation, we cannot totally exclude that such double detections exist. 

While planetary satellites have an identification based on the computation of their own ephemeris, no specific procedure is implemented to identify satellites of asteroids. As they share very similar coordinates, the satellites and the main body of the system can be given the same identifier. For instance, in \gdrthree, the dwarf planet (134340)~Pluto and its main satellite Charon are both identified as ``(134340)~Pluto''. 

If more than one SSO candidate satisfies the identification criteria, there is no obvious method to identify the correct object at this step of the processing. The best candidate is thus selected by calculating the quadratic distances between each observed transit and the corresponding SSO candidates. When there is more than one possible choice, the object minimising the distance above is selected. With the grid resolution chosen, this scenario is fortunately very unlikely. 

The validation of the identification process has shown that the rate of correct identifications is very close to $100\%$, with an uncertainty smaller than $1\%$. This mainly comes from uncertainties on the positions of some SSO candidates and from the presence of some unfiltered contaminants. The identification process successfully recovers 99.97$\%$ of the transits in the input list. The small fraction of non--identified transits, negligible in practical terms, is mostly due to minor differences in the adopted ephemerides that come from differences between osculating elements that are used to build the input list and those that are used to compute processing ephemerides. 

With the identification, a computation of the SSO ephemerides is performed for all the known sources. Some ancillary data, such as the distance and the apparent motion on the sky are computed, stored in an appropriate table, and are propagated to the pipeline where they remain available to other processing modules.

\subsection{Raw centroid processing}
\label{S:ccd_processing}

The distribution of the collected photo-electrons inside a pixel
window contains the raw information about the location of the
source. The determination of the average position of any source from this distribution, the ``centroid'', is 
carried out by IDU by means of a fit to a suitable model \citep{EDR3-DPACP-73}. The model is based on two assumptions: (1) the source is point-like, and (2) its image on the focal plane moves in the
along-scan direction at a rate that exactly matches the charge transfer in the CCD. 
The free parameters of the model are the mean position of the source (centroid) and its intensity (flux). 

In the case of SSOs,  hypothesis (1) is fulfilled in the very large majority of observing circumstances because only larger objects (order of thousands) can cause detectable signal distortions with respect to the point spread function (PSF) of the \textit{Gaia} instrument. Conversely, assumption (2) is in general not valid for SSOs because the proper motion of SSOs with respect to stars produces a systematic shift of the photo-electron distribution with respect to the scan motion rate. 

While a pure shift without smearing entails only a different but correct value of the centroid, the smearing introduces systematic biases both in centroid and flux determination because the signal is truncated at the window edge. 
The magnitude of the bias increases with the distance of the centroid from the centre of the window and depends on the velocity of the source. Details are given in the \href{https://gea.esac.esa.int/archive/documentation/GDR3/Catalogue\_consolidation/chap\_cu9val/}{online documentation}.

In order to mitigate the impact of the centroiding bias on the final data quality, IDU positions are rejected that due to their proximity to the window edges are expected to have a bias exceeding the formal error on centroiding. This filtering is performed just before the astrometric processing module discussed in the next section. As the effect of the shift accumulates along the transit, the last AF columns have a stronger rejection probability. Therefore, the nine AF positions are preserved for only 4\% of the transits.

\subsection{Processing of astrometry}
\label{S:astrometry_processing}

One of the most critical modules of the processing {pipeline} is devoted to process the astrometry, to determine {the} astrometric uncertainties, and to filter {the} positions that appear to be outliers. We describe the different procedures adopted for these tasks below.

\subsubsection{Coordinate transformations} \label{sect:coordinates}

We consider here the coordinates associated with a single CCD position as produced by IDU for an observed source in the AF instrument. These coordinates are expressed in the window reference aystem (WRS), and provide the pixel coordinate of the centroid of the SSO inside the transmitted pixel window and the OBMT reference time of the transit. 

As a first step, the epoch of the crossing of a conventional fiducial line on the CCD \citep{DR1-DPACP-14} is computed. This corresponds to the exact timing of mid-exposure, and is dependent on the location of the photocentre of the SSO inside the window, on the size of the window itself, the location of the window in the focal plane at the time of read-out of its reference pixel, and on some more technical aspects such as the binning and gating strategy.  The OBMT timing of the crossing of the fiducial line is effectively the AL coordinate in the WRS.

The WRS coordinates are then transformed to angular coordinates in the scanning reference aystem (SRS), whose axes are aligned to the AL and AC directions, with the origin at the centre of the focal plane \citep[for an overview of these reference systems, see Fig. 15 in][]{DR1-DPACP-7}. In this step, the geometric calibration of the focal plane is applied. In the processing cycle producing \gdrthree, the geometric calibration is among others dependent on the source colour, expressed by an effective wave number $\nu_\mathrm{eff}$. After having performed some tests, it was decided that assuming a solar spectrum for all SSOs (corresponding to $\nu_\mathrm{eff}=0.001561$~nm$^{-1}$) was an acceptable approximation.

A further transformation converts positions from the SRS reference system to the centre of mass reference system (CoMRS), non-rotating, with the origin at the centre of mass of \gaia. A last transformation produces the positions in the BCRS. In this step, the relativistic stellar aberration is removed; in other terms, the effect of the orbital motion of \gaia is suppressed. The outcomes are the position in equatorial coordinates, as seen from the centre of mass of \textit{Gaia} and the associated TCB. 

We stress again that due to the scanning motion of \gaia, the TCB of an observation is directly linked to the position of the source on the focal plane in the AL direction. The uncertainty on the provided TCB is $\sim$1~ms, during which an asteroid moves by no more than $\sim$200 $\mu$as (a very high value that is only reached by some near-Earth Objects). This is negligible with respect to the error budget illustrated below.

\subsubsection{Astrometric uncertainties} \label{sect:cu4sso_error_astrom_systrand}

We consider a simplified error model that separates uncertainty sources that are uncorrelated across a transit from one AF CCD to the next (random component) to uncertainties that are considered not to vary along a transit (systematic component). This scheme is represented by the complete covariance matrix $W$ of the transit,
\begin{equation}\label{eq:cu4sso_error_astrom_systrand}
    \vec{W} = \left( \begin{array}{cccc}
       \vec{W}_1 & \vec{0}   & \cdots & \vec{0} \\
       \vec{0}   & \vec{W}_2 & \cdots & \vec{0} \\
       \vdots      & \vdots      & \ddots & \vdots    \\
       \vec{0}   & \vec{0}   & \cdots & \vec{W}_9
     \end{array} \right) +  \left( \begin{array}{cccc}
       \vec{W}_\text{s} & \vec{W}_\text{s} & \cdots & \vec{W}_\text{s} \\
       \vec{W}_\text{s} & \vec{W}_\text{s} & \cdots & \vec{W}_\text{s} \\
       \vdots             & \vdots             & \ddots & \vdots    \\
       \vec{W}_\text{s} & \vec{W}_\text{s} & \cdots & \vec{W}_\text{s}
     \end{array} \right),
\end{equation}
where $\vec{W}_n$ is the covariance matrix of the right ascension and declination of the AF$n$ position, which we call
the random uncertainty of the AF$n$ position, $\vec{W}_\text{s}$ is a constant covariance matrix throughout the transit, which we call the systematic uncertainty of the transit, and $\vec{0}$ is a $2\times2$ matrix of zeros. 

The random component {incorporates} the uncertainty from the centroiding, a term that we call {\sl \textup{excess noise}}, and a contribution from the attitude. Readers familiar with the general astrometric processing by $\gaia$ should note that this is conceptually similar to the excess noise defined in AGIS \citep{EDR3-DPACP-128}, but its formulation is different. These terms are quadratically summed to obtain the total random uncertainty.

The uncertainty on the centroiding is initially provided as uncorrelated errors in AL and AC, but after transformation to the equatorial coordinate system, the corresponding uncertainties on right ascension and declination become highly correlated. However, taking the correlation into account, the user can recover the precise AL component of the uncertainty.
Whereas uncertainties in right ascension and declination are typically about 500~mas, the real uncertainty in AL is often smaller than 1\,mas.

In AL, uncertainties are at mas level and show the extreme precision of \gaia. In AC, the situation is more complex because we lack knowledge about a precise position. For sources with $G>13$, all pixels are binned to a single sample, and the position given corresponds to the centre of the transmitted window.  The transmitted window is determined by the on-board software at the beginning of the transit, so that the object is in one of the two central pixels, and is propagated to the next AFs such as to keep a non-moving object in the centre. Due to its motion, the SSO drifts away from the centre of the window. Therefore, for AF CCDs that are reached by the signal later on during the transit, the SSO can be anywhere in the window rather than in one of the central {pixels}. By assuming the dispersion of a rectangular distribution over the complete transmitted window as value for the uncertainty, it is clear that for early AFs, this is an overestimation. However, this approximation of the AC error model is not expected to impact the exploitation of the astrometry (e.g. for orbit computation) as the AC uncertainties remain $\sim$2--3 orders of magnitude larger than the AL uncertainties (further details are provided in Sect.~\ref{S:zigzag}). 

For objects $G<13$, 2D windows were downlinked to Earth, and a 2D centroid fitting was possible. In this case, the uncertainty in the AC direction is comparable to (but still larger than) the uncertainty in the AL direction.

The second random component to the astrometric uncertainty, the excess noise, was determined by analysing the orbit post-fit residuals of asteroids. It essentially affects bright objects ($G<12$), so that its origin could be linked to partially resolved shape effects (Sect.~\ref{S:astrometry_performance}). For \gdrthree, this additional uncertainty is given by
\begin{equation}
  \begin{cases}
    \epsilon_\mathrm{ex,AL} = 0.72 \times e^{-0.63578\,(G - 10)}\ \mbox{ mas}, & \mbox{in AL},   \\
    \epsilon_\mathrm{ex,AC} = 1.5 \times e^{-0.32673\,(G - 10)}\ \mbox{ mas}, & \mbox{in AC},
  \end{cases}
\end{equation}
where $G$ is the preliminary magnitude available {from} the pipeline input.

Finally, the uncertainty on the attitude associated with the AGIS solution is derived by analysing the dispersion of epoch positions in a magnitude range in which centroiding errors are negligible. It contributes both to the random and systematic errors. The values in Table~\ref{tab:cu4sso_error_astrom_} are the default ones, corresponding to regular period of operations (occasional time intervals with larger uncertainties are present).

\begin{figure}
\centering
\includegraphics[trim= 20 0 50 0, clip,width=0.48\textwidth]{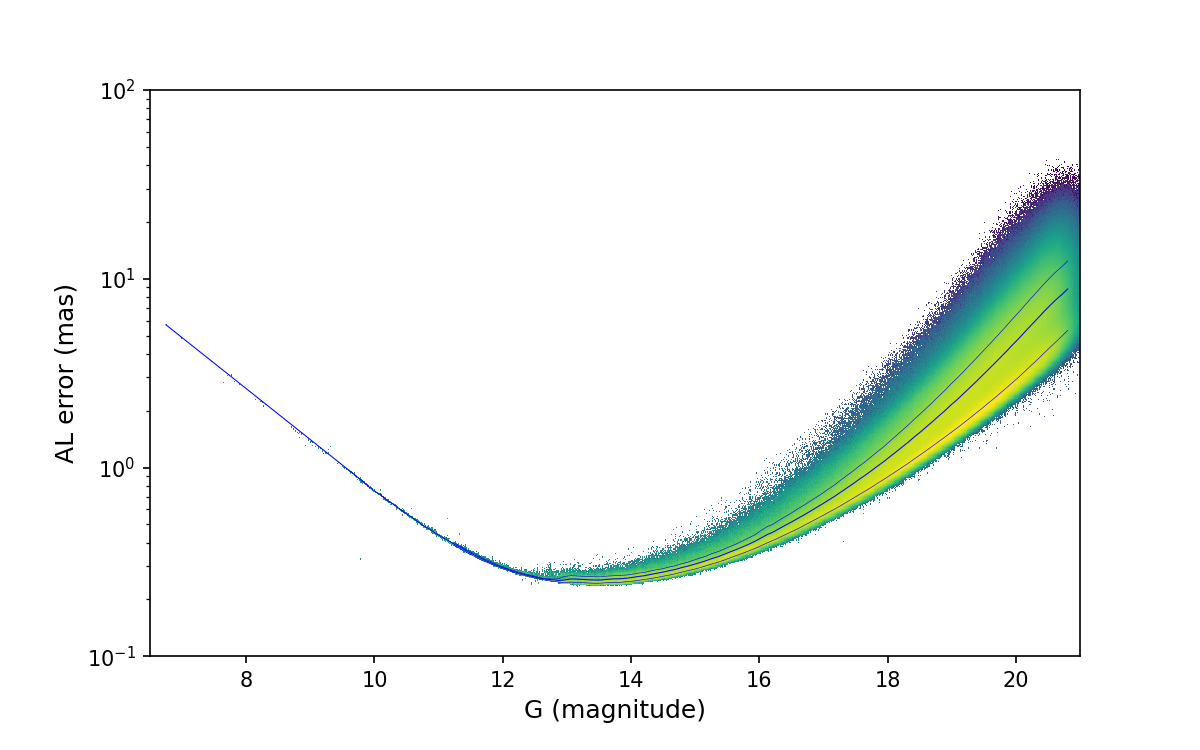}
\caption{Error model in the AL direction for the SSO astrometry in \gdrthree as a function of the G magnitude. The total error is represented as given by the squared sum of the random and the systematic component. The colour represents the data density (yellow, lighter colour: higher density). The thick line and the two thin lines on each side are the quantiles corresponding to the mean and the 1-$\sigma$ level.
}
\label{F:errormodel}
\end{figure}

The resulting uncertainties for all the observations published {in \gdrthree} are distributed as illustrated in Fig.~\ref{F:errormodel} as a function of the G magnitude for the AL direction. The dominating excess noise in the branch of G$<$12 and the average uncertainties of always $<$1~mas for 10$<$G$<$18 and $<$10~mas at G$\sim$21 are striking. We show below in more detail how the {post-fit residuals to orbits} represent this quality of the data. 

\begin{table}[h]
  \caption{Default uncertainties from the AGIS attitude (mas). \label{tab:cu4sso_error_astrom_}}
  \centering
  \begin{tabular}{l l l}
   \hline\hline
    & AL & AC \\
   \hline
    Contribution to the systematic uncertainty     &  0.051    &  2.3      \\
    Contribution to the random uncertainty         &  0.20   &  1.37    \\
   \hline
  \end{tabular}
\end{table}

\subsubsection{Filtering of positions}

\begin{figure*}[h]
\centering
\includegraphics[trim=0 0 0 0,clip,width=0.694\textwidth]{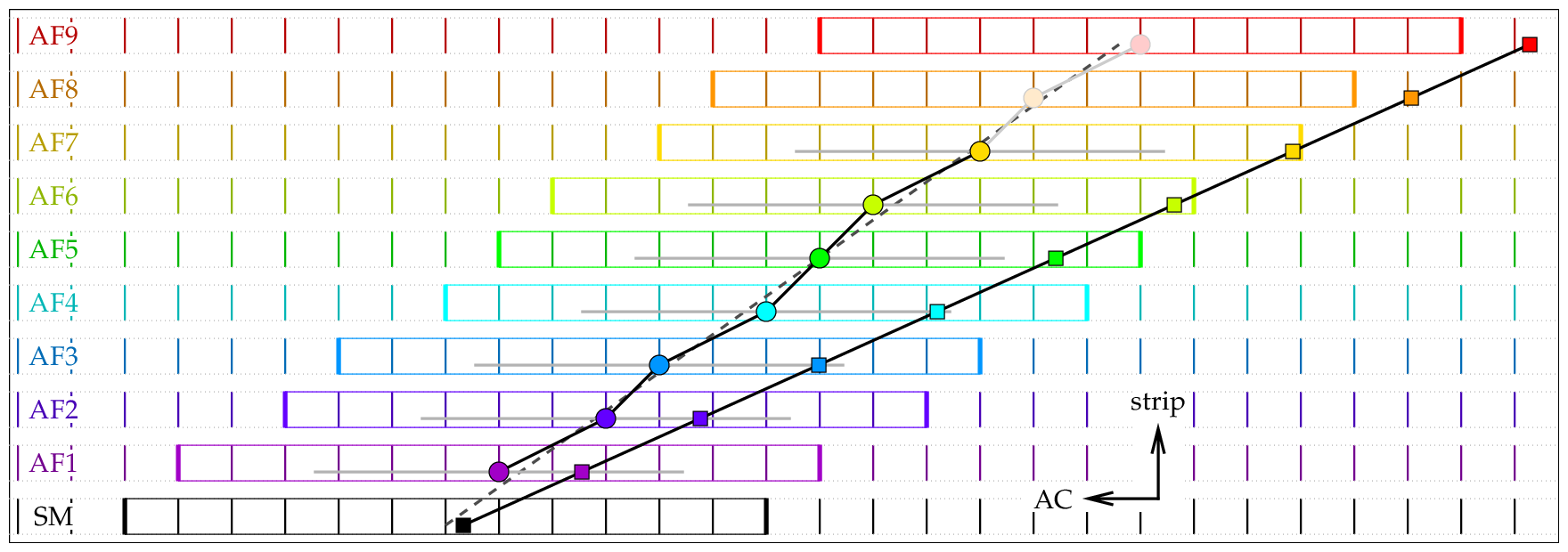}
\includegraphics[trim=0 0 0 0,clip,width=0.296\textwidth]{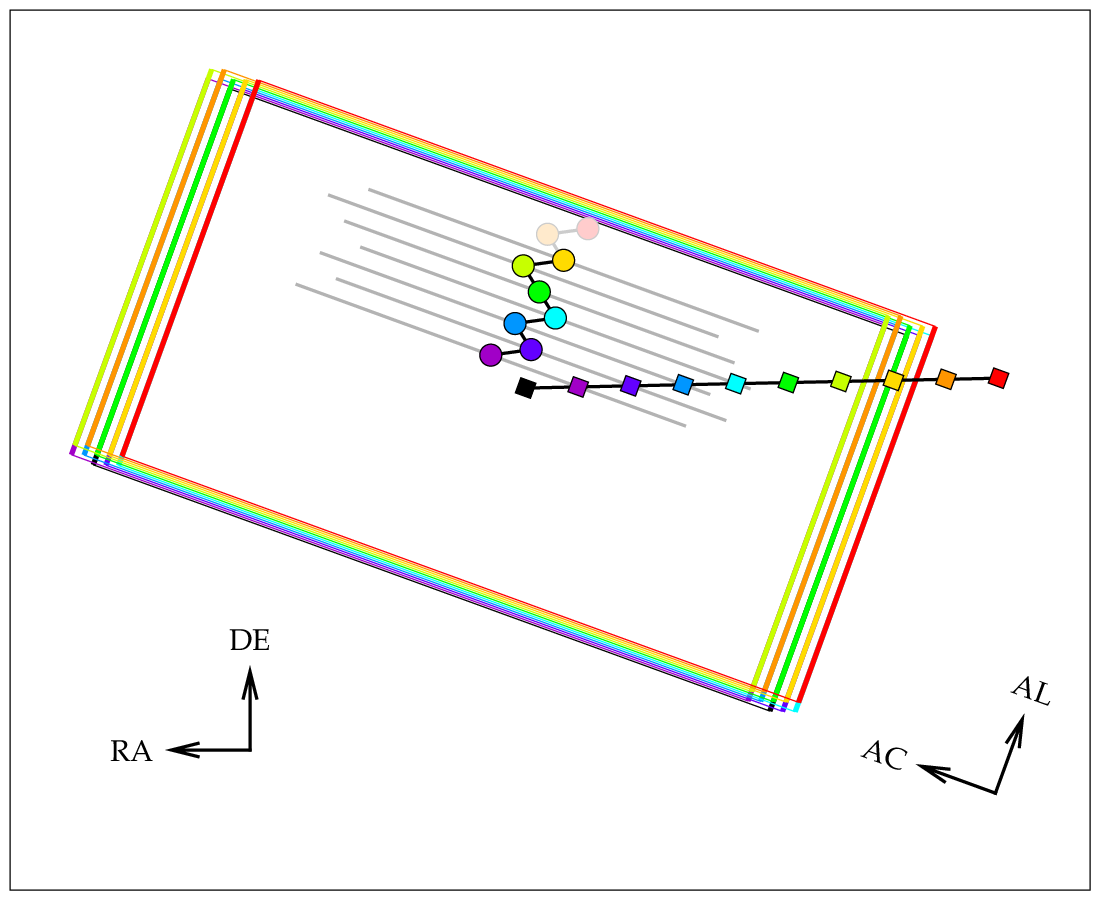}
\caption{General illustration of the motion of an asteroid on the focal plane (left panel) and how this is translated into a position by the on-board windowing strategy. The horizontal axis represents the AC direction, and the vertical coloured bars are the pixels in the AC direction. The vertical axis represents the different CCD strips (their separation is conventional in this scheme). The coloured squares connected on the black line represent the real positions of the asteroid in the different CCD strips of the SM and AF instrument.
The rectangular coloured boxes encompassing 12 AC pixels represent the assigned windows. The dashed black line represents the motion that a hypothetical star would have, starting from the same position as the asteroid in SM. The assignment of the windows in AF is such that their AC positions, corresponding to their {centres} (coloured circles), closely follow the dashed line. This AC position is transmitted to the ground. The horizontal light grey lines represent the uncertainties on the positions in AC, which correspond to the dispersion of a rectangular distribution over the assigned window. The right panel shows how this translates into positions {on the sky plane}. }
\label{F:onetransit}
\end{figure*}
Filtering the {outliers, which are most probably observations unrelated to SSOs, or data affected by anomalies, is} an important and delicate task that has been applied both at the level of individual positions (individual CCDs) and at the level of complete transits. The filter definition has required several iterations to optimize the rejection parameters and obtain the cleanest data set, minimising at the same time the number of rejected {genuinely} good positions.
The whole filtering procedure is described in detail in the documentation of \gdrthree, therefore we do not repeat it here. We just recall the broad test categories that have been implemented.
\begin{itemize}
    \item Test for anomalies at CCD level: {For example, samples eliminated or set to zero, attitude of poor quality, close proximity of stars, or missing data.} Two important rejections occur at this stage: all positions by the Sky Mapper (SM) CCD columns that are of lower quality are rejected, and all positions for objects fainter than $G=22$ beyond a reasonable range for accurate astrometry. 
    \item Test for the position uncertainty, within predefined limits (function of magnitude). This is based on the study of the error distribution (following the model illustrated above) and on the identification of {clear outliers.} 
    \item Test on residuals with respect to the fit of a linear motion during a transit. 
\end{itemize}

The two dominating rejection reasons (8.6$\%$ of the positions) are the lack of attitude data and unrealistically large or small uncertainties. 
All the criteria above are responsible for the elimination of a fraction of positions that is considered as outliers. An extensive analysis has shown that all transits with $\ge$2 remaining positions can be considered with high confidence as associated with a real SSO. Transits that at the end of the filtering have a single position are rejected.
At the end of the filtering process, 10.9$\%$ of the positions are rejected.

The astrometric processing is agnostic of the association of transits in bundles. It operates only at transit level and treats all transits individually and independently.

As in \gdrtwo, the asteroid astrometry at CCD level of \gdrthree is provided in the table \texttt{gaiadr3.sso\_observation} of the \gaia archive. It contains all the data required for its exploitation.

\subsubsection{Difference in the processing of astrometry with respect to DR2}

A single important difference should be considered by the data users that intend to exploit the astrometry at its full accuracy. In \gdrtwo, the relativistic light bending was eliminated from the observations by applying a correction that assumed the source to lie at an infinite distance. After consulting potential users, we decided not to apply any light bending correction in \textit{Gaia} DR3 and let the users include its effect in the models they use in their orbital fitting programs. Depending on the solar elongation, the difference of positions due to light bending can be about a few milliarseconds, which means that it is relevant to exploit the full accuracy.

Two others choices were made that impact the data volume, but not {the usage of data}. 
First, bright SSOs have residuals higher than the expectation based on formal uncertainties (see Eq.~\ref{eq:cu4sso_error_astrom_systrand}). In \gdrtwo, we took the simple approach to exclude all SSOs with G$<$10. For \gdrthree, the higher uncertainty is introduced in the error model by the excess noise (Sect. \ref{sect:cu4sso_error_astrom_systrand}), and bright objects are preserved in the output.
Finally, a small number of sample windows are truncated. It is not straightforward to provide a quantitative estimate because 0.5\% of the transits has a truncation flag set. Of these, only $\sim$10\% are probably really truncated, however. Their positions were discarded in \gdrtwo, but are preserved in \gdrthree\ because a dedicated study has shown that the quality of their astrometry is not degraded. 

\subsubsection{Interpretation of positions during a transit}
\label{S:zigzag}

The users of asteroid astrometry will see that the positions provided along a transit for an asteroid in general follow a zigzag pattern on the sky whose average displacement is related, but does not correspond exactly, to the asteroid proper motion. In the scheme of Fig.~\ref{F:onetransit} (left panel), we explain this effect. It results from the on-board windowing strategy. 

When a source enters the focal plane and is detected in a Sky Mapper CCD (SM),{ a window is assigned to the source} such that it is centred in the window within one pixel. In the case of a star, its image drifts across the subsequent window strips due to the satellite spin (dashed line). Due to precession, it will also drift in the AC direction by a few dozen pixels over the entire transit. Its trajectory is therefore not exactly vertical in the plot. 

The windows assigned in the SM strips are propagated through the AF strips assuming that the motion to be tracked corresponds to that of a star.  Since window shifts can only be performed by integer pixels, the shifts from one strip to the next will alternate between two adjacent integer values, causing a kind of zigzag motion of the assigned windows in AC direction. 

In general (i.e. {for $G>13$}), windows are fully binned in AC direction, and the only astrometric information we have in AC direction is the window position and that the object is somewhere inside the window.  Thus, the best estimate of the position in AC is the window centre, causing the derived positions to follow the same zigzag motion as the window.  The uncertainty on the position is consequently given as the dispersion of a rectangular distribution over the assigned window (shown as the horizontal grey bar in the figure).

As the AC motion of the real asteroid is totally independent {of} the window propagation, its signal can be severely truncated {and the asteroid can} leave the window. The derived position is thus affected by a large bias. In the {scheme,} this is represented by the fading of the circles when the asteroid reaches the window edges. AF7-9 positions will probably be rejected as unreliable by the astrometric processing. 

After they are transformed onto the sky {(right panel),} the different windows nearly overlap in AL, but show shifts in AC aligned on integer pixels. In the figure, a small AL motion (coming from the asteroid proper motion) is present in addition to the zigzag motion. It should be emphasised that after rotation to RA-Dec, the zigzag motion may visually appear in both coordinates, and that the uncertainties will seem to be large both in right ascension and declination, and thus (only apparently) mask the extreme precision of the position, which is still present in the AL direction.

While it can be important to understand the origin of this peculiar pattern, its influence on practical applications (e.g. orbit fitting) is negligible, as the spurious AC motion and its fluctuations will be zero within the given uncertainties in AC. Conversely, the AL position is very accurate and also represents the proper motion of the asteroid in that direction with high precision.

\subsection{Orbit computation}
\label{S:orbit_processing}

The computation of orbits from the Gaia astrometry alone has not been run with the core of the pipeline, but as a post-processing task. In the \gdrthree{} archive, orbits are published in a specific table named \texttt{gaiadr3.sso\_orbits}. 

Orbital fitting is performed based on \gaia-centric astrometric data in right ascension and declination at CCD level, using the corresponding error model with non-diagonal {covariance} matrix as weights, as derived in Sect.~\ref{sect:cu4sso_error_astrom_systrand}. Only known asteroids are considered. 

\begin{figure*}[ht]
\centering
\includegraphics[width=1.0\textwidth]{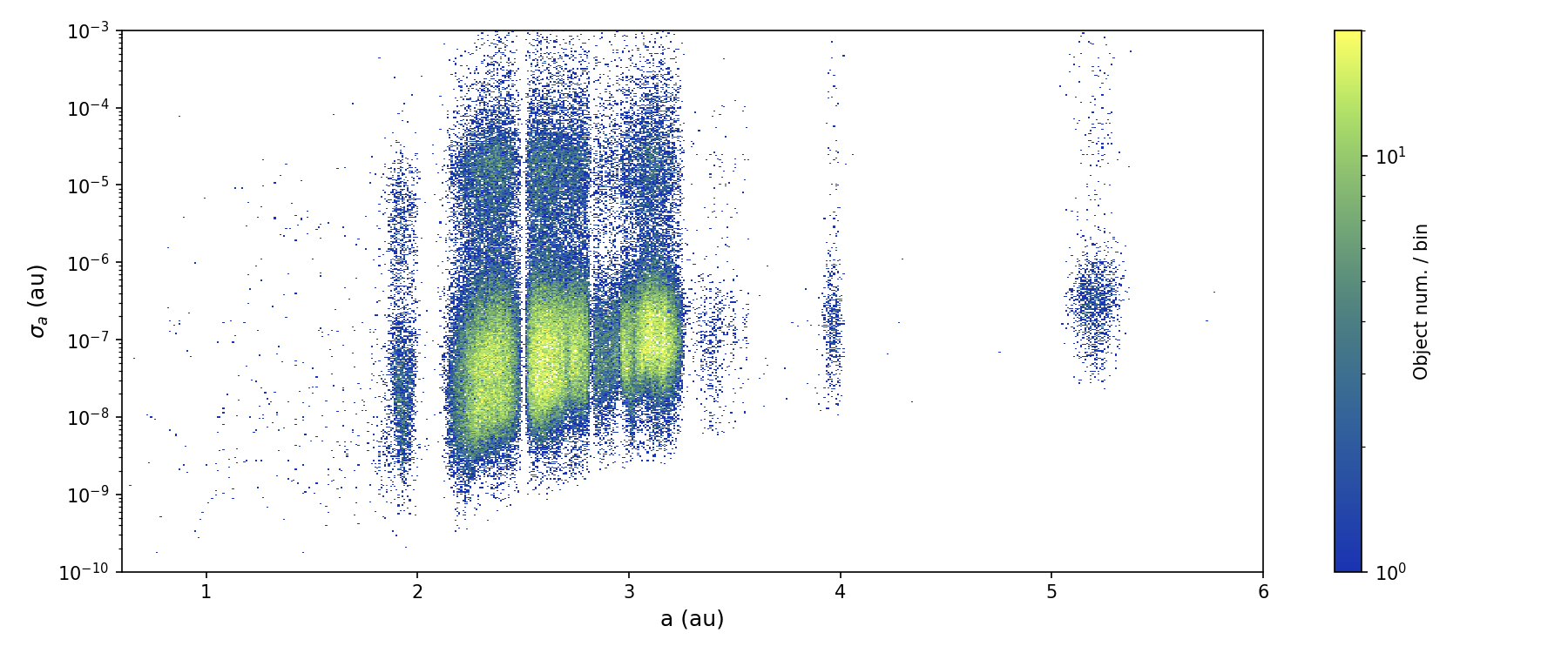}
\caption{Computed uncertainty of the semi-major axis $\sigma_a$ for the orbits published with \gdrthree as a function of the semi-major axis itself. Colour represents the density of objects computed in the bins (600$\times$500 bins in the  axis range). 
}  \label{fig:sigma_a_vs_a}
\end{figure*}

For this task, we adopt the usual procedure consisting in determining the corrections to the initial orbital elements by solving a linear system of equations for each asteroid, while minimising the residuals between the computed and the observed positions.
To determine the computed values as well as the required partial {derivatives}, the equation of motion for each source, including the relativistic contribution, is integrated simultaneously with the variational equations \citep{Beutler2005, Pontriaguine1969}.

The positions and velocities of the planets of the Solar System are obtained from the highly accurate dynamical model INPOP10e (consistently with the whole Gaia software framework), which includes the eight planets, the dwarf planet Pluto, and a selection of 343 asteroids \citep{2013arXiv1301.1510F, arXiv:2203.01586}. The heliocentric positions of the asteroids are computed from numerical integration 
including perturbations from eight planets and Pluto; relativistic corrections are made with the {parametrised post-Newtonian (PPN)} formulation.

Outlier rejection at the observation level has been implemented as in \citet{Carpinoetal2003}. We included a test on tolerance and on the maximum number of iterations. 

The reference epoch for the orbit solution can in principle be chosen arbitrarily and could be the same for all asteroids; however, a more precise result is obtained with a reference epoch halfway on the observational arc. This value is provided in the \gaia archive.

The final output of the procedure is the new improved state vector in the ICRF3 system at the reference epoch, 
together with its corresponding {covariance} uncertainty matrix.
These quantities are also transformed into heliocentric elliptical orbital elements with the associated {covariance} matrix at the reference epoch (see Sect.~\ref{sect:reference_frames} for the details about the reference systems used). 

Residuals are expressed in right ascension and declination,
but with correlation due to the orientation of the \gaia scanning plane during the measure. When the scanning direction on the sky is known, these residuals can also be expressed in the independent along-scan and across-scan directions, that is, in the (AL, AC) plane, without a correlation.  General validation is performed on the residuals expressed in the (AL, AC) coordinates, and more particularly, in the most precise AL direction (see Sect.~\ref{S:residuals}). 

The orbital fit implementation was validated by various comparisons of the state vector, orbital elements, and their associated {covariance} matrices. This includes internal and external data and codes, in particular, a comparison with orbits derived independently from the preliminary astrometry provided by IDT, the \texttt{astorb} and JPL databases, {covariance, and the} NIMA code (J. Desmars, priv. com.) for orbit determination and propagation. A final validation into the DPAC frame has been also performed by using the \texttt{orbfit} code and is discussed in \citet{DR3-DPACP-127}.

The number of known asteroids for which the orbit fit was run is 156\,801. The procedure did not converge for 451 sources. The final number of orbits is 156\,350. 

Finally, not all orbit solutions are meaningful because only a small number of astrometric data is available and the covered time span is limited. We decided to discard extremely uncertain orbits, defined as being based on $\le$20 observations, or those for which the astrometry covered $\le$60~days. Another 29 orbits were discarded by a direct threshold on relative uncertainty $\sigma_a/a>9~10^{-4}$. A total of 1609 orbits were finally rejected ($\sim$1\%). The final number of orbits is 154\,741.
Figure~\ref{fig:sigma_a_vs_a} shows the general distribution of the uncertainty on the semi-major axis for the published orbits. 

A total of 23\,327\,388 observations were processed, grouped as 3\,212\,676 transits. Out of this grand total of observations, 23\,031\,703 were not rejected by the fitting process. The rejected observations ($\sim$1.27\% ) mostly belong to the 1609 eliminated orbits. The set of retained orbits contains 23\,167\,198 observations grouped into 3\,192\,098 transits, so that the {proportion} of rejected observations from this set of objects {corresponds to only} $\sim$0.58\%. 
 




\subsection{Processing of photometry}
\label{S:photometry_processing}

G-band photometry is obtained from the signal measured by the unfiltered AF instrument, the as was same used for astrometry. For the SSOs, the G-band photometry is not derived in the frame of their specific processing modules, but similar as in \gdrtwo, by an upstream system (PhotPipe) that treats and calibrates photometry for all the sources observed by $\gaia$, producing fluxes and their errors for all the AF CCDs \citep{Riello2021A&A...649A...3R}.

In the frame of the SSO processing, only the weighted average of the fluxes at transit level is computed. In this process, measurements that are rejected by the astrometric processing (mentioned above) are also rejected. The rationale for {this} is the fact that problems with the signal that prevent the computation of an accurate position (most often truncation, attitude problems, or other anomalies) also produce problematic flux measurements. 
Other specific flags set by upstream systems are also reasons for flux rejections. The details of these rejection criteria are given in the \gdrthree \href{https://gea.esac.esa.int/archive/documentation/GDR3/Catalogue\_consolidation/chap\_cu9val/}{online documentation}.

The average G-band flux per transit was computed using the weighted average, where weights correspond to the inverse of each flux variance. The quadratic sum of the flux errors per CCD was used to obtain the flux error for the transit. The average \gmag{} -magnitude value was computed by using the average G-band flux and the \gmag{} -magnitude zero-point provided by the general photometric calibration.

The flux, its uncertainty, and the magnitude can be accessed in the table {\texttt{gaiadr3.sso\_observation}} in the fields \texttt{g\_flux}, \texttt{g\_flux\_error} and \texttt{g\_mag}. While it is recommended to use the flux uncertainty, conversion into error on magnitude can be performed by taking the zero-point error provided in Sect. 5.4.1.3.2 of the DR3 online documentation into account. The average value repeats identically for all observations in the transit. 
As a post-processing task, all magnitudes G$>$21 or with errors $>$0.2 were considered as unreliable and were eliminated.

\section{Evaluation of the sample of unmatched sources}
\label{S:unmatched_validation}

As explained in Sect.~\ref{S:input_list}, the search for unmatched sources was performed with orbital elements as known at the end of 2017. Meanwhile, many new orbital elements have been computed and published.  In early February 2022, we did a search to identify the unmatched sources using the most recent orbital elements published by the Minor Planet Center.

Because the list of unmatched sources contains 3541 transits, and the file with orbits contains about 1.2 $\times$ 10$^6$ orbits, this results in a total of $\sim$4 $\times$ 10$^9$ residuals that were to be confirmed, each requiring a numerical integration of the orbit with full perturbations by all major planets.  To reduce the number of computations to a manageable quantity, some optimisations were essential.

Therefore, we divided the sky into tiles of roughly one square degree each. In a first iteration, we computed geocentric ephemerides of all asteroids with a published orbit at one-day intervals, without perturbations, to determine the tile in which the asteroid is predicted to be for each \gaia transit of an unmatched source. Correspondence to the transit positions was tested in a tile set that included the central tile plus surrounding tiles (with at least one corner in common). This compensates for the parallax effect between \gaia and the geocentre, and for the absence of perturbations.  Potential identifications are all asteroids found in the tile set at the epoch closest to the epoch of the transit.  Far from the ecliptic, this can result in fewer than ten candidate identifications, which grows to $>$1000 identifications in the ecliptic region.

The second step was to compute the residuals between the \gaia position and the position from ephemerides for each candidate identification, but now \textit{Gaia}-centric, computed with full perturbations, and for the exact epoch.  Since the motion of an asteroid in the course of a transit is small compared to the uncertainty in position in AC, we used only one position per transit.  We applied this procedure three times, with three different values of the threshold for a tile to be a neighbour: 2 degrees, 5 degrees, and 10 degrees. In each iteration, only the still-unidentified sources were considered.  The last iteration did not give any additional identification.

Next, we defined some criteria to accept an identification.  The key point in this is to rotate the residual in right ascension and declination to the directions parallel and perpendicular to the line of variation.  The line of variation is defined as the line on the sky that results from varying the mean anomaly of the asteroid without changing the epoch or the other orbital elements.  The rationale behind this is that errors in the orbital elements will result in periodic errors in the position of the object, except for an error in the semi-major axis (or the period), which will result in a drift in mean anomaly, linear with time.  This will cause the error in mean anomaly at the epoch to have the largest impact in the error in position.  To find suitable thresholds, we performed a similar exercise using an orbital elements file from 2014, in which some of the now numbered minor planets still had rather poor orbits.  Thus, we were able to determine how residuals behave for identifications that are known to be correct compared to incorrect identifications.  For incorrect identifications, we found differences in motion that were evenly distributed between 0 and more than 500'' per day, while for correct identifications, we found that the differences rarely exceeded 2'' per day.  For correct identifications, however, positions may be off by up to 10\arcmin along the line of variation.  Perpendicular to the line of variation, residuals do not exceed 10''.  No incorrect identification had residuals smaller than 10'' perpendicular to the line of variation, with a difference in speed smaller than 2'' per day at the same time.  Thus, it is possible to distinguish between correct and incorrect identifications with {only a very small risk} of an incorrect identifications if the speed can be measured, that is, if there are at least two transits.

For unmatched sources with at least two transits, we therefore set up as limits 400'' for the residual along the line of variation, 20'' for the residual perpendicular to the line of variation, and 7''/day for both components of the difference in motion.  This last rather high value is justified by the fact that if an object has only two transits that occur less than 0.25 day apart, for instance, the uncertainty on the motion is larger than the expected differences in speed.  For unmatched sources with only one transit, we had to be more severe on the residuals on the positions because the speed could not be computed. We set 5'' as limit along the line of variation and 1.5'' perpendicular to the line of variation.  

With these criteria, we found an identification for 712 out of the 1320 unmatched sources.  Some of the unmatched sources turned out to be the same {asteroid}, so that in total, they represent only 567 different {asteroids}.  Strangely, the highest residual we found was only 13'', rather than the expected several hundred arcseconds.  This shows the lack of preliminary orbits in the 2022 Minor Planet Center orbital elements files.
The derived identifications are made available in the auxiliary data web page of \gdrthree.
   
\section{Astrometric performance}
\label{S:astrometry_performance}

\subsection{Orbit quality}
\label{S:residuals}

The large differences in both data volume and time coverage for each asteroid source means that the quality of orbits computed from \gdrthree astrometry varies from excellent to very approximate. This is shown by the values of the semi-major axis uncertainty $\sigma_a$ in Fig.~\ref{fig:sigma_a_vs_a}. 

\begin{figure}[h]
\centering
\includegraphics[width=0.475\textwidth]{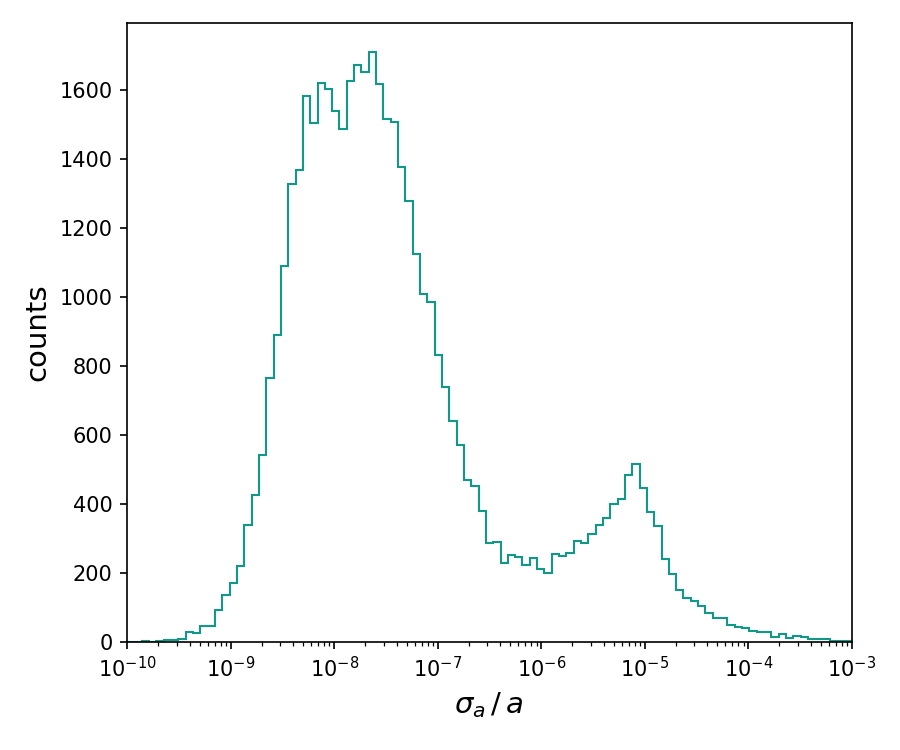}
\caption{Distribution of the relative uncertainty on the semi-major axis for the computed orbits in the range of $a$ between 2.2 and 2.5 au. The right peak of high values in general contains orbits obtained from a small number of observations over a short observational arc. }  
\label{fig:hist_sigma}
\end{figure}

A bimodal distribution of the uncertainty appears. This feature is more clearly visible from the frequency distribution of the relative uncertainty (Fig.~\ref{fig:hist_sigma}), in a limited range of semi-major axis (here from 2.2 to 2.5 au). 
This signature can also be detected in similar distributions obtained with ground-based observations { \citep{desmars03-orbital-uncertainty}}, reflecting a combination of short observation arcs and small astrometric data sets. This can be confirmed by the {statistics derived for} the two populations, assuming that they are separated at the threshold value $\sigma_a\,/\,a\ =\ 10^{-6}$. The average number of non-rejected observations is 147.5 and 79.1 for the lowest and highest uncertainty, respectively, while the arc lengths are 771 and 380 days.

An interesting question related to the quality of the orbital solution from astrometry in \gdrthree is how it compares to { the orbits} obtained from larger data volume and longer time-span, including the whole set of astrometric measurements available from the ground. In particular, we recall here that the time span of \gdrthree is just 34 months, which corresponds to the orbital period of an asteroid with a semi-major axis of slightly less than 2~au. To address this question, we have obtained the most recent orbital solution (updated on 7 March 2022) from the JPL database of orbits\footnote{\url{https://ssd.jpl.nasa.gov/tools/sbdb_query}}  for all objects for which we have an orbit. We also used the related {absolute magnitude $H$} provided by the same source to categorise the comparison with respect to the absolute brightness.

\begin{figure}
\centering
\includegraphics[trim=10 0 10  0,clip,width=0.49\textwidth]{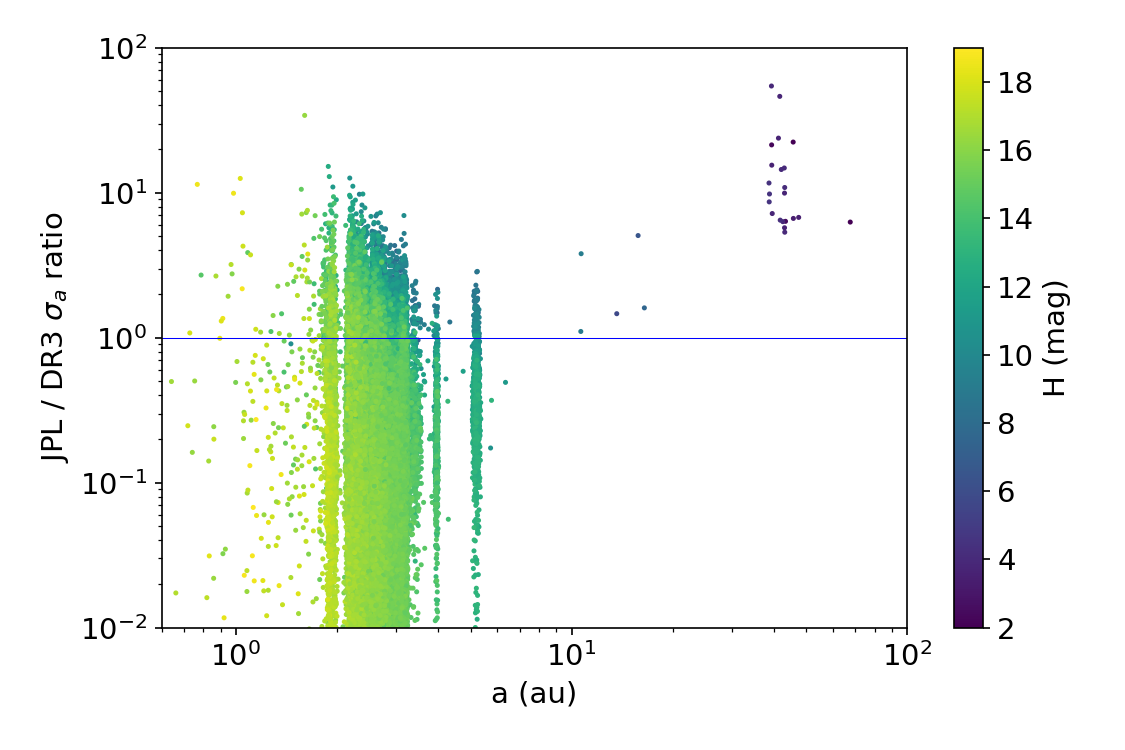}
\includegraphics[trim=10 0 10
0,clip,width=0.49\textwidth]{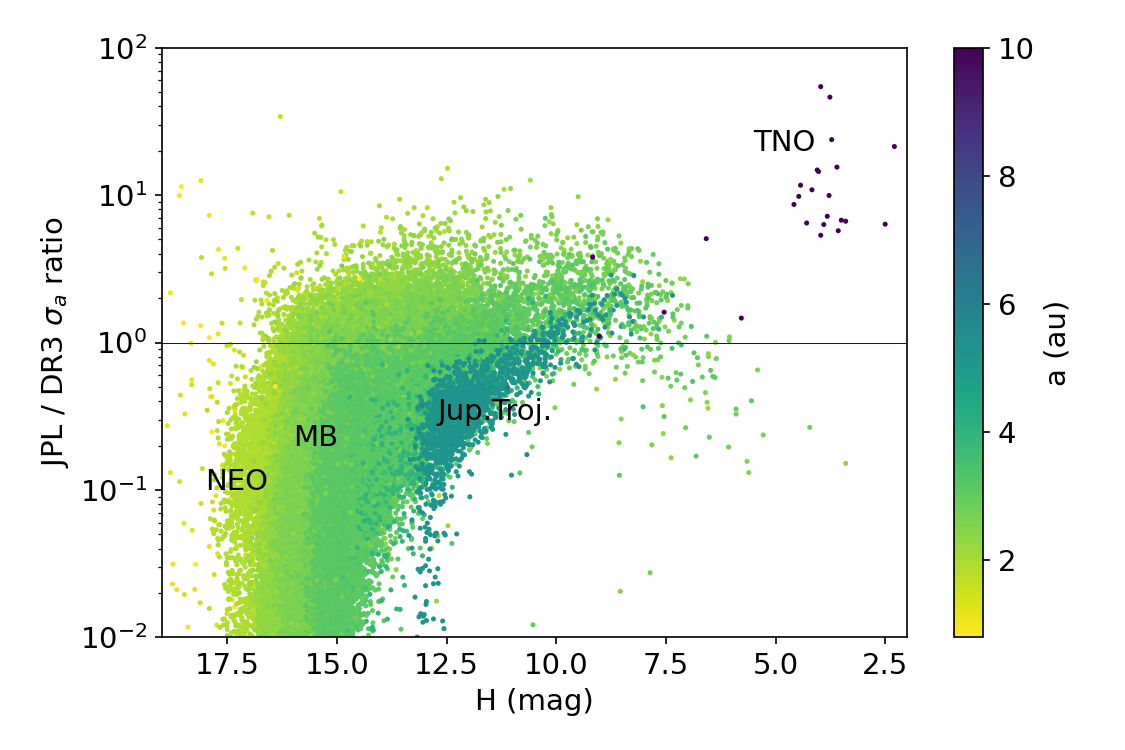}
\caption{Ratio {of $\sigma_a$ from JPL and from \gdrthree} as a function of semi-major axis (top panel) and of absolute magnitude H (bottom). Colour represents H (top) and semi-major axis (bottom). The horizontal line shows equal uncertainties (ratio=1).}
\label{fig:JPL_DR3}
\end{figure}

In Fig.~\ref{fig:JPL_DR3} we plot the ratio of $\sigma_a$ from JPL and the \gdrthree value as a function of semi-major axis and H. 
A general trend of increasing accuracy with brightness appears for {MBAs}. While most of the orbits lie below the line of equivalent accuracy, as expected from the long time coverage and data value for the general population, it is interesting to note that 8736 asteroids reach a better accuracy when \textit{Gaia}-only data are used, including some NEOs, several Jupiter Trojans, {and TNOs}. 
\begin{figure}[h!]
\centering
\includegraphics[width=0.465\textwidth]{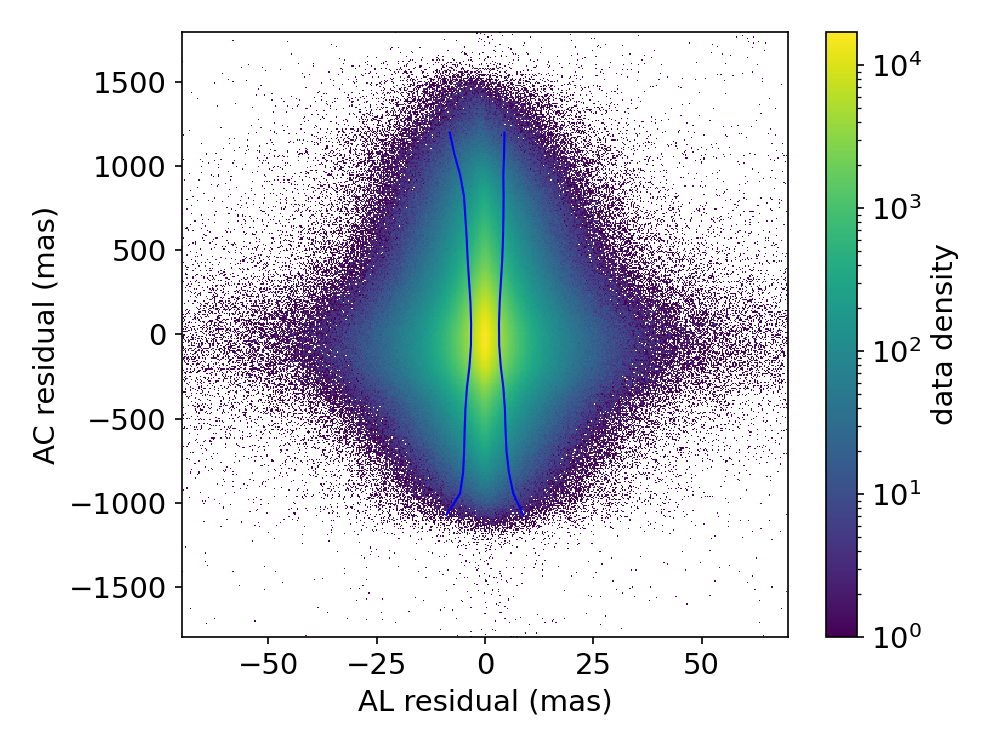}
\includegraphics[width=0.44\textwidth]{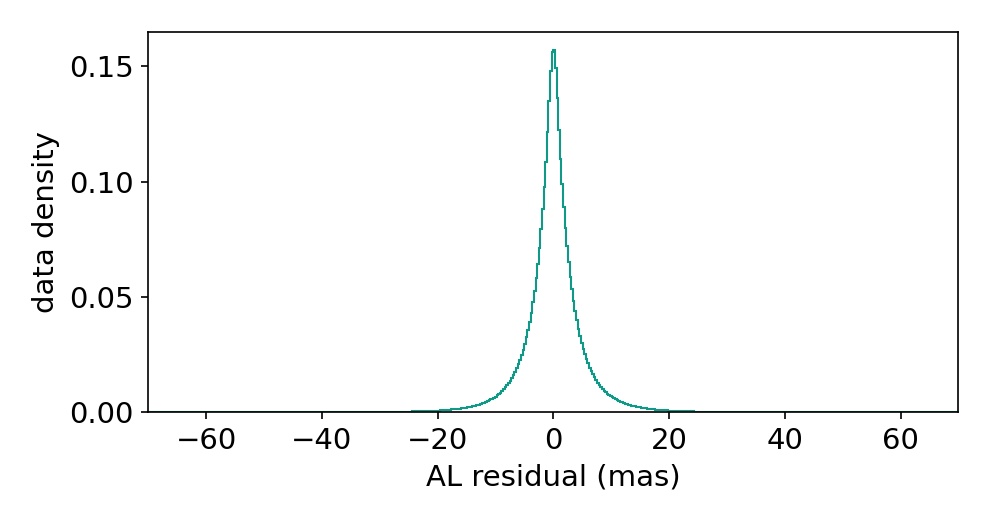}
\includegraphics[width=0.44\textwidth]{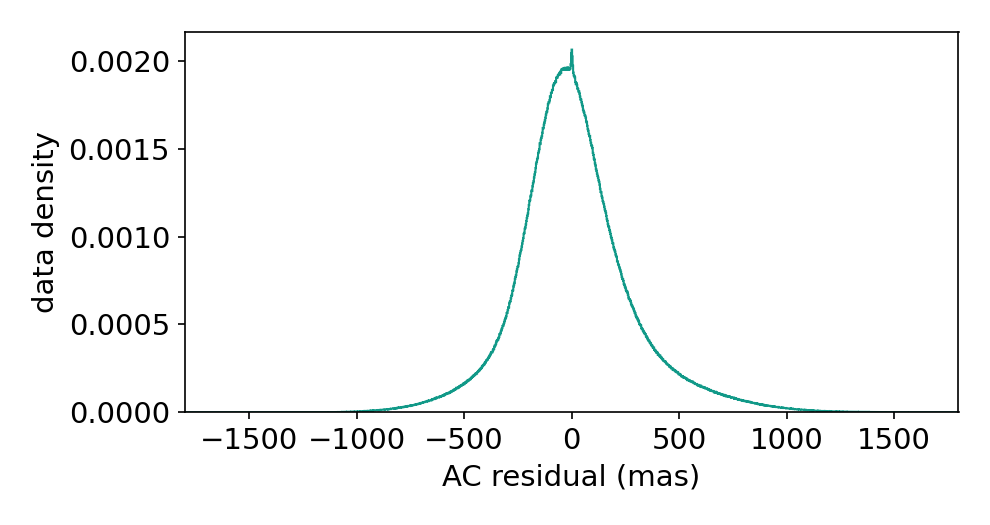}
\caption{Distribution of the post-fit astrometric residuals in the AL,AC plane. The vertical lines
mark the 1-sigma quantiles computed {for} bins of 90~mas on the AC axis. The frequency distributions along the {two axes} are reproduced in the middle and bottom panel (in AL and AC, respectively). The corresponding standard deviations are 5.15~mas and 270.14~mas.
}
\label{fig:ALAC_histo}
\end{figure}

The post-fit residuals to the orbit adjustment are a final sensitive test for the quality of the astrometry. Figure~\ref{fig:ALAC_histo} presents the global view of the residual dispersion. 
The distribution is strongly dispersed in the AC direction, as expected, due to the low accuracy of astrometry in the across-scan direction with respect to the AL direction. 

\begin{figure}
\centering
\includegraphics[width=0.5\textwidth]{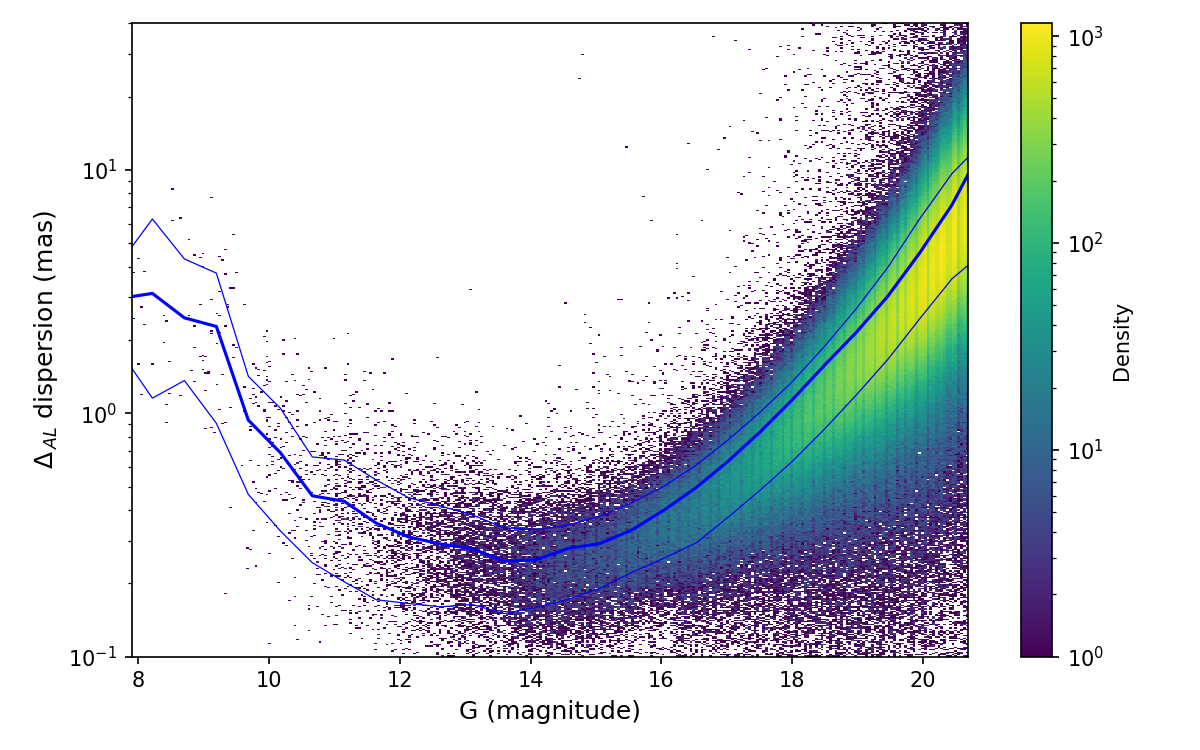}
\caption{Dispersion of post-fit residuals in AL for each transit. The colour represents the data density. The deep blue line shows the mean, and the {light blue} lines show  the quantiles corresponding to 1-sigma, computed over 50 bins in the interval of G magnitude {from 8 to 21~mag}.}
\label{F:res_std_AL}
\end{figure}

Extreme values of residuals correspond to the less accurate orbit of faint objects with fewer observations. Although the range of values appears large, the core of the distribution is very compact.
This is visible in the histograms of Figure~\ref{fig:ALAC_histo}, where the central peak appears. 
The global {histogram, however, does not convey complete} information on {the astrometric accuracy.} 
To highlight the quality of \gaia astrometry as a function of brightness, we chose the same approach {as was used} in \citet{Spoto18}, and we represent the dispersion (represented by the standard deviation) of the residuals over a transit (Fig.~\ref{F:res_std_AL}).   

\begin{figure}
\centering
\includegraphics[trim=20 0 40 0,clip,width=0.45\textwidth]{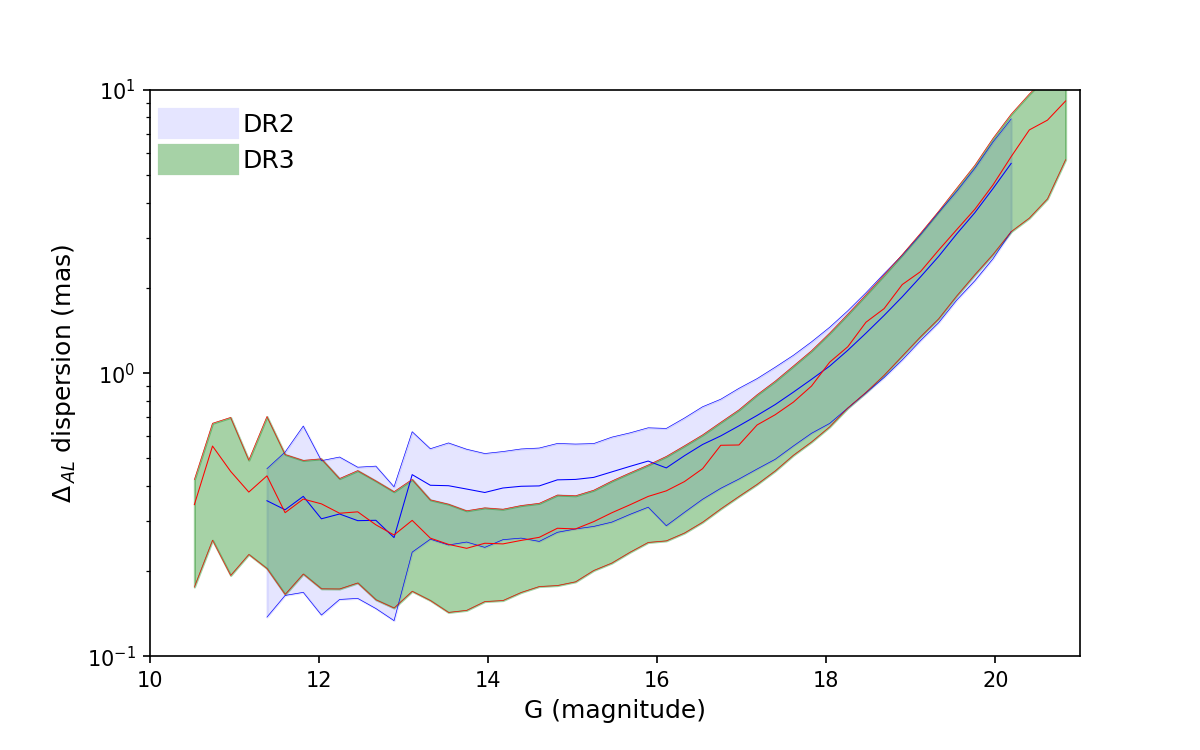}
\caption{Average and 1-$\sigma$ quantiles for the AL dispersion of the post-fit residuals for all transits in common between DR2 and DR3. {In contrast to} Fig.~\ref{F:res_std_AL}, the single data points are not represented. The background curve (light blue) represents the distribution computed for \gdrtwo.}
\label{F:AL_DR2_DR3}
\end{figure}

The average and quantile lines clearly illustrate the trend. The average value reaches a bottom plateau at the exceptional value of $\sim$0.25~mas over a range of three magnitudes, from {$G=12$ to 15~mag. Up to $G=17$~mag, the dispersion remains at submilliarcsecond level}. This is in general agreement with Fig.~\ref{F:errormodel}, although the orbital post-fit residuals can still reflect some systematic effects that are not yet taken into account in the modelling (see Sect.~\ref{S:subsubphotLutetiawobble}).  In this range, the accuracy of \gdrthree is clearly higher than that of \gdrtwo. 
Moreover, the transitions in the error values that were visible in \gdrtwo at $G=13$ and $G=16$~mag have now disappeared as a result of the increased global quality of the calibration. Figure~\ref{F:AL_DR2_DR3} shows that the improvement reaches {a factor of} $\text{about two}$ and appears fully consistent with the robust estimate of the standard deviation for the general astrometric processing of stars in Fig.~A.1 of \citet{EDR3-DPACP-128}.

\subsection{Shape and size effects}

\subsubsection{Binary systems}
\label{S:binaries}
As foreseen \citep{tangaGaiaUnprecedentedObservatory2008,pravecSmallBinaryAsteroids2012,tanga12-ocpd}, one of the most interesting applications expected for the accurate astrometry by \gaia is the possibility of detecting satellites of asteroids. As the orbital fit tends to converge to the trajectory followed by the centre of mass of the object, residuals can contain the signature of an asymmetric light distribution between the primary and the secondary component of an unresolved binary. The expected amplitude and periodicity of these residuals can fall in a range that is accessible to \gaia, and they can also cover a range of component ratios and sizes that is not accessible to other commonly used techniques 
\citep[including radar ranging, adaptive-optics imaging, and photometry of mutual eclipses or occultations;][]{2015-AsteroidsIV-Margot}.

The amplitude of the wobbling $w$, that is, the maximum distance between the system photocentre and barycentre seen by an observer, can be easily estimated by assuming spherical components of identical albedo and bulk density, characterised by a diameter {ratio $k=D_2/D_1$, hence a mass ratio $q=k^3$} \citep{hestrofferGaiaMissionAsteroids2010a}. With this notation, the ratio of their illuminated surfaces is proportional to $q^{2/3}$ and determine the position of the photocentre. It is then easy to show {that}
\begin{equation}
    w = \left( \frac{1}{1+q^{-2/3}} - \frac{1}{1+q^{-1}}\right) a,
    \label{E:wobbling}
\end{equation}

\noindent where $a$ is the semi-major axis of the mutual orbit. In this approximation, no phase effects are introduced so that the individual photocentre offsets for each component are neglected. 
\begin{figure}
\centering
\includegraphics[width=0.5\textwidth]{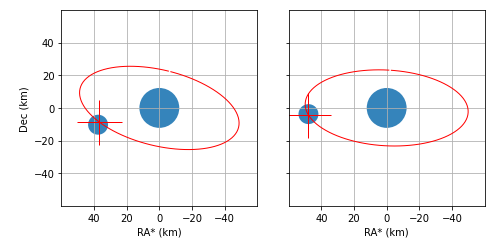}
\caption{Size and relative position of the primary and secondary components of (4337)~Arecibo as derived from the occultation (in blue) of 19 May 2021 (left), and 9 June 2021 (right) on the plane of the sky in the equatorial reference. The red crosses and the orbit are derived from the model described in the text.
}
\label{F:arecibo_occ}
\end{figure}
\begin{figure*}
\centering
\includegraphics[width=0.85\textwidth]{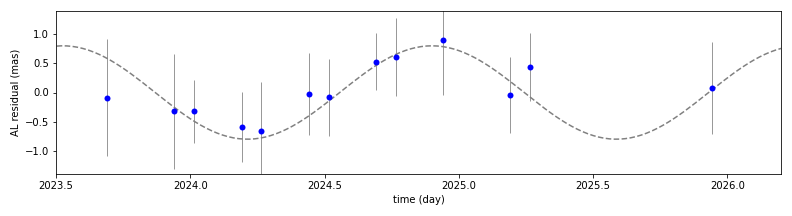} \includegraphics[width=0.85\textwidth]{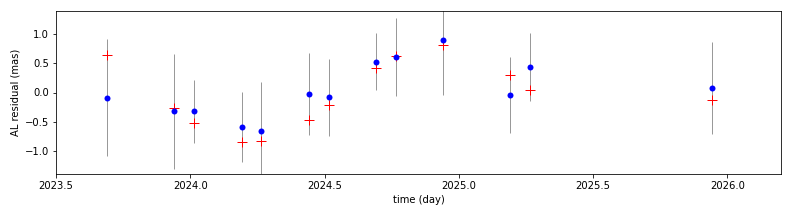}
\caption{Residuals to the orbital fit of (4337)~Arecibo (blue dots). They are obtained from the average of single-observation residuals over each transit. The error bars are given by their standard deviation. In the top panel, the dashed grey line is not a model fit, but a simple overplotting of a sinusoid of the period derived by photometry, adjusted in amplitude (0.8~mas) and phase to the data. In the bottom panel, the same data {are shown} with the residuals predicted by the optimised binary model described in the text (red crosses). }
\label{F:arecibo_residuals}
\end{figure*}
\begin{figure}[h!]
\centering
\includegraphics[width=0.5\textwidth]{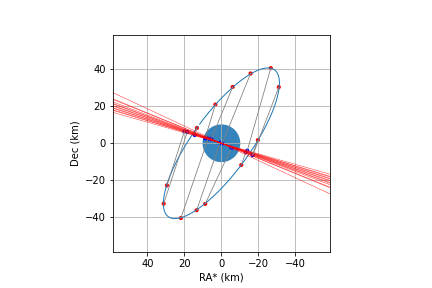}
\caption{Orientation of the modelled orbit, projected on the equatorial reference (RA* indicates that the factor $cos(Dec)$ is included), at the mid-epoch of the sequence of observations exploited to model the binary (4337)~Arecibo. The {positions} of the satellite at the epoch of each observation (red dots) are shown, together with their projection on the instantaneous direction of the scan (AL, red lines). The measured photocentre wobbling is proportional to {the} component of the binary separation in the AL direction.}
\label{F:arecibo_AL}
\end{figure}

A tempting opportunity to look for binary-related photocentre wobbling in \gaia data came from the discovery of the {binarity of the MBA} (4337)~Arecibo through two lucky occultation events (Fig.~\ref{F:arecibo_occ}) 20.78 days apart in May and June 2021 \citep{gaultNewSatellite43372022}. The two occultation chords for each event provided diameter estimations from a circular fit of the components ($D_1$=24.4$\pm$0.6~km, $D_2$=13.0$\pm$1.5~km; $k_{occ}$=0.55) and two accurate astrometric measurements of their relative positions, with an apparent separation of the components of 25 and 32~mas. In the absence of any additional information, it is not possible to derive {information about} the mutual orbit from these two relative positions. At the given apparent separation, however, it is clear that the object does not appear as resolved to \gaia. The signal of the two components falls within one pixel (60~mas AL). However, given the extreme centroiding accuracy in the AL directions, perturbations corresponding to the photocentre wobbling can be expected.

\gdrthree contains astrometric data for (4337)~Arecibo in 38 transits. The G magnitude is present for 36 transits, so that a photometric inversion was attempted with the genetic algorithm used in \citet{Cellinoetal2019}. This provided a period of 32.972823~hours and pole ecliptic coordinates ($\lambda, \beta$)=(271$\deg$, 68$\deg$). Given the long time span covered by the observations (881 days), the period appears rather well constrained (at a level probably better than 1\,s), while the typically estimated uncertainty of the pole direction is $\sim$10$\deg$. Moreover, the rotation period is compatible with a preliminary unpublished light curve obtained from ground-based observers in the weeks following the occultation (32.85$\pm$0.38~hours; Behrend, R. et al., private communication). We stress here that we do not expect to directly find indications of the binary nature of the asteroid in the sparse photometry alone.

In the \gaia data set, we looked for the longest possible sequence of consecutive transits and found one {composed of} 13 transits over 2.3 days, 12 of which are consecutive (corresponding to six rotations of \gaia). During this short time span, the orientation of the satellite orbit with respect to the observer does not change appreciably. This sequence is therefore a good candidate to search for any periodicity in the residuals that could suggest the presence of {wobbling}. In addition, during a single transit, the astrometry is expected to be affected by a systematic displacement, if present, by the same amount. It is thus possible to exploit the average of all residuals collected over a single transit for a more robust estimate. We also assume that the standard deviation of the residuals represents the typical uncertainty. 

As the AL direction conveys the accurate astrometry, only AL residuals were taken into account. Therefore, any measured wobbling would be caused by the component of the two-dimensional photocentre-barycentre shift (as it appears to the observer, on the sky plane) {in the} direction of the scan (AL) for each transit.

The analysis of the average residuals as a function of time within the selected sequence clearly suggests a systematic fluctuation. Figure~\ref{F:arecibo_residuals} shows that fluctuations {occur} on a timescale compatible with the rotation period. Based on the considerations above, the compactness of the system is suggestive of a satellite revolution period that is synchronous with the primary rotation, which would explain the compatibility between the astrometric and the photometric periods.

With the elements above, we can now show that \gaia is able to provide an orbital solution for the system. We assumed that the most robust parameter {\sl \textup{a priori}} available for the model is the rotation period, and based our analysis on the relative astrometry derived from the occultation and the AL transit residuals, with their error bars. Only the occultation chords, but not the derived relative astrometry that we exploit, are sensitive to the absolute sizes $D_{1,2}$. For the same reason, while the wobbling recorded by \gaia in the residuals depends on the size ratio $k$ through Eq.~\ref{E:wobbling}, the occultation results {do not}.

To model the {wobbling,} we exploited Eq.~\ref{E:wobbling} and the position angle of the scan, available with the astrometric data, to compute its component projected in AL. The model also requires  the ephemeris of the asteroid as ancillary data to correctly reproduce the observation geometry, including the light--time delay to be considered for each observation epoch. 

We defined a target function as the squared sum of the O-C of the model-derived astrometry with respect to the measured astrometry. We considered the pole coordinates ($\lambda, \beta$), the semi-major axis $a$, and the initial rotation phase at an arbitrary reference epoch (set to the first occultation event) as free parameters of the model. A downhill simplex algorithm (Nelder-Mead) was run for its minimisation. Our result provides the values 
\begin{equation*}
a=49.9\pm 1.0~km,\ (\lambda,\beta) = (261\deg\pm 3\deg, 60\deg\pm 3\deg).
\end{equation*}

The pole coordinates remain compatible with those derived by photometric inversion that were used as initial conditions. 
The fit to the occultation data is strikingly good (Fig.~\ref{F:arecibo_occ}), although the final result shows a small discrepancy in the occultation astrometry, especially for the first event. However, the chords of the first event are very similar to each other, and errors related to shape assumptions {(spherical shape hypothesis)} could show up at the milliarcsecond level.

\begin{figure*}[t!] 
\centerline{
    \includegraphics[width=55mm]{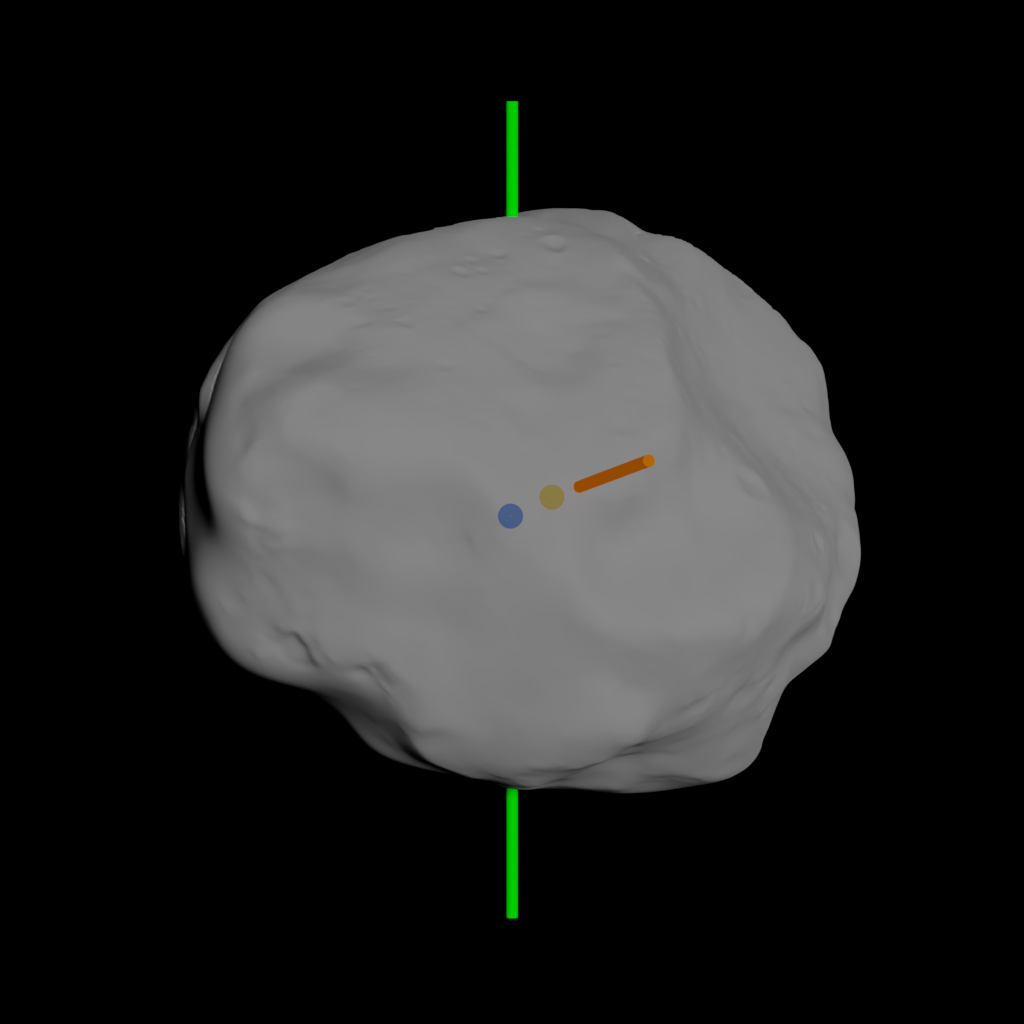}
    \hspace{0.5cm}
    \includegraphics[width=55mm]{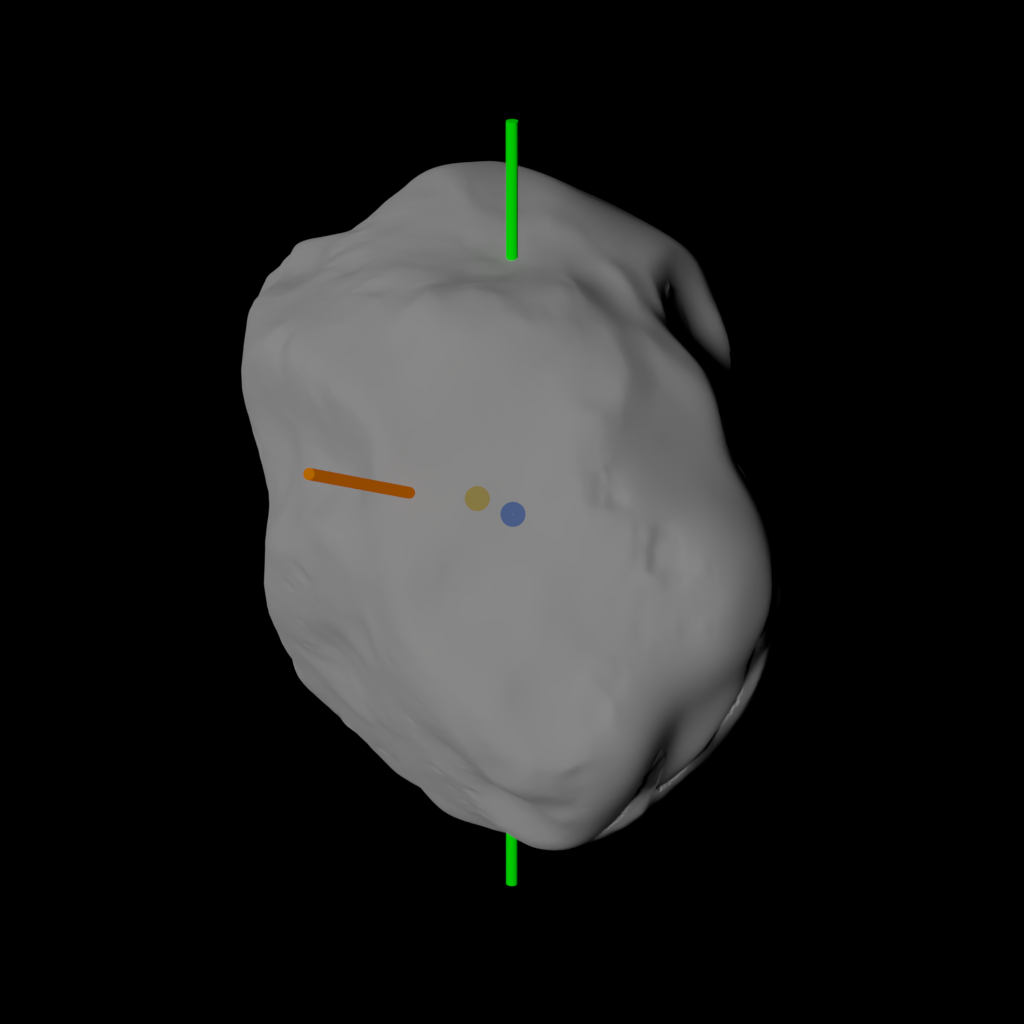}
    \hspace{0.5cm}
    \includegraphics[width=55mm]{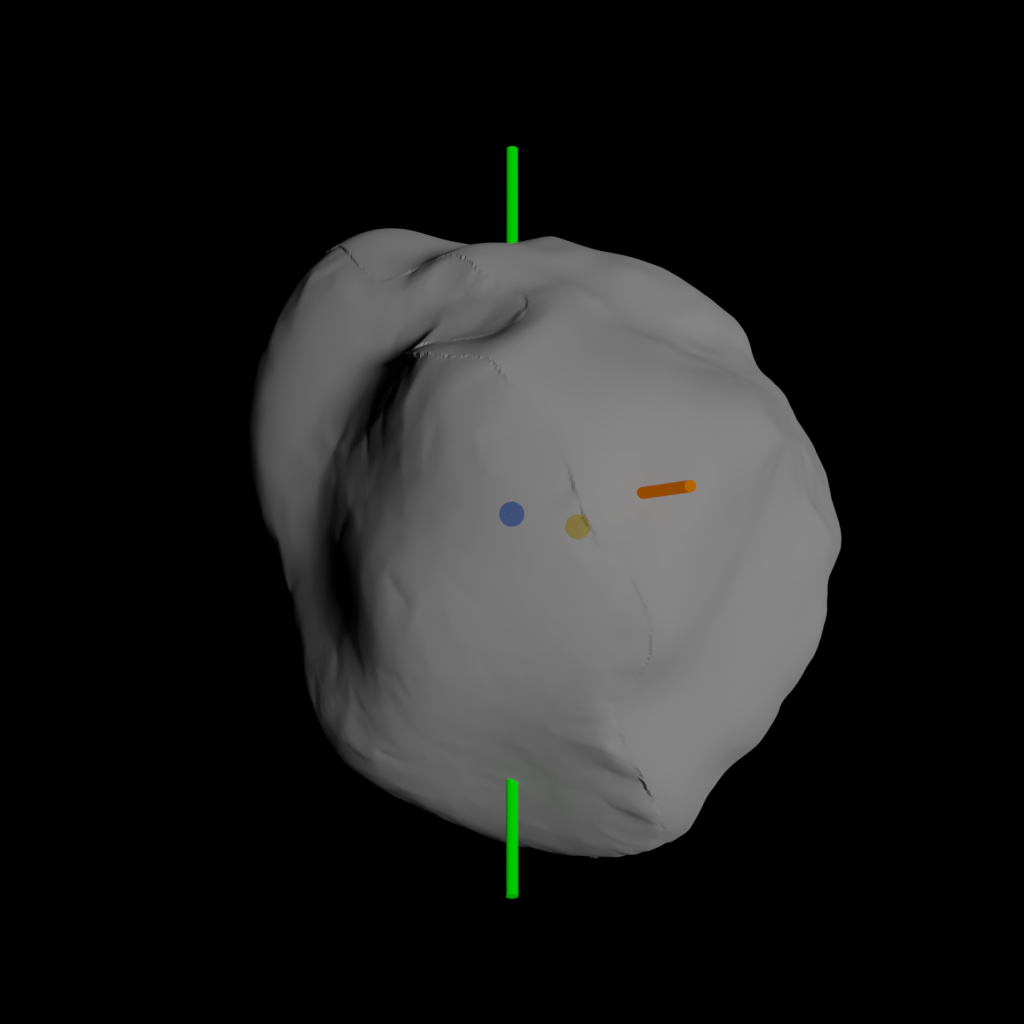} }
  \caption[photocentre wobble for (21)~Lutetia]
  {Photocentre and barycentre positions of asteroid (21)~Lutetia (brown and blue bullets) as projected in the direction of Gaia (normal to the image plane) for three epochs. The $z$-axis of the equatorial coordinate system is denoted by the green bars. North points to the top. The Sun illuminates the asteroid from the directions denoted by the red bars. From left to right, the solar phase angle takes the values of 16.7, 24.2, and 20.9 degrees, and the photocentre-barycentre offset takes the values of 7.42, 5.96, and 10.91~km.}
\label{fig:Lutetia_photocentre}
\end{figure*}
\begin{figure*}[h] 
  \centerline{
  \includegraphics[width=0.4\textwidth]{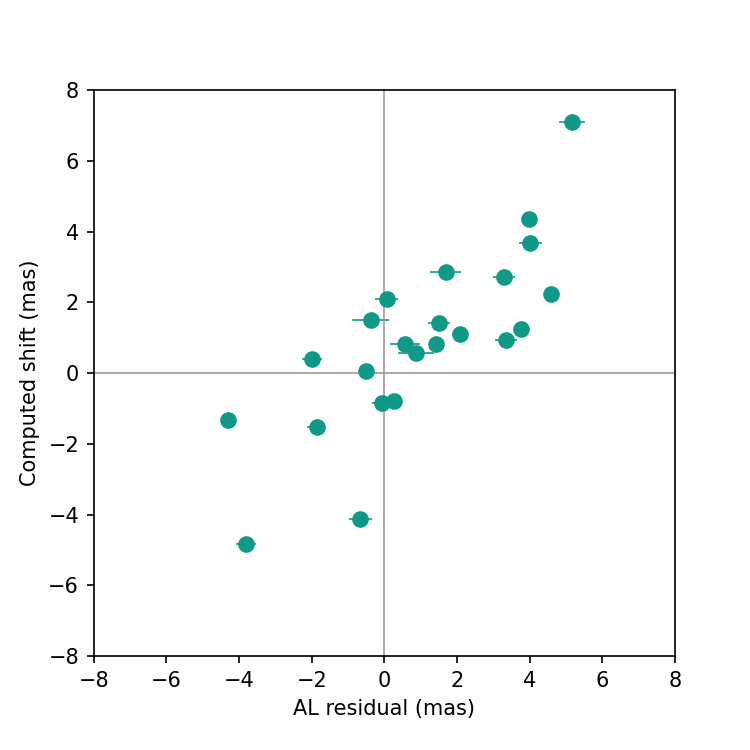}
  \includegraphics[width=0.4\textwidth]{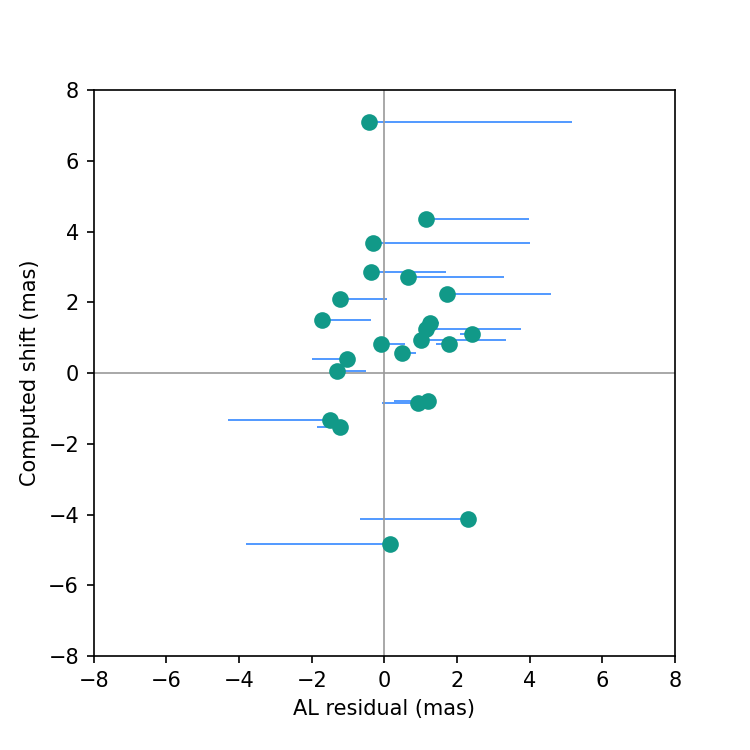}}
  \caption{Comparison between the average transit residuals for (21)~Lutetia in the AL direction (x-axis) and the computed photocentre shift from the light scattering model (y-axis; left). The error bars represent the standard deviation of the residual across the transit (often smaller than the symbol size). The right panel shows the equivalent plot after correcting for the for the computed shift. The lines represent the displacement of each data point with respect to the left panel.}
\label{fig:Lutetia_shift}
\end{figure*}

The wobbling amplitude following {Eq.~\ref{E:wobbling}} could reach 8.5\% of the object separation, or about 2.7\,mas at the distance of \gaia (2.24~au) for the observations exploited here. This amount is 44\% of the apparent radius of the primary component. It is therefore much larger than a photocentre shift due to shape effects alone for the phase angle at the same epoch (14.3$\deg$).

The measured wobbling is strongly reduced by the projection in the AL direction (Fig.~\ref{F:arecibo_AL}). However, our fit attempts show that it cannot be fitted unless{} a value of $k$ is assumed that is lower by $\sim35\%$ with respect to the nominal value derived by occultations ($k_w = 0.35\,k_{occ}$). With this correction, the model appears to reproduce the observations pretty well (Fig.~\ref{F:arecibo_residuals}, bottom panel).

The apparent contradiction between $k_{occ}$ and $k_w$ can have many origins that are all equally interesting, implying that the mass of the companion is lower than what is estimated from the geometric size ratio provided by the occultations. A first possibility is a non-spherical shape of the components with a flattening that is more pronounced in the case of the satellite. In this case, because the occultation chords for the second event appear to constrain mainly the equatorial radius of the two components, $k_{occ}$ would not represent the volume ratio well (the first event does not constrain the size ratio well in any case). A second possibility, relevant for the formation mechanisms, is that the companion could have a lower bulk density. In either case, the mass ratio $q$ would be reduced. 

Of the two options, shape flattening is likely to be favoured because two additional other elements support it. First, the thermal infrared diameter \citep{NEOWISE} is $32\%$ smaller than the surface-equivalent diameter from the occultations. Therefore, the equatorial size constrained by the occultations would rather be a maximum shape extent, while the polar radius could be smaller. Second, when we apply Kepler's law to our best-fitting parameters, the density is too low when the nominal occultation diameter is used. The size must be reduced by $19\%$ to increase the density to a minimum value 1~g/cm$^3$. A low density like this is expected for this object, which belongs to the Themis family, whose potentially water-rich, highly porous members are expected to have a density below 1.3~g/cm$^3$ \citep{marssetCompositionalCharacterisationThemis2016}.

Future observations, by \gaia and from the ground, should be able to better characterise this system and also investigate the possible role of surface scattering and photocentre shift of each of the components.

\subsubsection{Photocentre wobble for (21) Lutetia}
\label{S:subsubphotLutetiawobble}

The decrease in accuracy of \gaia astrometry towards the bright magnitudes \citep[also seen in \gdrtwo][]{Spoto18} is suggestive of the fact that shape and size effects could affect the centroid determination and degrade the performance. This effect was also present in the astrometry of the \gaia precursor HIPPARCOS \citep{hestrofferPhotocentreDisplacementMinor1998}.

While a detailed and systematic search of this effect in \gdrthree asteroid data is beyond the scope of this article, a single case can already provide interesting indications and can in particular show whether a discrepancy between the photocentre and the centre of mass of the asteroid is detectable in the astrometric residuals with respect to the fitted orbit. Asteroid (21)~Lutetia with its detailed shape model (sect.~\ref{subsubphotLutetiaSteins}) is especially well suited for this study. Lutetia can be approximated as an ellipsoidal asteroid with total axial dimensions of ($121 \times 101 \times 75$)~km and a mean radial distance from the geometric centre of 49~km. By using the high-resolution non-convex shape model resampled at five-degree resolution, we computed the photocentre-barycentre offset for the 23 epochs of the GDR3 photometry \citep[for the computation of the photocentre, see for example ][]{Muinonen2015}. 

Example offsets are illustrated in Fig.~\ref{fig:Lutetia_photocentre} for the illumination and observation geometries in three representative epochs, corresponding to phase angles of 6.7, 24.2, and 20.9 degrees. The corresponding photocentre offsets were 7.42, 5.96, and 10.91~km. At the distance of \gaia at the epoch of the observations (3.15, 2.45, and 2.78~au, respectively), the angular offsets are 3.4, 3.3, and 5.4~mas, respectively. 

We exploited the position angle of the scan to project the computed offset of each epoch in the AL direction, deriving a prediction for the offset that \gaia should have measured. We then compared it to the orbital post-fit residuals of \gaia observations, always in the AL direction. In this case, the residuals are represented by the average of the residuals of the individual observations of each transit. The orbital fit was obtained by the same approach as used in Sect.~\ref{S:yarko} and included all the astrometry available at MPC. 

The result is illustrated in Fig.~\ref{fig:Lutetia_shift}. It shows a clear correlation between predictions and observations (left panel), with a certain scatter. 

We then subtracted from the \gaia astrometry the computed photocentre offset, and then applied the same orbital fitting procedure, with the subsequent analysis of AL residuals. As expected, now the distribution of residuals is much more compact (right panel), and the correlation with the computed shift has disappeared. The final residuals are distributed around zero with a standard deviation of 1.2~mas. This remaining scatter is relatively large with respect to the brightness of (21) Lutetia, observed by $\gaia$ at G$\sim$13.\ It suggests that a margin of improvement to the photocentre model probably exists. Concavities at large phase angles (particularly relevant in the third panel of Fig.~\ref{fig:Lutetia_photocentre}), where shadowing effects can enhance the offset by a large amount, play a relevant role. The correct modelling of the light-scattering properties of the surface is probably critical in these situations. Albedo variations are also not considered, but might contribute to the residuals.

Despite these limitations, this example provides the compelling evidence that an asteroid in the 100~km class can exhibit a rather large photocentre offset that is strongly dependent on shape details and reaches an order of 20\% of the average radius at the phase angles of the observations by \gaia. At the level of sensitivity of \gaia astrometry, it should be possible to recover this effect in \gdrthree on asteroids that are probably two to three times smaller. 


\subsubsection{Pluto and Charon}
\label{S:plutocharon}
We present here the peculiar case of an emblematic resolved {binary system}: the Pluto and its major satellite Charon. The Pluto system is regularly seen by the \gaia optical system, in the same way as any other SSO. Pluto is a bright source of magnitude $G \approx 14.5$ and is easily detected. The observations were processed in the same way as for any other SSO. {However, the} largest satellite of Pluto, Charon, is at a brightness $G \approx 16.5$~mag and separated on the sky by at most $1 \arcsec$ from Pluto.  \gaia can detect both provided the projected separation on the scan direction (AL direction) is larger than at least $0 \farcs 25,$ so that the on-board detection could resolve the system into two independent sources. Over the time interval of DR3, 17 resolved and 6 unresolved passages were expected. As explained in \ref{S:SSO_identification}, there is no particular procedure in the identification step to flag each of the components of a resolved system. They are matched to the system, and all observations go through the pipeline. Only at the very end of the astrometric solution can we see how many have survived and assign the solutions to either Pluto or Charon. In this particular case, the G magnitude  allows us to obtain a flawless identification, but the comparison to the computed positions would work as well.

\begin{figure}
\centering
\includegraphics[width=0.45\textwidth]{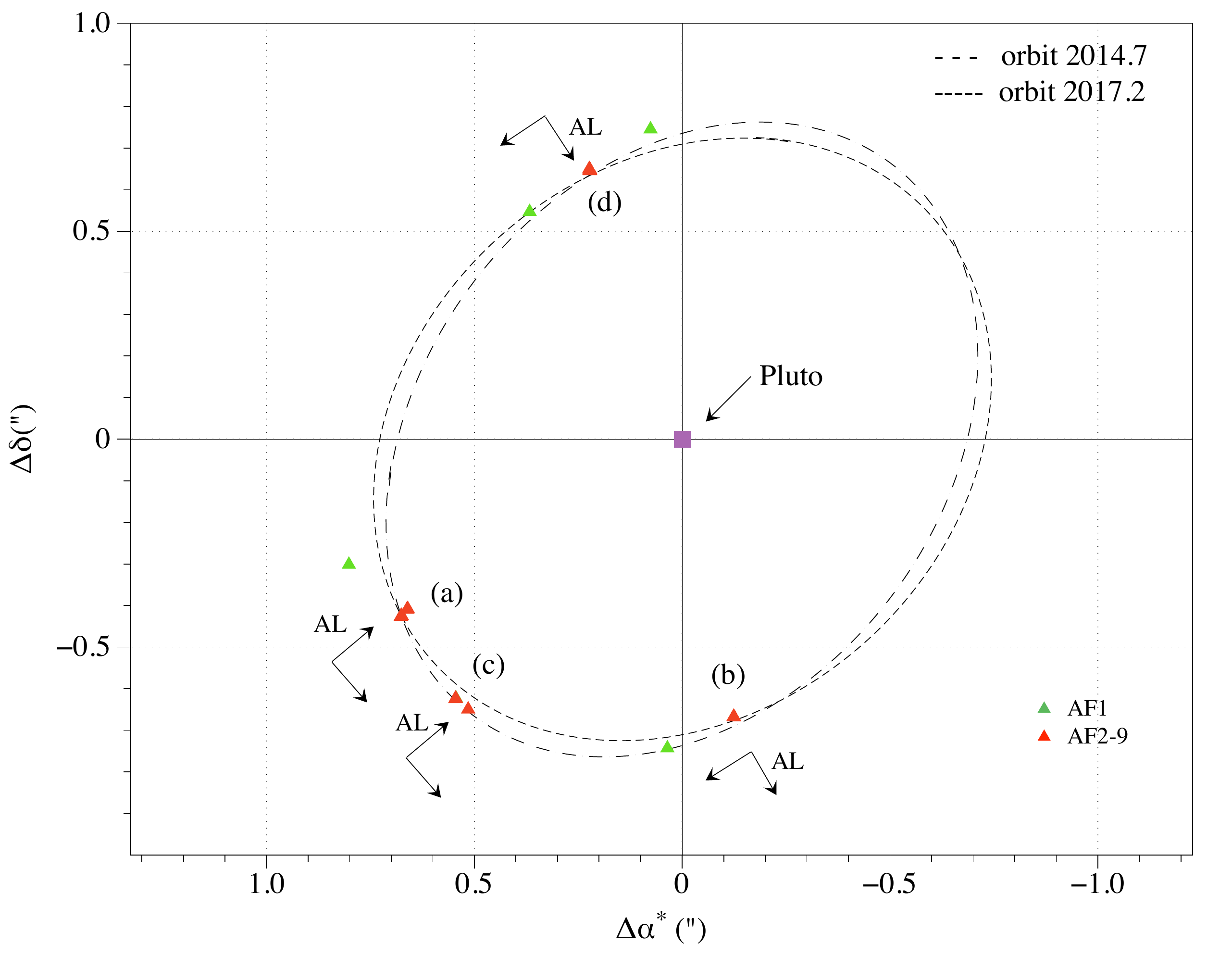}
\caption{Relative positions of Charon with respect to Pluto from \gdrthree, with the computed apparent orbits at the epoch of the first (a) and last (d) observation. The red triangles show the positions in AF2-AF9 and the green triangle shows AF1. The arrows indicate the AL-AC directions (see text for the details).}
\label{F:pluto_charon}
\end{figure}

We found that six transits were successfully resolved for Pluto and Charon, thus providing 44 accurate absolute positions for each body. We plot in Fig.~\ref{F:pluto_charon} the relative positions of Charon {as referred} to Pluto for these six transits, together with an outline of the apparent orbit at the times of the first (a) and last (d) transit. Pluto is at the centre of the plot. The (a) and (d) observations combine transits from both fields of view, while a single transit is available in the (b) and (c) passages. The green triangles are only for AF1 and were singled out because of a frequent offset in the relative positions for these observations (in the AC direction, therefore {irrelevant} for the astrometric accuracy of these measurements). The red triangles in (a) are stacked for 13 observations and {in (d) for 14 observations}. There are 6 observations for (b) and (c) in the upper triangle, including one AF1. The lower (c) triangle is AF8, and it is offset compared to the other six in this transit. The arrows indicate the direction of scan (AL) and its perpendicular in the AC direction. The successful resolution of the binary system occurs when Pluto and Charon are favourably oriented with respect to AL, that is, when their separation projected on AL is large. We have an excellent accuracy in this direction, and as expected, we find almost all the points on or very close to the computed orbital path at the relevant epoch, knowing that (c) is in September 2015, and (d) is much later, in March 2017. 

\subsection{Yarkovsky effect detection}
\label{S:yarko}
The Yarkovsky effect is the most important non-gravitational {perturbation} acting on small Solar System bodies~\citep{vok2000}. This secular perturbation produces a drift in the semi-major axis of the objects, changing the orbit of small asteroids over millions of years. It is now considered {to be} responsible for their migration from the main belt to the near-Earth region, and it represents the key to understanding the evolution of asteroid families~\citep{bottke01,spoto15,novakovic22}.

The Yarkovsky effect is proportional to the inverse of the diameter (larger on smaller bodies) and it depends on several physical quantities, such as the thermal inertia, the {Bond} albedo, the density of the object, and the rotation period, which {are usually unknown}. As a consequence, different methods have been developed to directly measure the Yarkovsky effect from the astrometry~\citep{farnocchia13,delvigna18,greenberg20}. 

These methods can easily lead to false detections, especially when the astrometry contains errors that are usually hidden in the low quality of the data. To avoid this possibility, a detection is usually considered valid if {the following conditions are met}:
\begin{itemize}
    \item The signal-to-noise ratio (S/N) of the Yarkosky measurement is higher than $3$.
    \item The ratio of the expected value and the actual value is lower than $2$, where the expected value is an approximation of the value that we would expect to find for an asteroid of the same size as the one we consider. 
\end{itemize}
For a complete description of the validation methods, we refer to~\citep{farnocchia13,delvigna18}. 
The Yarkovsky effect has so far been measured for $234$ asteroids (source JPL Small-Body Database\footnote{\url{https://ssd.jpl.nasa.gov/tools/sbdb\_lookup.html\#/}}), and all of them belong to the NEO population. The explanation is easy: we need accurate orbits and small objects, and these two characteristics are usually easier to be found in the NEO population. NEO orbits are better studied because of the {impact hazard, and radar observations can also be performed at their close approaches to the Earth. A combination of all these points makes their orbits more accurate and less prone to errors.}

As already mentioned at the beginning of the section, the Yarkovsky effect is also the main key for understanding the evolution of asteroid families. Families are generated by past collisions between asteroids. The orbits of the smaller members of the families moved from their initial configuration {because they were} perturbed by the Yarkovsky effect. A measurement of this effect gives us the age of the family, which corresponds to the time of the initial collision~\citep{spoto15}. The Yarkovsky effect has so far never been measured in the {main asteroid belt}, mostly because we lack accurate observations for MBAs. 

The orbits and their uncertainties were obtained as a result of the validation procedure. An orbit determination fit, independent from the one presented in Sect.~\ref{S:orbit_processing}, was performed with a modified version of the \texttt{OrbFit} software\footnote{\url{http://adams.dm.unipi.it/orbfit/}}. More information about the independent validation of \gdrthree Solar System observations can be found in \citet{DR3-DPACP-127}. 

\begin{figure}[h]
\centering
\includegraphics[width=0.475\textwidth]{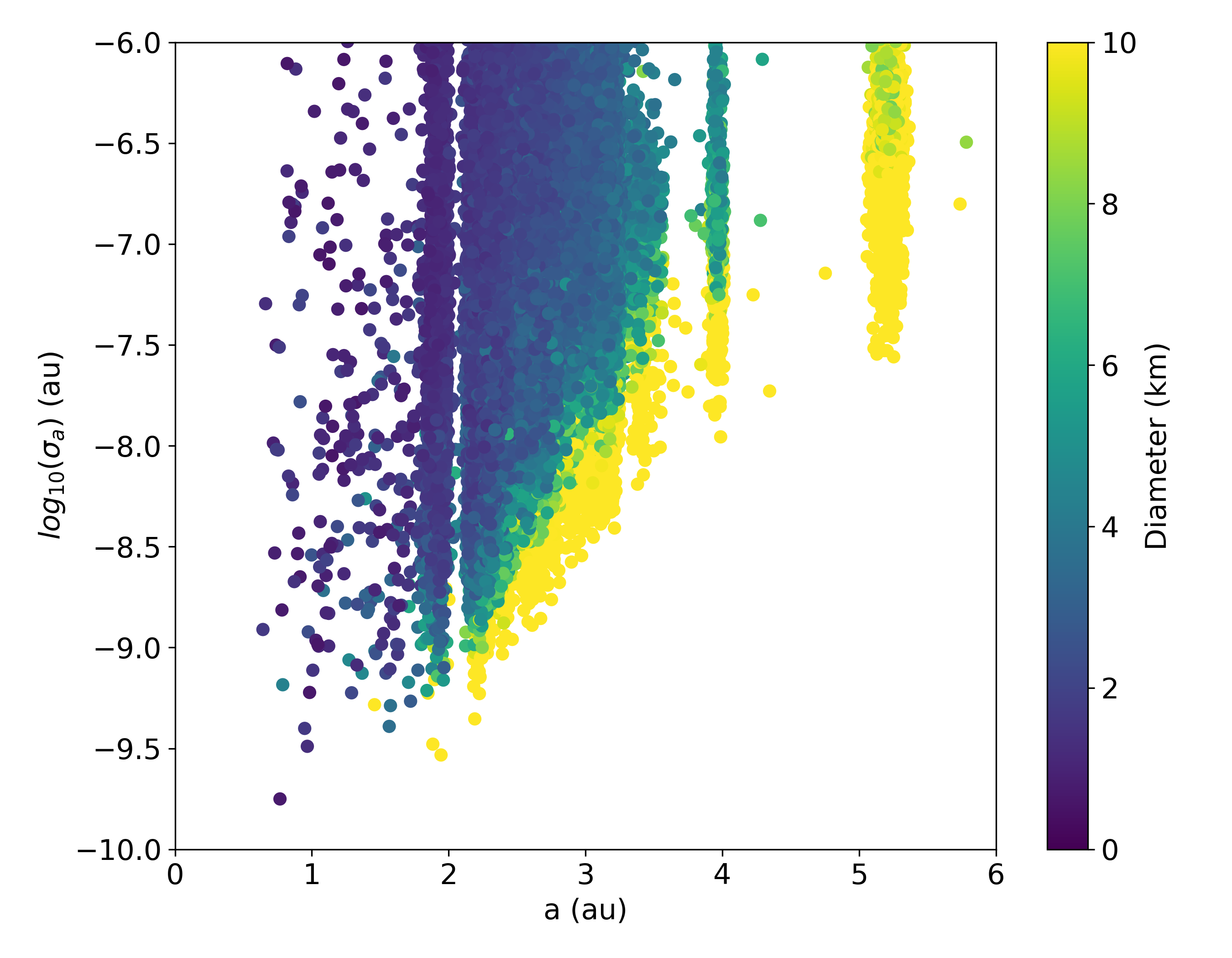}
\caption{Semi-major axis (au) vs $\log10$ of the semi-major axis uncertainty (au) for the objects for which the orbit determination procedure converged during the validation process. The colour bar represents the diameter estimate for each object from JPL SBDB.}
\label{fig:WP948_a_sigmaa_D_plot}
\end{figure}

The results in Fig.~\ref{fig:WP948_a_sigmaa_D_plot} were obtained using \gdrthree observations alone. They show the semi-major axis and its corresponding uncertainty $\sigma_a$. The colour bar represents the estimate of the diameter for each object from the {JPL Small-Body Database (SBDB)}. $\sigma_a$ represents a measure of the quality of the orbit. It is clear that the orbits of larger objects are better constrained in the main belt, while it is easier to find very good orbits among smaller {NEOs}, even using \gdrthree alone. 
The $34$ months of observations covered by \gdrthree are still a too short time interval to detect the Yarkovsky effect for {MBAs}. We do not expect to be able to use \gdrthree observations alone, but it is worth { noting} that some objects still reach extremely small orbital uncertainties, as shown in Fig.~\ref{fig:WP948_a_sigmaa_D_plot}. 

\begin{figure}[h]
\centering
\includegraphics[width=0.475\textwidth]{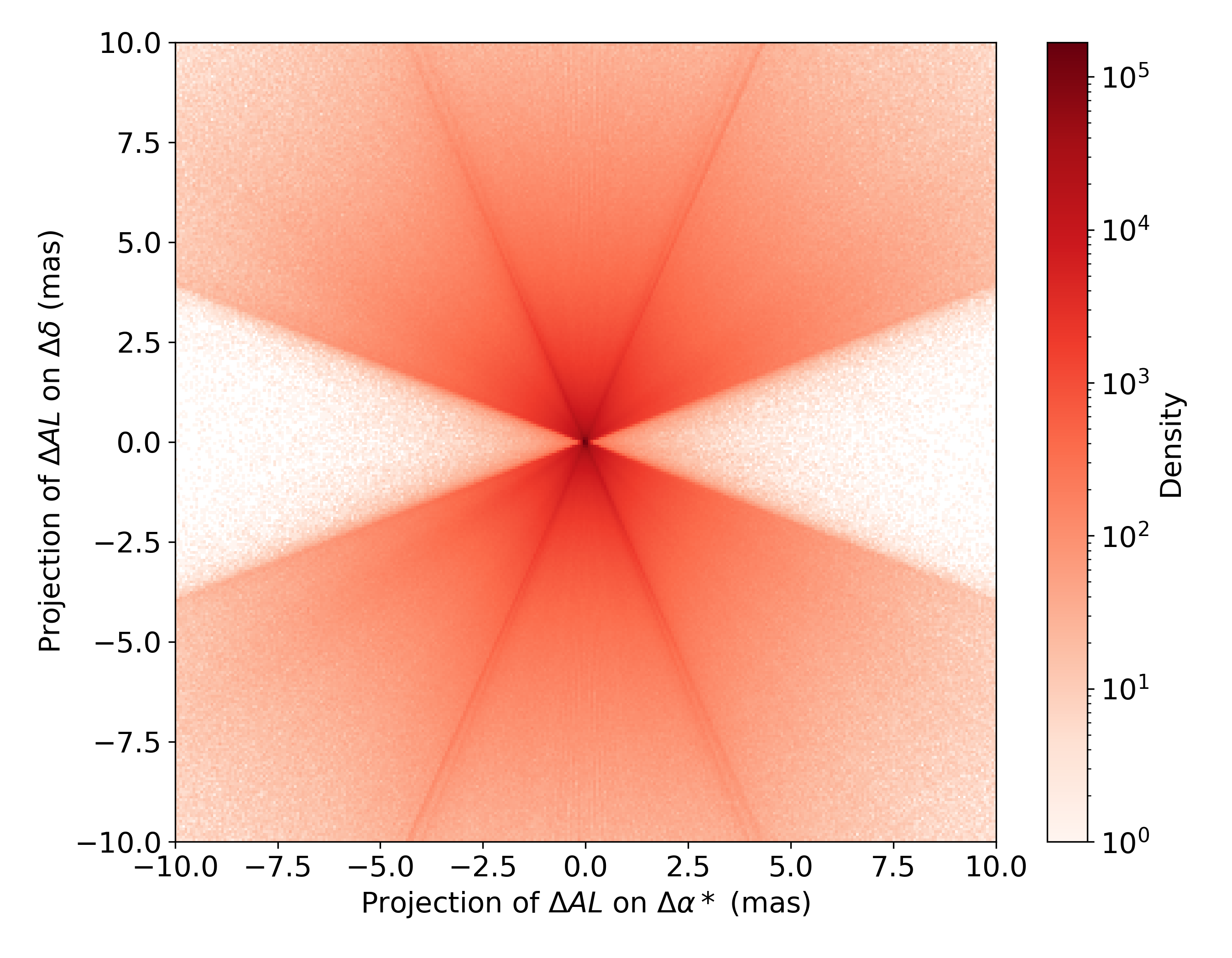}
\caption{Density plot of the projection of the $AL$ post-fit residuals in the $(\Delta\alpha \cos\delta, \Delta\delta)$ plane. The projection represents the quality of \gdrthree observations compared to the typical {sky-plane} residuals for ground-based observations. The diagonal most prominent lines correspond to the EPSL period.}
\label{fig:WP948_DeltaAL_projection}
\end{figure}

To fully exploit the data, we combined \gdrthree observations with all the available observations from the Minor Planet Center, \url{https://minorplanetcenter.net/}. The ultra-accurate \gdrthree observations, combined with the numerous ground-based observations, represent for {the main belt} the equivalent of having very accurate radar measurements for {the NEOs}. 
Figure~\ref{fig:WP948_DeltaAL_projection} shows the accuracy of the along-scan post-fit residuals as a projection on the $(\Delta \alpha \cos\delta,\Delta \delta)$ plane. In the projection, two main diagonal lines appear to be more dense: they correspond to the initial part of \gaia operations, when an ecliptic pole scanning law (EPSL) drove the observations. 

Figure~\ref{fig:WP948_DeltaAL_projection} also highlights the accuracy of \gdrthree data when compared to the ground-based post-fit residuals, which are typically about $500$ mas (two orders of magnitude higher than $Gaia$). 

\gdrthree contains $447$ {NEOs}. Of these, $432$ can be considered as small objects with a diameter smaller than 5 km, and $197$ are small and have a very good orbit uncertainty (the uncertainty on the semi-major axis is smaller than $3\times10^{-9}$~au). In the latter list, $24$ objects already have a good measurement of the Yarkovsky effect in literature. Most of the times, this was obtained through radar data. 

We take as an example the case of (3200) Phaethon, the parent body of the Geminid meteorite shower. Phaethon has already a well-established measurement of the Yarkovsky effect. From the JPL SBDB {$A_2 = -5.56 \pm 0.68$ $10^{-15}$~au/d$^2$}, where the value was obtained using $5090$ optical observations and $\text{eight}$ radar observations. We wish to show that {by removing radar observations and adding \textit{Gaia} observations,} we are able to find a very similar result. The goal is not to prove that we can neglect radar observations when working with NEOs, but to show the strength of \textit{Gaia} observations when radar is not available, for example in the main belt. We used a total of $6723$ observations. This total includes $356$ observations from \gdrthree. We fit the observations to determine the six orbital parameters as well as the $A_2$ parameter defining the Yarkovsky effect, {as in~\citet{farnocchia13} and \citet{delvigna18}}. We used the INPOP10e ephemerides~\citep{FiengaEtAl2016} to be consistent with the \gaia framework, a gravitational model including the $\text{eight}$ planets, $23$ massive asteroids, and a relativistic model as already described in~\cite{delvigna18}. We obtain a value of
$A_2 = -6.10 \pm 0.75$~au/d$^2$, which is in the 1$\sigma$ interval with respect to the JPL solution. This result is extremely important because we were able to fit \gdrthree with ground-based observations, we did not use radar observations, and we did not have to manually modify the existing astrometry to obtain a meaningful result. 

A second example we wish to present is a case for which the Yarkovsky effect could not be detected without \gaia astrometry. This is asteroid (1620) Geographos. Using all the available observations ($5242$ optical observations, $105$ observations from \gdrthree, and $\text{seven}$ radar observations) and the same method as described above, we find a value of {$A_2 = -3.25 \pm 1.01$~au/d$^2$}. It is clear that we are just above the S/N level of $3,$ and \gaia observations allowed the detection. 


\section{Photometric performance}
\label{S:photometry_performance}

The brightness of an SSO measured at any given epoch depends on the observing circumstances. In addition to the distance from the Sun and the observer, which can be easily taken into account if the orbit of the object is known, they include the rotation of the body around its spin axis, and the so-called aspect angle, namely the angle between the line of sight of the observer and the orientation on the celestial sphere of the object spin axis (the asteroid pole). Moreover, the illumination conditions at the epoch of observation are critically important. All these parameters vary over shorter and longer timescales and mean that any photometric measurement of an asteroid is an event that is cannot be repeated in practice.
In addition, the measured brightness of an SSO observed at any given epoch also depends upon a set of constant physical parameters of the object, including its shape, spin period, surface geometric albedo, and light-scattering properties \citep{kaasalainenAsteroidModelsDiskintegrated2002}.
The mechanisms of single and multiple scattering of sunlight incident onto the surface of an SSO determine the intensity of the flux of scattered sunlight that is measured from different directions. These mechanisms depend in a complicated way upon poorly known properties of the object surface, including particle sizes, shapes, and optical constants (composition), as well as the volume density and surface roughness of the surface regolith layer.

For the sake of simplicity, the light-scattering properties of an asteroid surface can
be described in terms of the dependence upon one single parameter, namely the phase angle. This is defined as the angle between the
directions to the Sun and to the observer as seen from the target body. 
\citet{Muinonenetal2015} showed that this can be a reasonable approximation for objects with 
symmetrical shapes and surfaces that scatter incident sunlight according to a Lommel-Seeliger surface reflection coefficient.
It is clear, however, that in the real world, we can expect far more complicated situations.

Based on the above considerations, the validation of sparse SSO photometric measurements is not straightforward. 
Except for a handful of objects that were observed in situ by space probes, it is practically impossible to make a sufficiently 
accurate a priori computation of the expected magnitude of any given 
asteroid observed at an epoch corresponding to given observing circumstances. For this reason, the validation procedures 
developed for \gdrthree SSO photometric data are based on two basic tools: an analysis of the phase-magnitude dependence 
of SSOs in the DR3 catalogue, as explained in Sect.~\ref{subsubphasemag},
and an analysis of the test results of inverting sparse photometric data for a sample of SSOs, as explained in Sections 
\ref{subsubphotinv} and \ref{subsubphotLutetiaSteins}.
The procedures and results described in the previous sections represent a generalisation and extension of procedures that were adopted in the validation of DR2 data \citep{Cellinoetal2019}. In the case of DR3, however, the available data are much better in quantitative and qualitative respects.

\subsection{Magnitude - phase relation}
\label{subsubphasemag}
Asteroid magnitudes are subject to significant shape-dependent periodic variations that are due to rotation 
around the spin axis.\begin{figure*}[t!] 
\centerline{
   {\includegraphics[width=75mm]{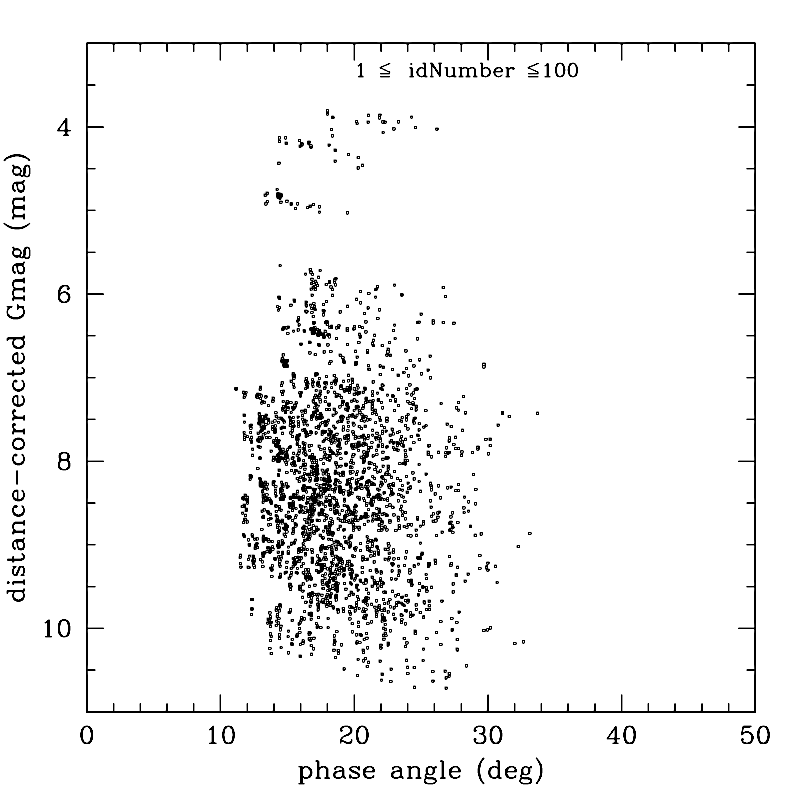}}
   \hspace{1cm}
   {\includegraphics[width=75mm]{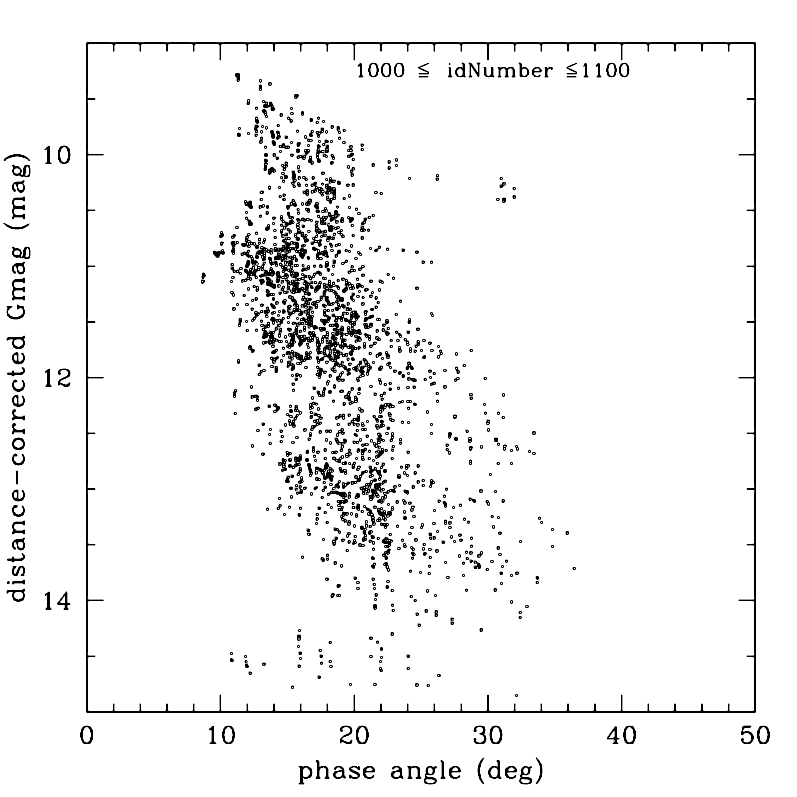}}}
 \centerline{
    {\includegraphics[width=75mm]{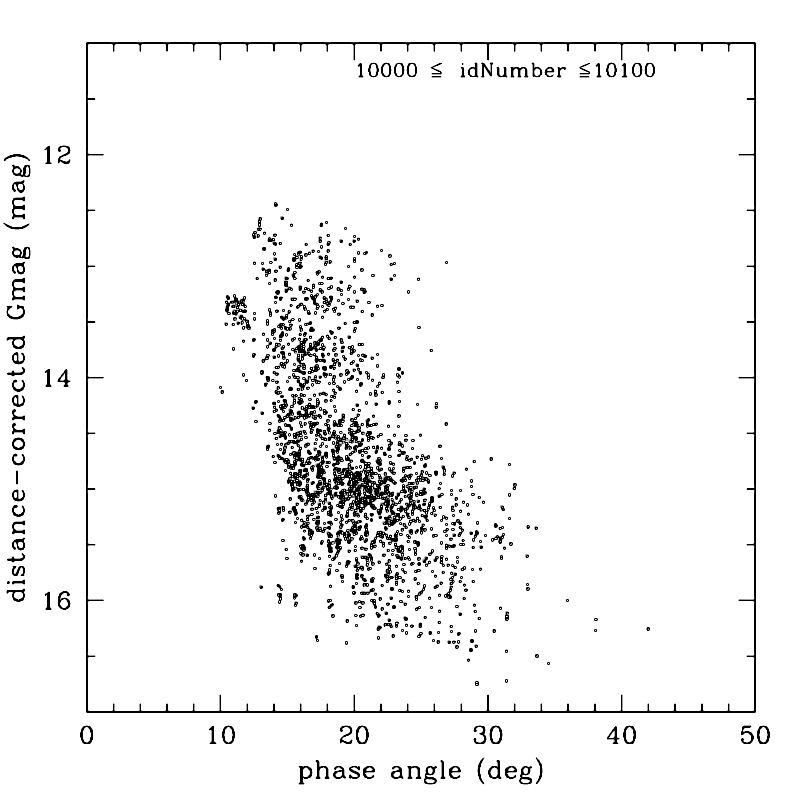}}
   \hspace{1cm}
    {\includegraphics[width=75mm]{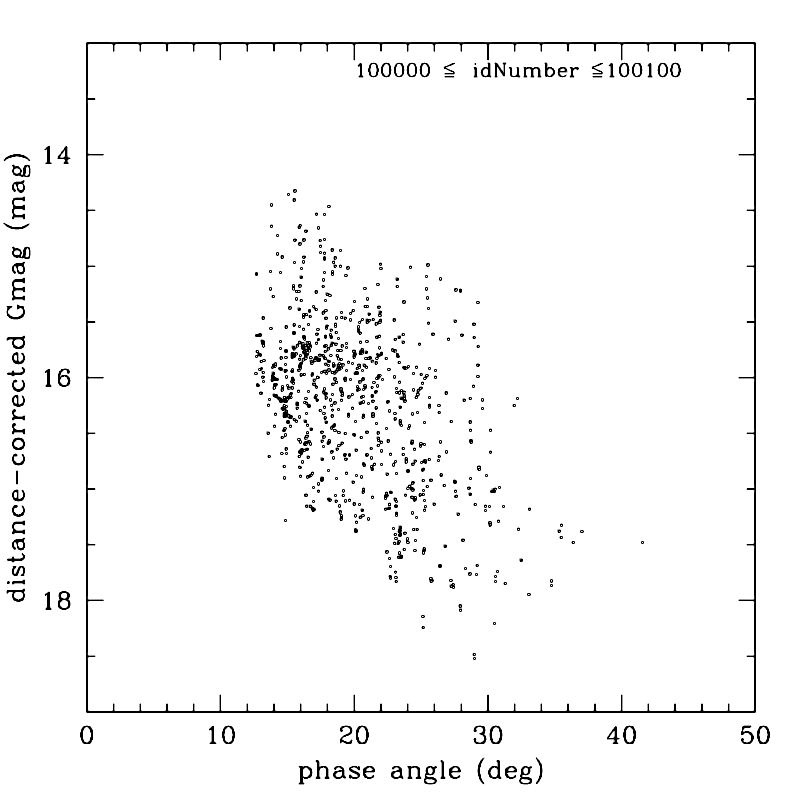}}}
\caption[Resulting phase - magnitude relations for selected samples of asteroids.]
{Computed phase - \gaia magnitude data for the set of asteroids numbered from $1$ to $100$ that
passed the filtering criteria described in the text (top left). The top right panel shows the same, but for objects numbered from$1000$ to 
$1100$. The bottom left panel shows the same, but for asteroids numbered from $10000$ to $10100$. The bottom right shows the
same, but for asteroids numbered from $100000$ to $100100$.}
  \label{fig:phase-magsubsamples}
\end{figure*}
\begin{figure*}[t!] 
\centerline{
   {\includegraphics[width=75mm]{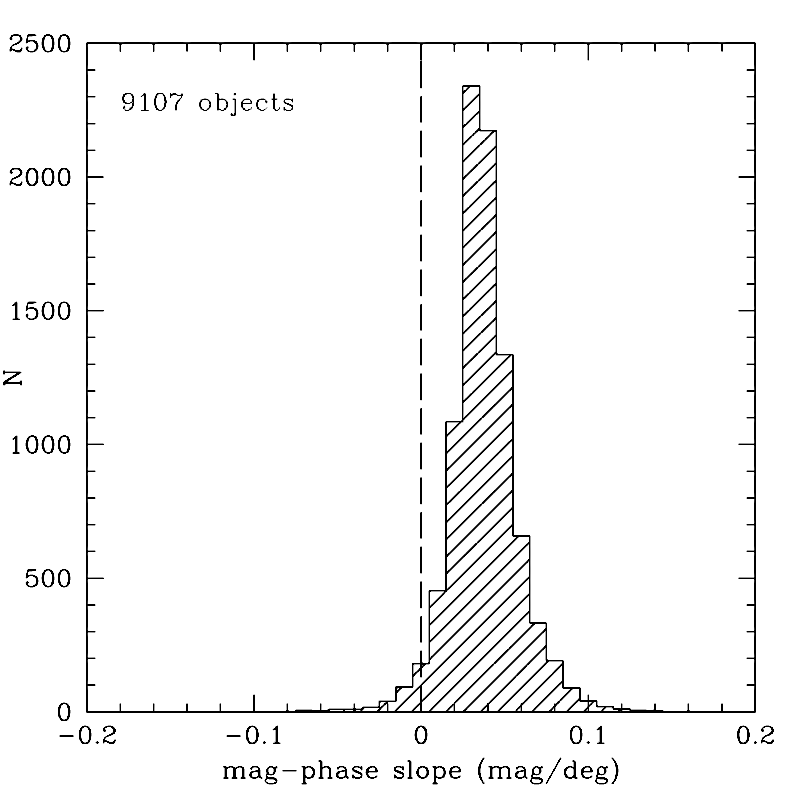}}
   \hspace{1cm}
   {\includegraphics[width=75mm]{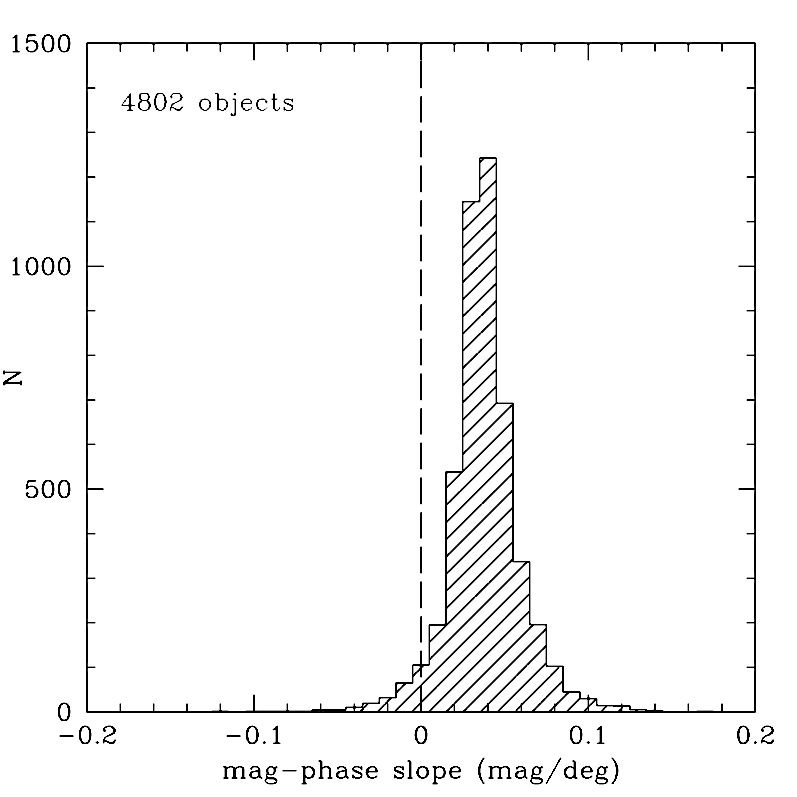}}}
 \centerline{
    {\includegraphics[width=75mm]{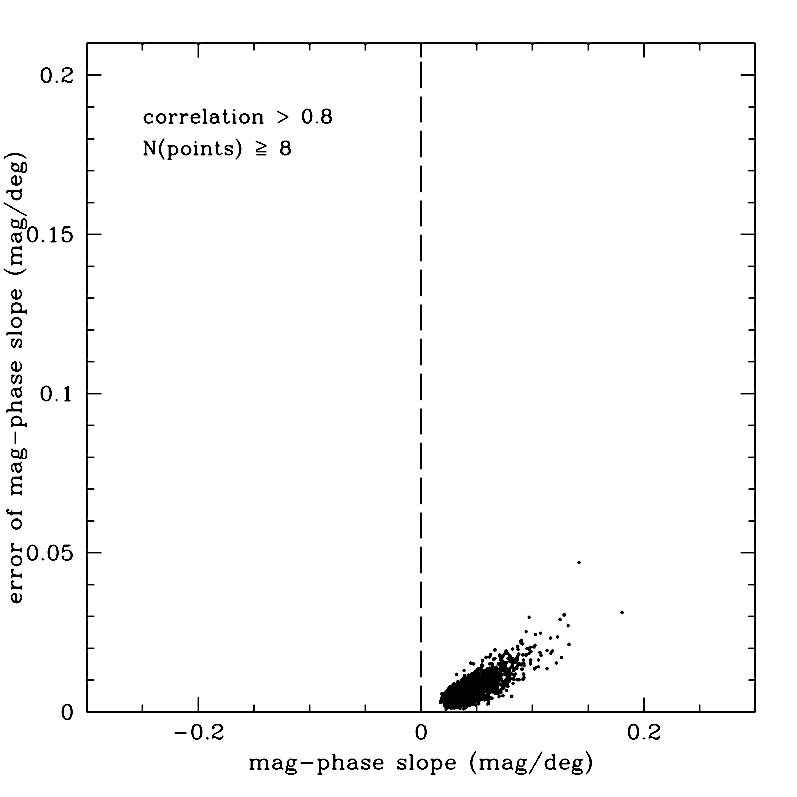}}
   \hspace{1cm}
    {\includegraphics[width=75mm]{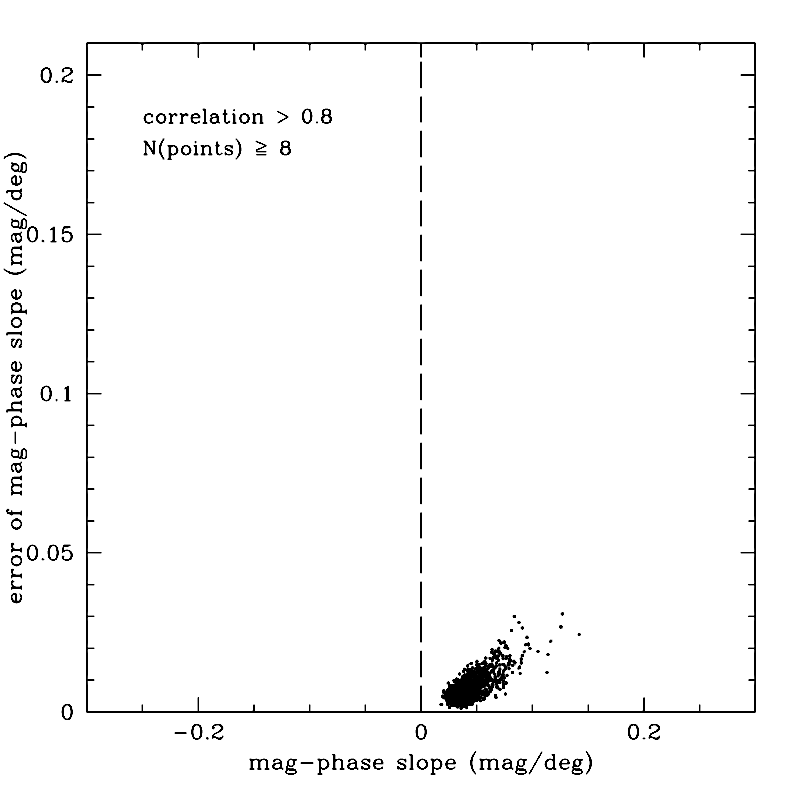}}}
\caption[General analysis of the phase - magnitude relation for two samples of available data.]
{Histogram of the computed slopes of the linear fits of the phase - magnitude for asteroids
numbered from 1 to 10000 (top right). The top right panel shows the same, but for objects numbered from $100000$ to 
$110000$. The bottom left panel shows the slope error vs{\it } slope for asteroids numbered from 1 to 10000 after 
those with values of the resulting linear correlation $<0.8$ and/or a number of accepted transits $<8$ were removed.  
The bottom right panel shows the same, but for asteroids numbered from $100000$ to $110000$.}
  \label{fig:phasemaghistog}
\end{figure*}
In classical ground-based studies, the data normally consist of full photometric light curves obtained 
at different phase angles. The phase-magnitude relation is then derived by considering only magnitude values that were taken at the
maximum (or mean) {brightness} value of each light curve in order to avoid to mix magnitude data corresponding to different cross 
sections of the rotating body. 

Equally importantly, the photometric data in ground-based studies are generally collected during one single apparition of an 
object, namely during a relatively short interval of time (several weeks), during which the object is seen in a nearly 
constant geometric configuration. Only the illumination conditions, described by the phase angle, vary with time.
{Recent examples of the derivation of phase-magnitude relations are presented by \citet{Albinoetal2019} and \citet{mahlkeAsteroidPhaseCurves2021}.}

In the case of \gaia data, however, we deal with sparse measurements taken during a considerable interval of time and
covering an interval of phase angles that for main-belt asteroids generally ranges from $10^\circ$ to $30^\circ$.
The phase-magnitude relation derived by these data is intrinsically noisy because it includes measurements taken at epochs
corresponding to different illuminated cross-sections. This is due both to differences of rotational phase around the spin axis and 
to differences of the orientation of the body with respect to the line of sight when data are taken at epochs that are sufficiently
distant in time. In other words, the phase-magnitude curves shown in the \gdrthree database are contaminated by magnitude 
variations that are not uniquely due to differences in phase angle. This is also a commonly encountered situation for the majority of ground-based data, with the exception of targeted campaigns.

These problems can be partially overcome when we consider the statistical behaviour of a large number of objects. This is made possible by exploiting the fairly large \gdrthree database.
This allows us to adopt some filtering procedures 
aimed at mimicking the procedures that are traditionally adopted in ground-based studies more closely.
In particular, of all the available magnitudes that were measured within the same day, generally 
corresponding to two or more consecutive detections in the two {FOVs} of Gaia, only the brightest recorded magnitude was kept in the analysis for each object. This was done to limit the noise that is uniquely due to the rotation of the object around its spin axis. This allowed us to mimic the procedures used in the analysis of full light curves taken from the ground more closely, as explained above. Moreover, transits for which the apparent \gaia magnitude had a nominal error $\ge 0.05$~mag were not taken into account. 
It was also decided to discard all objects for which the interval of phase angles covered by the observations was exceedingly narrow, {$\le 5^\circ$}. As a next step, all asteroids were discarded for which the number of accepted magnitude measurements was smaller than 
a limit related to the phase angle interval covered by the observations. This limit was set
to 4 when the covered phase angle interval was larger than 9 deg, it was set to 5 when the phase angle interval was between 6 and 9 deg, and it was set to 7 when the covered phase angle interval was between 5 and 6 deg. 

The apparent magnitudes of objects that passed these filters were converted into unit distance from the
Sun and from \gaia. A linear least-squares fit of magnitude versus phase was finally computed.
The derived phase - magnitude relations exhibit the {typical behaviour} of asteroids. They are characterised by an 
overall linear trend in the interval of phase angles covered by \gaia observations. 
Some examples of the obtained phase - magnitude relations for different subsets of 100 accepted objects chosen in 
different magnitude regimes (corresponding to objects numbered in different intervals of identification number) {are shown in 
Fig.}~\ref{fig:phase-magsubsamples}. 

The plots shown in Figure \ref{fig:phase-magsubsamples} correspond to the overlapping of data corresponding to many different
objects, and a general assessment of the statistical behaviour of the whole sample of 
objects passing our selection filter is difficult to derive. 
We therefore performed a more extensive statistical analysis of the behaviour of two much larger samples of objects, considering their number identifier as a proxy for brightness: 
the first sample included asteroids numbered from 1 to 10000. Of these, 9107 passed our filtering procedure and were accepted 
for computation of the phase-magnitude relation. The second sample was chosen to be that of asteroids numbered from 
100,000 to 110,000. Of these, 4802 were accepted for the phase - magnitude analysis. The decreasing number of accepted 
asteroids in a fainter magnitude regime is related to the decreasing number of acceptable transits for increasing faintness. 
The two samples allowed us to compare the phase-magnitude behaviour of objects in two {clearly} different magnitude regimes. 
Some results are shown in Figure \ref{fig:phasemaghistog}. They display some general properties of the two considered 
samples. 

We found that the correlation of the obtained linear fits was quite variable, with a sharp maximum for values 
between 0.8 and 0.9, but far lower values were also included. This was an expected consequence of dealing with data that,
even after the filtering procedures described above cannot completely remove the effects caused by the sparseness in time.
The nominal error of the computed linear slopes was predominantly about $0.01$ mag/degree, but higher values
(but very rarely higher than $0.03$ mag/degree) were found in a non-negligible number of cases. 
{A histogram} of the slopes of the obtained linear fits is shown for the two considered samples in the top panels
of Figure \ref{fig:phasemaghistog}. 
The variety of obtained slopes {generally agrees} reasonably well with typical values mentioned in the literature using 
ground-based data. These typical values range mostly from $0.01$ ti $0.04$~mag/deg, {as discussed for instance in \citet{Albinoetal2019} and \citet{Muinonenetal2010}}. 
However, some cases of negative slopes (corresponding to bodies whose brightness would increase for increasing
phase angle, a clearly aberrant result) are found for a minority of cases. Some very high values for the slope
($\ge 0.06$ mag/deg) are also found. The existence of these aberrant cases is very likely {to be due to} insufficient 
removal of transits corresponding to an exceedingly high variety of observational circumstances, and/or to objects with 
an insufficient number of available measurements. As shown in the bottom panels of Figure \ref{fig:phasemaghistog}, which show a
slope - error (slope) plot for each of the two considered samples, but after objects for which the
linear correlation of the phase - magnitude data turned out to be $<0.08$ mag/deg and/or the number of accepted 
magnitude measurements was $<8$ were removed, the resulting linear fits are well compatible with typical ground-based values. 
All the negative values of the slope disappear, while the number of high positive values reduces to just a few per sample.

To summarise, we conclude that a preliminary analysis of the phase-magnitude relation using the SSO photometric data available 
in \gdrthree indicates that in the vast majority of cases, the observed phase-magnitude behaviour of the SSOs is nicely 
compatible with the expectations when the unavoidable noise arising from the use of limited numbers of 
sparse photometric data that are taken in a range of epochs that can correspond to non-negligible differences in observing circumstances are taken into account.
We also emphasise that in our analysis, we found evidence of differences in the behaviour of objects belonging 
to different taxonomic classes, in agreement with current ground-based evidence. In particular, according to 
\citet{BelskayaShevchenko2000}, dark asteroids tend
to have steeper phase-magnitude slopes than moderate-albedo asteroids \citep{mahlkeAsteroidPhaseCurves2021}.
In this respect, we note that Figure \ref{fig:phasemaghistog} suggests a small shift of the peak (from $0.03$ to $0.04$ mag/deg) 
in the histogram for the objects numbered above 100,000. The reason might be that fainter 
objects tend to be located preferentially in the outer region of the asteroid main belt, where dark bodies are predominant.
The distribution of photometric slopes among objects belonging to different taxonomic classes and orbiting at different 
heliocentric distances will be better analysed in future data releases.

We are aware that the analysis of the phase-magnitude relation described in this section must be considered
very preliminary. Recently, a deep analysis of phase-magnitude data published in \gdrtwo has been carried out by
\citet{Martikainenetal2021}, who used realistic asteroid shape models and computed very good fits of phase-magnitude data against 
the ($H,G_1,G_2$) phase function developed by \citet{Muinonenetal2010}. We expect that a similar analysis
will be performed as soon as the \gdrthree data will become public.

In this respect, we expect that the future availability of much larger 
numbers of transits per object will allow us in future data releases to produce cleaner phase-magnitude plots and to
analyse the correlations of the photometric behaviour of asteroids with their spectroscopic 
properties and the albedo in more detail and to develop a new taxonomy based on \gaia spectroscopic data.

\subsection{Photometric inversion}
\label{subsubphotinv}

Using magnitudes reduced to unit distance, we can analyse the variation in the brightness of any given object that is measured at different epochs. For a set of sparse photometric measurements of the same object, it is convenient to work in terms of brightness differences with respect to one of the available measurements (usually the first measurement). In this way, any dependence upon constant light-scattering properties of the surface can be removed, assuming that the surface has homogeneous properties. This reduces the time-dependence of the measured brightness data to a function of the following physical properties:
\begin{enumerate}
\item The rotation period $P$. This is the rotation of the object around its spin axis that continuously modifies the cross section of 
the illuminated area seen by the observer, depending upon the object shape.
\item The overall shape, which is described by a number of unknown parameters.
\item The orientation of the spin axis with respect to the line of sight of the observer (the pole of the object). This corresponds to two unknown parameters, namely the ecliptic longitude and latitude of the pole itself.
\item The dependence of the brightness upon the illumination conditions, which is described in terms of the phase angle 
(see \ref{subsubphasemag}). A simple dependence upon the phase angle summarises for the sake of simplicity the overall effect of the mechanisms of single and multiple scattering of the sunlight incident onto the surface.
These mechanisms determine the intensity of the flux measured by the observer in different observing circumstances, with a complicated dependence upon poorly known surface properties, including albedo, texture, and roughness. These can be assumed to be a constant but unknown function of the phase angle for any given object.
\end{enumerate}

Based on these considerations, it is in principle possible to develop numerical codes to determine the unknown physical 
parameters of an object whose brightness has been measured by \gaia in a sufficiently large number of observed transits. 
A code like this, based on a genetic algorithm, has been developed for the purposes of \gaia data processing. It will be used to produce 
results of SSO photometric inversions in future \gaia data releases. The algorithm assumes that the shapes of the object are
triaxial ellipsoids, described by two parameters (axial ratios), and that there is a linear variation in magnitude as a 
function of the phase angle (see \ref{subsubphasemag}).
This algorithm has already been adopted in preliminary analyses
of \gdrtwo data \citep{Cellinoetal2019}, and has been used to validate \gdrthree photometric data, as
explained in what follows.

A preliminary step was identifying objects with reliable predictions of the brightness at any given epoch of 
observation. We profit in this way from a detailed knowledge of the physical properties characterising the object. In 
principle, the best possible validation test must consist of comparisons between expected and measured magnitudes for a 
set of objects for which our knowledge of their physical parameters is extraordinarily accurate, being based on in situ
{\em } measurements carried out by space probes. The number of these objects is unfortunately extremely small, 
and they deserve a separate treatment. It is therefore necessary to take advantage of larger data sets of ground-based asteroid photometric data. Decades of ground-based photometry, mostly at visible wavelengths, have produced large 
catalogues of asteroid light curves. The rotation periods of about 10 000 of these asteroids have been derived with good accuracy. A smaller number of these objects have been observed in a variety of observation
circumstances sufficient to derive accurate estimates of the spin axis direction for them. In many cases, more than one 
pole solution is found to be compatible with available data. In a large number of cases, an overall shape, derived using 
complex algorithms of light-curve inversion, has also been obtained. Shape details, however, are of limited importance 
in our case, because we compute photometric inversion using a simple triaxial ellipsoid model, which is only a first 
approximation of what the real shape of an asteroid can be. In our analysis, we used currently available catalogues 
of asteroid rotation periods and pole coordinates. In particular, we massively exploited the database of
asteroid models from inversion techniques (DAMIT), which is publicly available \citep{durechDAMITDatabaseAsteroid2010}
\footnote{\href{https://astro.troja.mff.cuni.cz/projects/damit/}{https://astro.troja.mff.cuni.cz/projects/damit/}}. 
For objects not included in DAMIT, we took the periods from the asteroid 
light curve database catalog, available at the web site of the NASA Planetary Data System.

We limited our analysis to the brightest asteroids, numbered from 1 to 1000, for which at least $30$ accepted transits were available in the \gdrthree database. In particular, we accepted only transits for which the nominal error of the G magnitude was not larger than $0.02$ mag because the results of photometric inversion can be negatively affected when data of insufficient quality are used. In any case, our sample of asteroids includes objects that are sufficiently bright to have systematically smaller magnitude uncertainties than the above limit. The resulting sample used for our photometric inversion test includes $430$ asteroids. 

We note that the results of our photometric inversion algorithm are also able to distinguish between prograde and retrograde
rotation, according to IAU criteria. In particular, objects found to have a retrograde rotation have negative assigned values of
their estimated $P$ value, while the determined spin axis direction, expressed in ecliptic longitude 
and latitude, has always an assigned positive value for the latitude of the pole.

\begin{figure}
\centering
\includegraphics[width=0.5\textwidth]{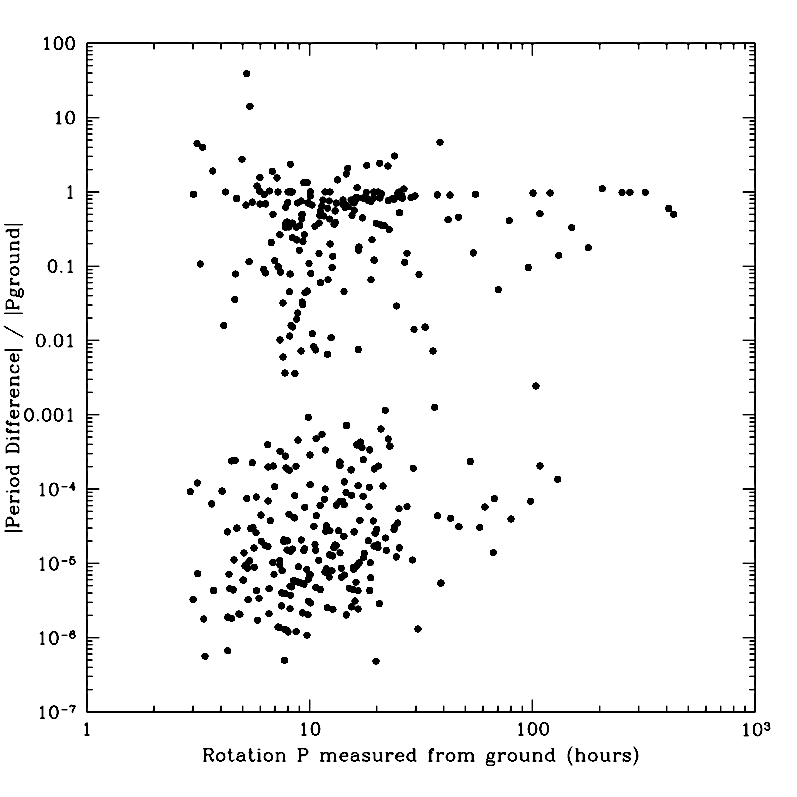}
\caption[Results of the determination of the spin period $P$ from inversion of \gdrthree data]
{Relative difference between the absolute value of the difference between the resulting $P$ solution and the $P$ value determined by ground-based observations in units of the ground-based $P$ value for each object of the considered sample.}
\label{Fig:relPPdiff}
\end{figure}

\begin{figure*}[t!] 
\centerline{
   {\includegraphics[width=75mm]{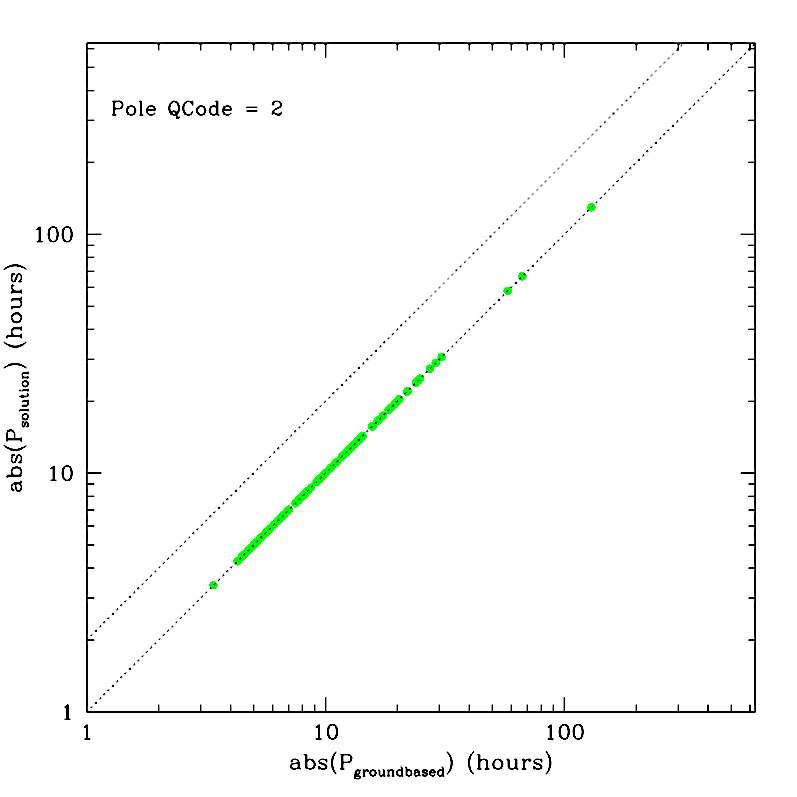}}
   \hspace{1cm}
   {\includegraphics[width=75mm]{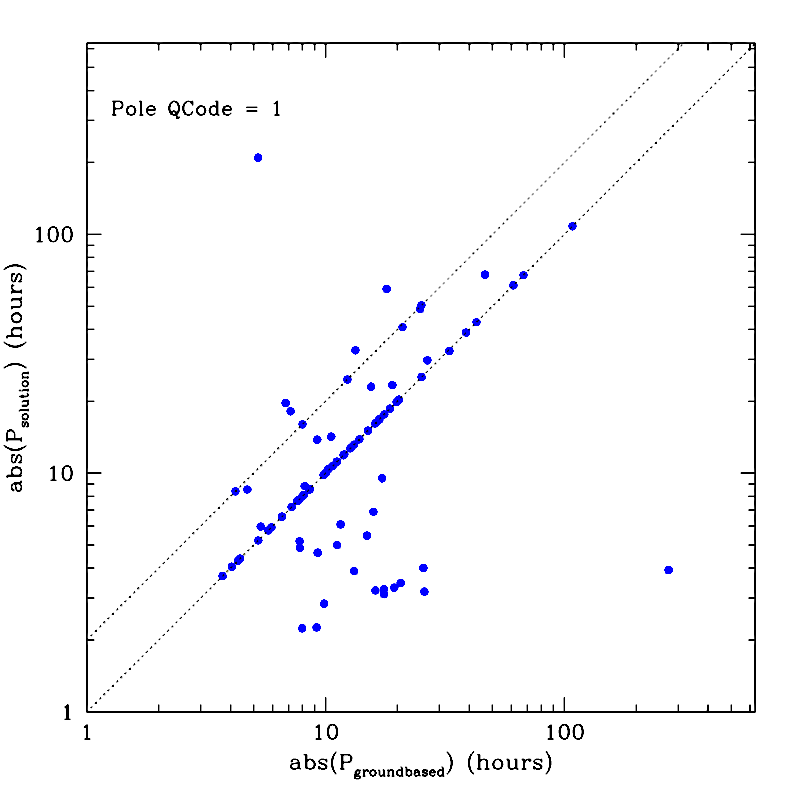}}}
 \centerline{
    {\includegraphics[width=75mm]{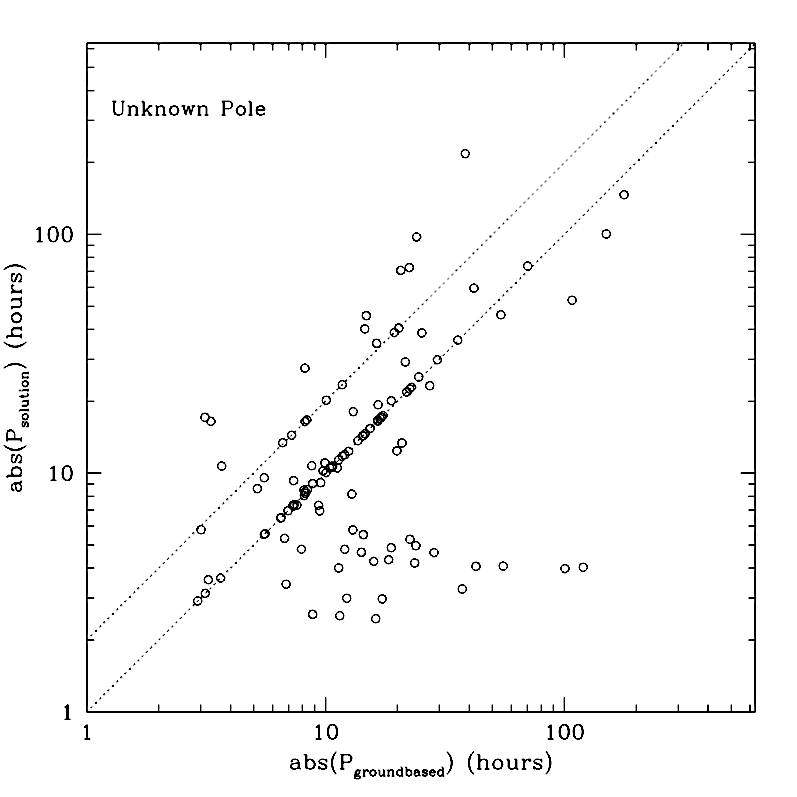}}
   \hspace{1cm}
    {\includegraphics[width=75mm]{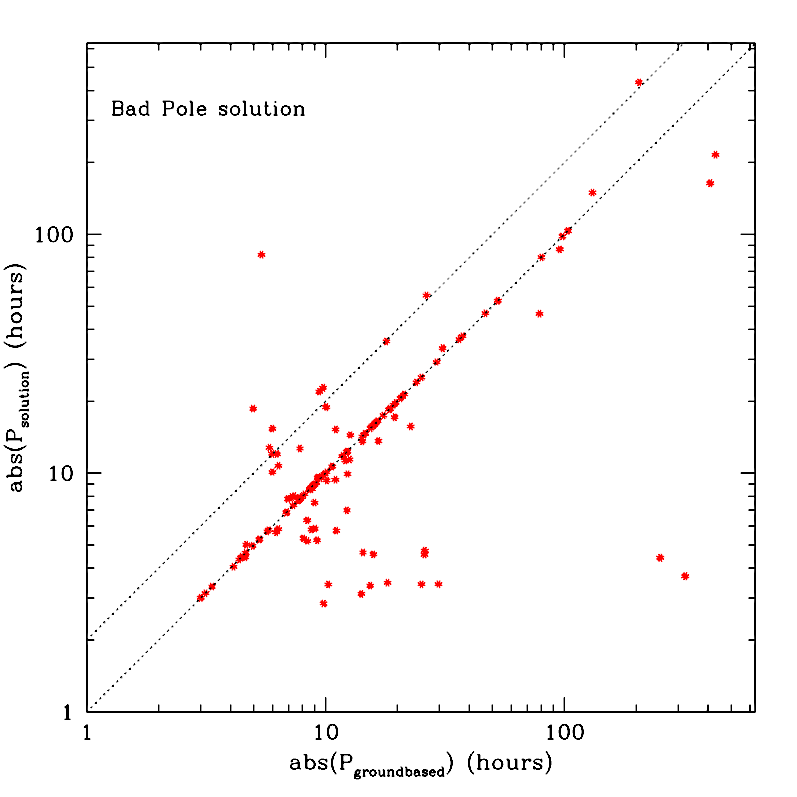}}}
\caption[Period solution vs. ground-based period solutions for different pole quality code solutions.]
{Comparison of \gdrthree inversion solutions for the spin period $P$ and ground-based $P$ determinations for all asteroids of our sample with a pole quality code = 2 (best inversion solutions; top panel). The top right panel shows the same, but for objects with a pole quality code =1.
The bottom panels show the same as the top panels, but for objects with a quality code solution = 0 (corresponding to an unknown pole solution from ground-based data) and for objects with a pole quality code = -1 (corresponding to complete disagreement with any existing ground-based pole solution). The upper line represents periods derived from \gdrthree photometry that are exactly twice those obtained from the ground. }
  \label{fig:PPQC}
\end{figure*}

It is important to note that the adopted algorithm for photometric inversion produces for each object a set of 
15 different inversion solutions for each object.
The reason is that an intrinsic property of the adopted genetic algorithm is that it does not always converge to a unique 
final solution. This happens because the evolution of the population of possible solution parameters can often evolve along a dead branch, leading to a poor-quality solution. $\text{Fifteen}$ different genetic solutions have been found to be a 
reasonable compromise between the need to obtain a good-quality solution and the need of minimising the CPU execution time.

We stress that at this stage, only the best obtained solution for each object, namely the solution that produces the smallest residuals
with respect to the measured magnitudes at the available transits,
was considered and compared with ground-based results. This choice has some 
consequences, because in several cases, more than one inversion solution gave equivalent 
residuals. We defined as equivalent residuals those that differ by no more than $0.0015$ mag. As a consequence, the inversion solution is not unique in some cases. In these cases, the nominally best solution 
may not correspond to the rotation {period and pole} listed in ground-based catalogues, whereas another equivalent 
solution that is not considered for the moment would correspond to these values. For this reason, the results presented here are conservative. 

We recall that for the purposes of inversion of a set of sparse photometric measurements, it is of 
paramount importance to have a good sampling of the possible observing circumstances for any given object. This means that 
the data should include measurements that adequately sample the whole interval of 360 deg in ecliptic longitude. This is 
especially important for the determination of the spin axis orientation (the asteroid pole). However, this is not yet the case 
for the data that are available in \gdrthree. For each object, large gaps 
exist in the interval of the covered ecliptic longitudes. The publication of photometric inversion of \gaia data is 
scheduled as an end-of-mission task. The results of the preliminary inversion attempts presented here must be considered as 
no more than a useful tool for the scientific validation of \gdrthree data. We expect to obtain much better inversion solutions 
in future data releases as new measurements will become available. 

We considered as a successful determination of the rotation period $P$ an inversion solution for which {the} absolute value of the difference between the resulting $P$ solution and the $P$ value determined by ground-based 
observations, expressed in units of the absolute value of the ground-based $P$ value, expressed in hours, is not higher 
than $0.001$. This criterion takes 
into account the fact that for fast rotations of just a few hours, small errors in $P$ lead to strong 
differences in the rotational phase of the object, which is computed at epochs differing by few years. On the other hand, in the case of 
very long rotation periods, longer than several dozen or hundreds of hours, it is not reasonable to impose a
required accuracy of about a few seconds, for example, on the determination of the period. 

Based on our adopted criterion, we obtained the correct $P$ solution for $229$ out of $430$ asteroids of our sample. 
These results are shown in {Fig.~\ref{Fig:relPPdiff}. This figure} shows an interesting feature: in addition to the $229$ cases of correct $P$ determinations,
a significant number of cases exist,for which the $P$ value determined by photometric inversion is 
nearly exactly twice the $P$ value determined from ground-based photometry. This feature is not entirely unexpected considering the assumptions of the adopted model. In particular, the algorithm
assumes that over a full rotation of the object, the brightness reaches two maxima and two minima, which are expected
{to be} (nearly) equal. In the real world, however, and in particular when an object is
not strongly elongated and the maxima and minima tend to be shallow and/or strongly asymmetric, the 
inversion algorithm might derive a rotation period that can be twice the correct one. Moreover,
at least in some cases, a derived double period might be indicative of a photometric behaviour that is dominated not by the shape, 
as assumed by the inversion model, but by variation in surface albedo. 
This is the case, for instance, of the large asteroid (4) Vesta \citep{CellinoVesta}, which is known
to have a light curve producing only one maximum and minimum per cycle.

It seems therefore that the results shown in Fig.~\ref{Fig:relPPdiff} are very encouraging when the limits of the adopted shape model, the small minimum number of 
accepted transits per object, the
still limited variety of observing circumstances, and the conservative criteria of definition of the inversion solution are taken into account. Ambiguous cases are not considered, together with the the non-negligible number of solutions corresponding to a $P$ twice as large
as ground-based determination.

To determine the pole, the situation is intrinsically more complicated. In the vast majority of cases, two or more
different pole solutions per object are listed in the literature. Moreover, in our assumption of a triaxial ellipsoid shape model, 
an ambiguity of 180 degrees in the determination of the ecliptic longitude of the pole may be present in some cases due to 
the symmetry of this shape model. In particular, a $180^\circ$ ambiguity on the longitude of the pole can be triggered by
a low orbital inclination of the object {or} by an unfavourable distribution of the observations in ecliptic 
longitude. Moreover, simulations have shown that a low ecliptic latitude of the pole of the object tends to
make {the photometric inversion more challenging with} the assumed shape model \citep{SantanaRos2015}. 
We limited our analysis to a comparison between the obtained pole solutions and the DAMIT database of asteroid poles, which is
considered to be the most accurate list of ground-based asteroid period and pole determinations. 
Based on the results of our analysis, we decided to subjectively define a small number of  pole solution quality classes (QC). 
In particular, we assigned pole QC~=~2 to pole solutions that differed by no more 
than about 10 deg (separately) in ecliptic longitude and latitude with respect to one existing DAMIT pole solution, and we 
also imposed that the rotation period determination was correct. We assigned pole QC = 1 to objects for which the best pole solution was 
not so close to a DAMIT pole solution, independently of the obtained period. We assigned pole QC = -1 in situations in which the 
obtained pole solution had little to do with any existing DAMIT solution. Finally, we assigned pole QC = 0 to objects for which no 
DAMIT pole solution exists. 

Figure \ref{fig:PPQC} shows plots of the {period} determined by inversion of \gdrthree data versus the $P$ value known
from ground-based determinations for objects belonging to different pole solution QCs. The figure shows that except for the case of pole solution QC~=~2, for which a (nearly) perfect agreement with ground-based {period}
solutions is imposed by definition, when objects belong to different pole solution QCs, the relation between the QC of the pole solution and the success in the period determination by inversion of \gdrthree data is not always obvious. In particular, poor pole determinations (red points in the bottom right panel of {Fig.}~\ref{fig:PPQC}) do not
correspond to a larger fraction of {erroneous period} determinations. We note, however, that the fraction of cases in which
the period from \gdrthree photometric data inversion tends to disagree with ground-based period determinations tends
to increase when we consider objects for which no pole solution from ground-based data is available (see the bottom left
panel in {Fig.}~\ref{fig:PPQC}). This might be considered as an indication that asteroids for which no ground-based pole
determination is available may well be challenging cases, for which even the ground-based period solutions could be more
uncertain and possibly incorrect. The relatively high number of objects with long
rotation periods among those determined by ground-based data should be noted in particular, and for which ground-based data are so far insufficient
to derive a pole solution. Long rotation periods correspond in many cases to relatively large uncertainties in the 
determination of the period.

Based on the results shown in this and the previous section, we conclude that \gdrthree photometric data
are of a good if not excellent quality because we successfully produced a correct inversion of a large number
of objects using a clearly simplistic shape model and relatively small numbers of measurements covering a still partial 
fraction of the possible observing circumstances. We limited our analysis for the moment to measured \gdrthree magnitudes 
with nominal errors not exceeding $0.02$ mag. The results of our photometric inversion attempts seem to be very encouraging.
In principle, we cannot rule out the possibility that in some cases, the
error bar of some \gdrthree magnitude might be higher than the nominal value. In some cases, this might
affect the results of photometric inversion negatively, and might explain some incorrect results.

More convincing tests should be based on the analysis of data of 
asteroids for which our knowledge of the rotation period and spin axis direction are of the best possible reliability.
This is the case of a very small number of objects that were visited in situ by space probes. The results for some of
them, for which we have a reasonable number of \gdrthree observations, are shown in the next section.

\subsubsection{Photometry of (21) Lutetia and (2867)~\v{S}teins}
\label{subsubphotLutetiaSteins}

\begin{table}[t!]
  \caption[Rotational parameters and taxonomical classes of (21)~Lutetia and (2867)~\v{S}teins]
    {Rotation period ($P$, in hours), the ecliptic longitude
      and latitude of the pole ($\lambda$ and $\beta$, in degrees),
      and the Tholen taxonomical classes for asteroids
      (21)~Lutetia and (2867)~\v{S}teins. For the latter, there are
      two possible pole solutions
    I and II. Numbers in parentheses depict the published uncertainty in 
    units of the last digit shown.}
\centering
\begin{tabular}{ccc}
\hline\hline
                                 & (21)~Lutetia           & (2867)~\v{S}teins \\ 
\hline 
$P$                           & 8.168270(1)           & 6.04681(2) \\ 
$\lambda$, $\beta$ & 52.2(4), -7.8(4)        & I: 96(5), -85(5)\\ 
                                 &                               & II: 142(5), -83(5) \\
Class                         & M                            & E \\        
\hline
\end{tabular}
\label{tab:cu4sso_LutetiaSteins} 
\end{table}

\begin{figure*}[t!] 
\centerline{
    \includegraphics[width=79mm]{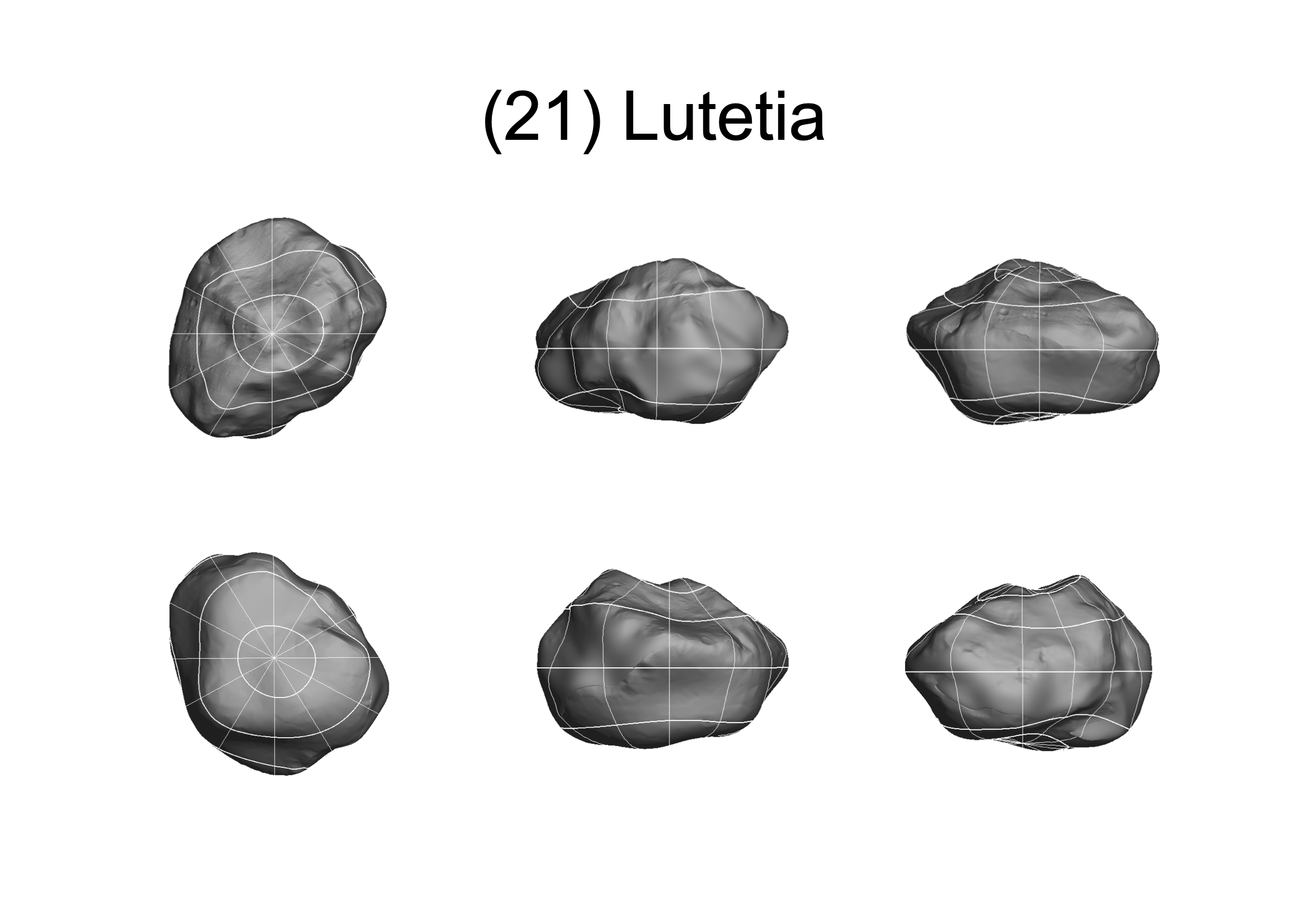}
    \includegraphics[width=79mm]{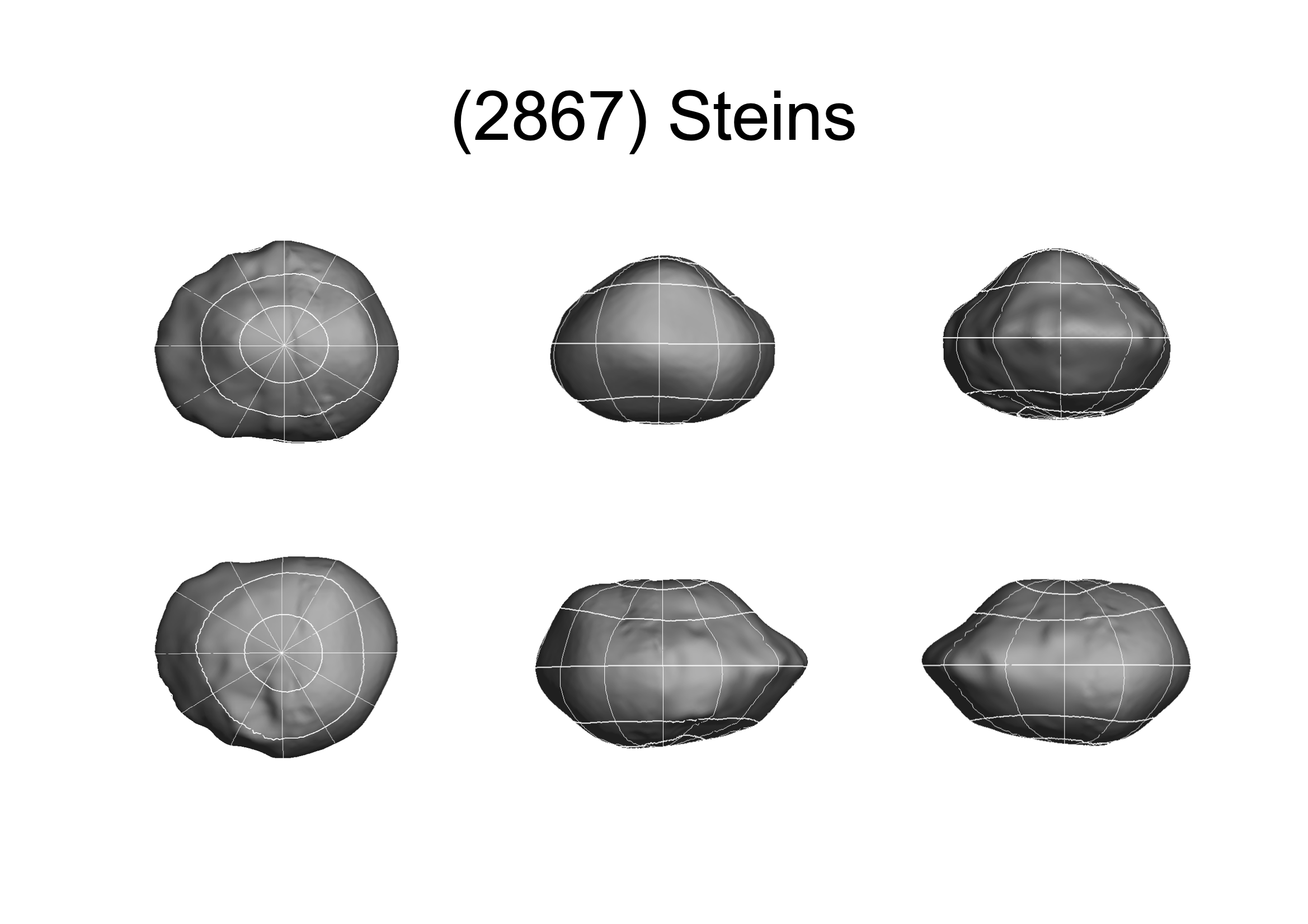}}
  \caption[High-resolution shape models of (21)~Lutetia and (2867)~\v{S}teins]
  {Shape model of asteroid (21)~Lutetia (left).
    Shape model of asteroid (2867)~\v{S}teins (right). The models are
    based on in situ{\em } images obtained by the Rosetta space
    mission. In the panels, the top and bottom plates on the left
    correspond to polar views along the $z$-axis
    (axis of rotation). The top (bottom) plates in the middle and to
    the right correspond to viewing along the $x$-axis ($y$-axis).}
  \label{fig:cu4sso_LutetiaSteinsShapes} 
\end{figure*}

\begin{figure*}[t!] 
\centerline{
   {\includegraphics[width=75mm]{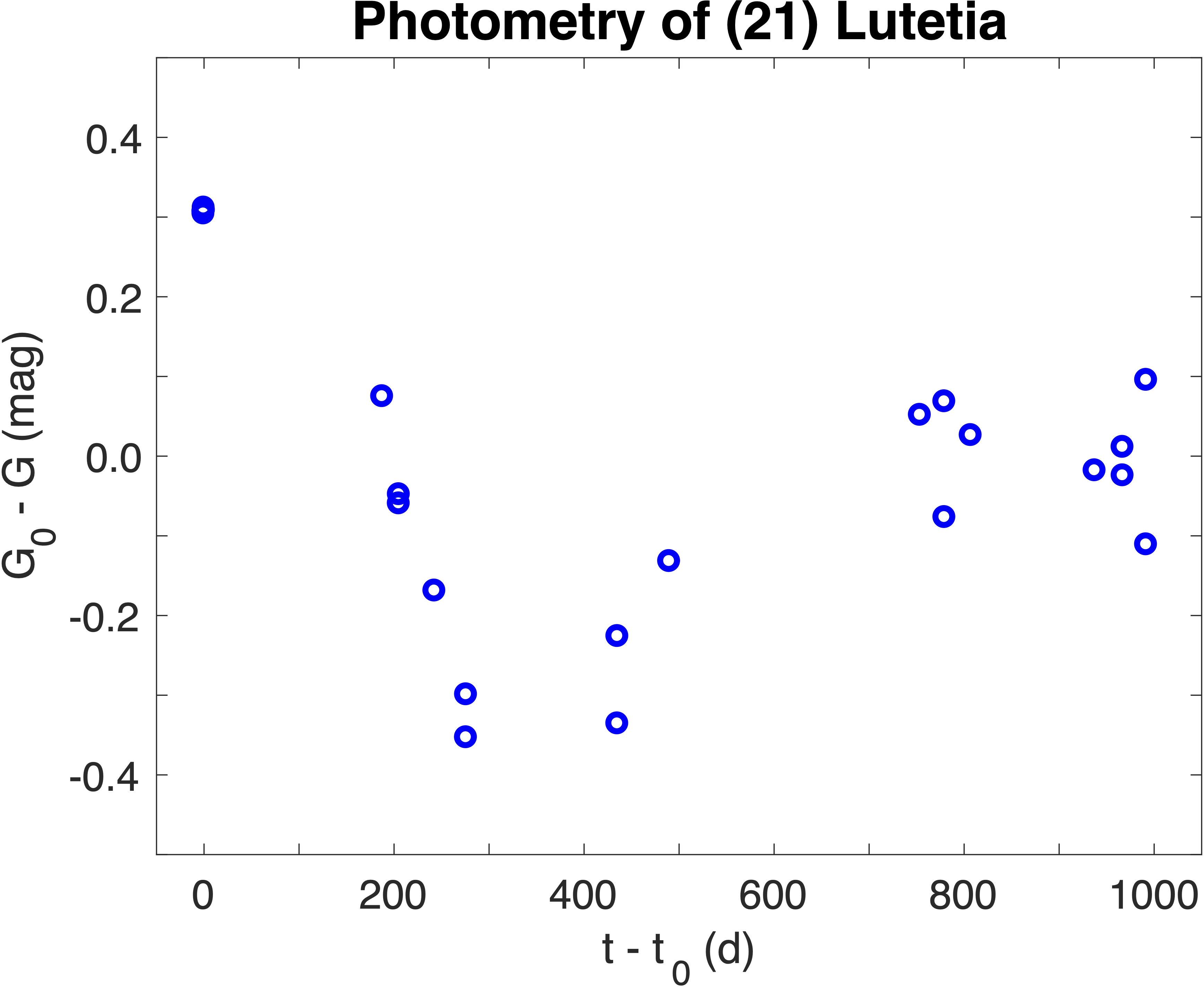}}
   \hspace{1cm}
   {\includegraphics[width=75mm]{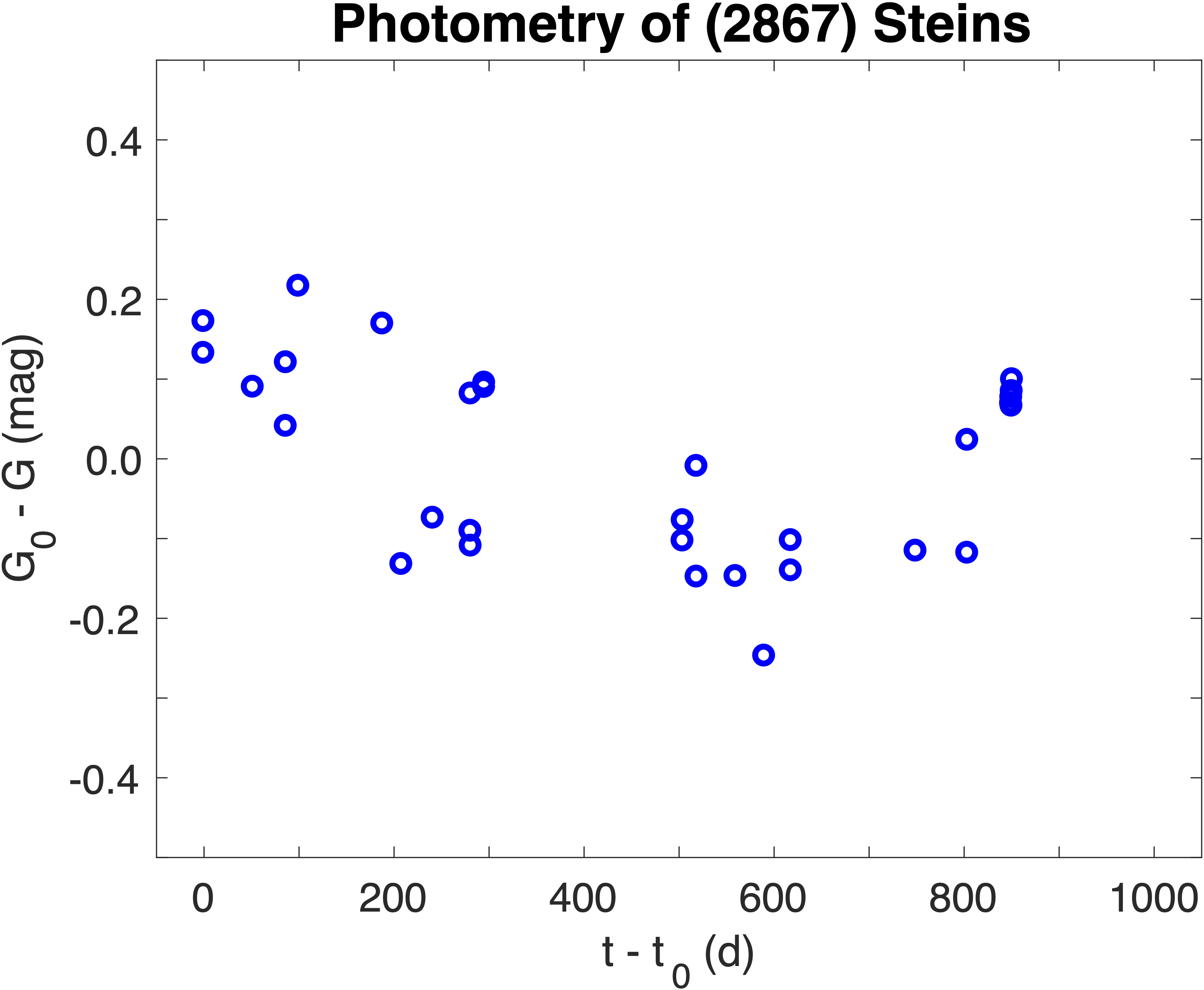}}}
 \vspace{0.5cm}
 \centerline{
    {\includegraphics[width=75mm]{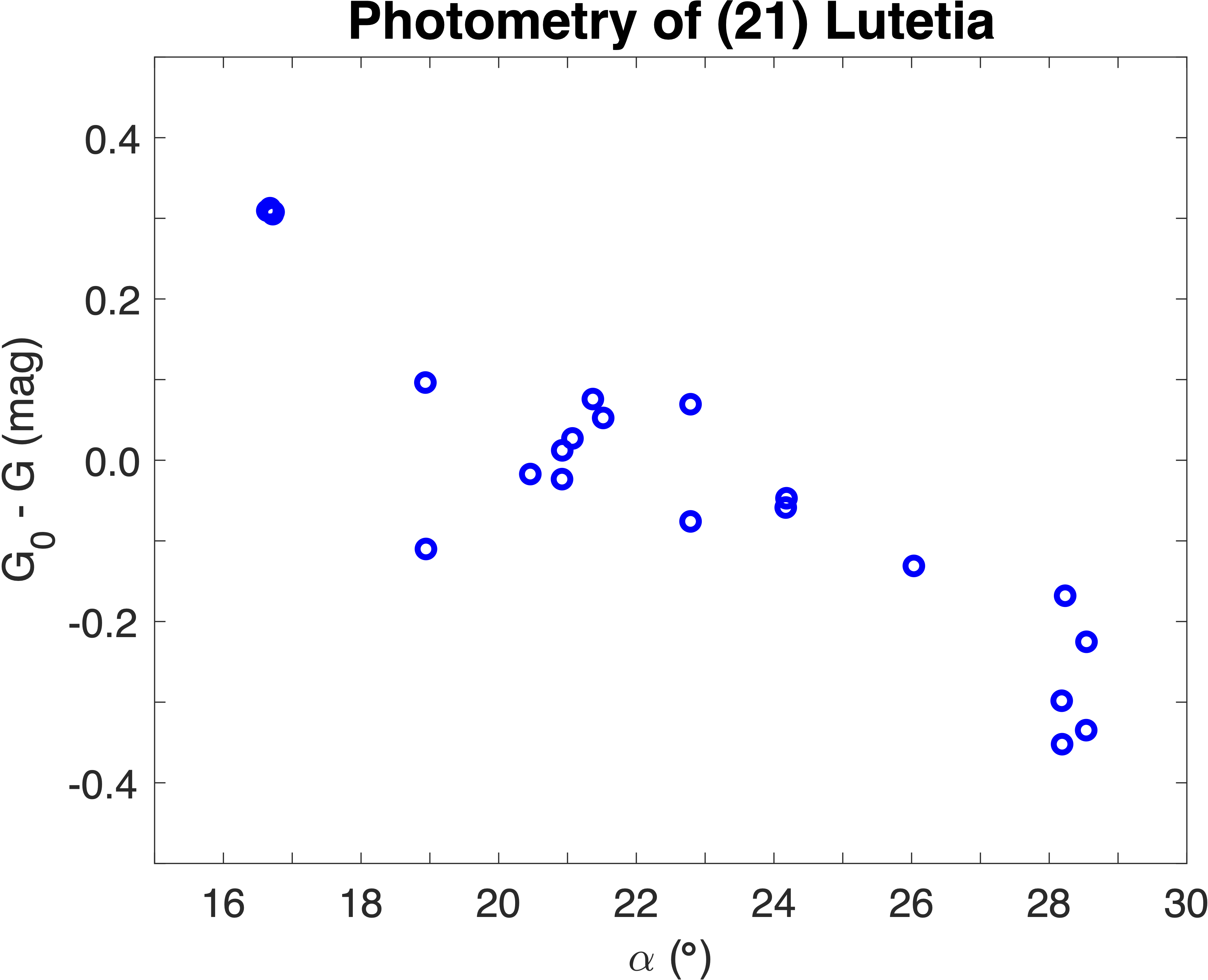}}
   \hspace{1cm}
    {\includegraphics[width=75mm]{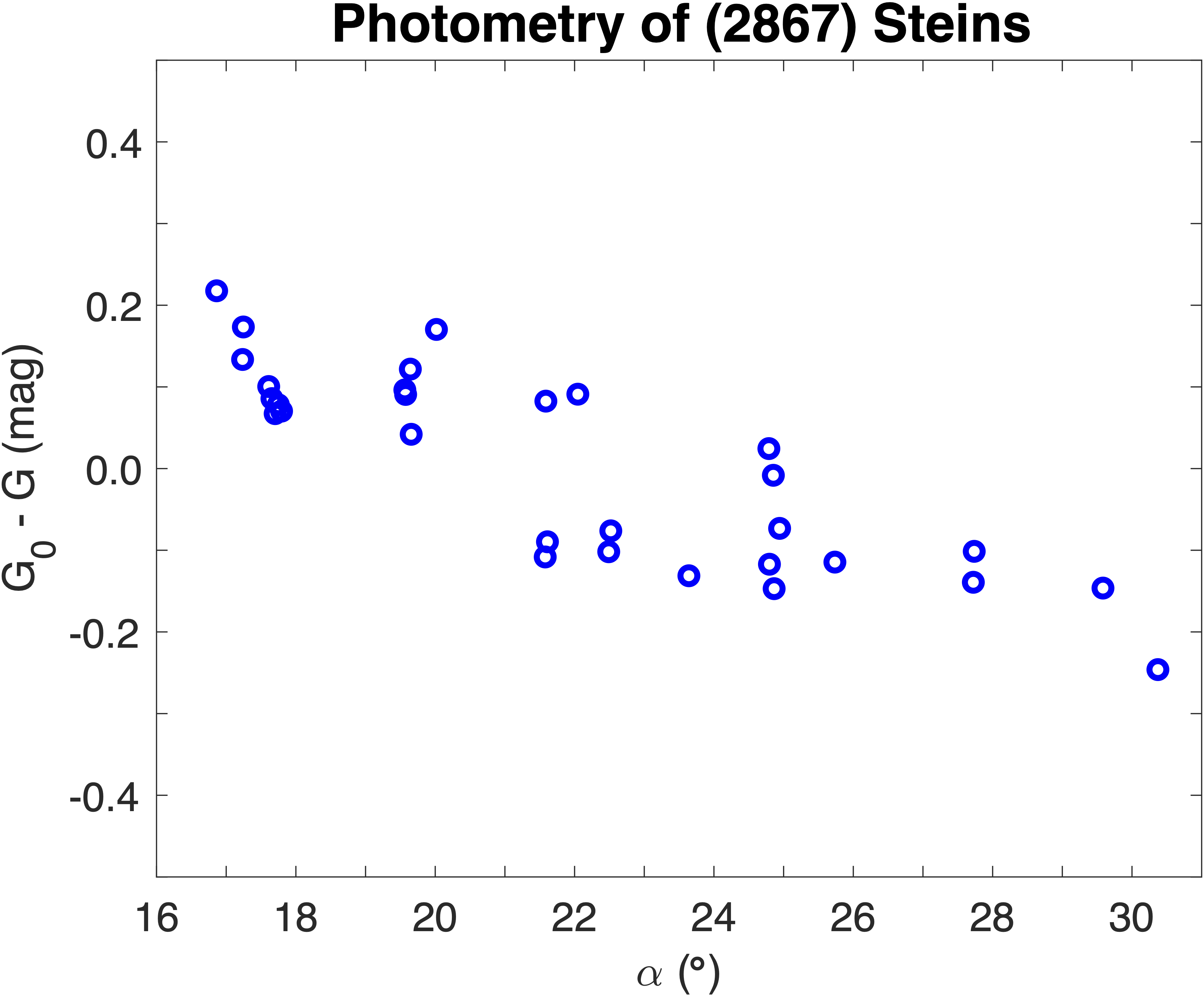}}}
\caption[Measured magnitudes of (21)~Lutetia and (2867)~\v{S}teins.]
{Observed $G$ magnitudes of asteroid (21)~Lutetia included
  in \gdrthree, relative to the mean of the magnitudes ($G_0$), as a function of days after
  the first observation ($t_0$) (top left). The top right panel shows the same as the left panel for (2867)~\v{S}teins.
  The bottom panel shows the same as the top panel, but the magnitudes are depicted against the
  phase angle. }
  \label{fig:cu4sso_LutetiaSteinsG}
\end{figure*}

Our validation consists of comparing the observed {\gaia}
photometry to photometry computed for known asteroids that have
accurately determined rotation periods, pole orientations, and
high-resolution shape models. \gdrthree\ contains photometric data
for more than a dozen asteroids studied by space missions. By using
the shape models, rotational parameters, and taxonomical
classifications of (21)~Lutetia and (2867)~\v{S}teins, which are asteroids that were visited
by the ESA Rosetta space mission, we studied whether it is possible to
validate the \gdrthree SSO photometry. We note that (21)~Lutetia and
(2867)~\v{S}teins were assessed earlier in the documentation of
\gdrtwo.

For (21)~Lutetia and (2867)~\v{S}teins, the Planetary Data System
provides the shape models illustrated in
Fig.~\ref{fig:cu4sso_LutetiaSteinsShapes} as well as the rotation
periods and pole orientations described in
Table~\ref{tab:cu4sso_LutetiaSteins} \citep{FarnhamJorda2013,
  Jordaetal2012, Farnham2013, Sierksetal2011}. The table also includes
the Tholen taxonomical classes of the two asteroids.

The \gdrthree photometric measurements for these two asteroids are plotted
as a function of the observation epochs, expressed in days after
the first observation, and against the phase angle in degrees, in
Fig.~\ref{fig:cu4sso_LutetiaSteinsG}. Both phase-magnitude relations
show a decreasing trend in disk-integrated brightness with increasing
phase angle. Furthermore, the apparent slope of decreasing brightness
is steeper for Lutetia, in agreement with Lutetia and \v{S}teins being
lower-albedo M-type and higher-albedo E-type objects in the Tholen
taxonomy, respectively.

\begin{figure*}[t!] 
  \centerline{
    \includegraphics[width=75mm]{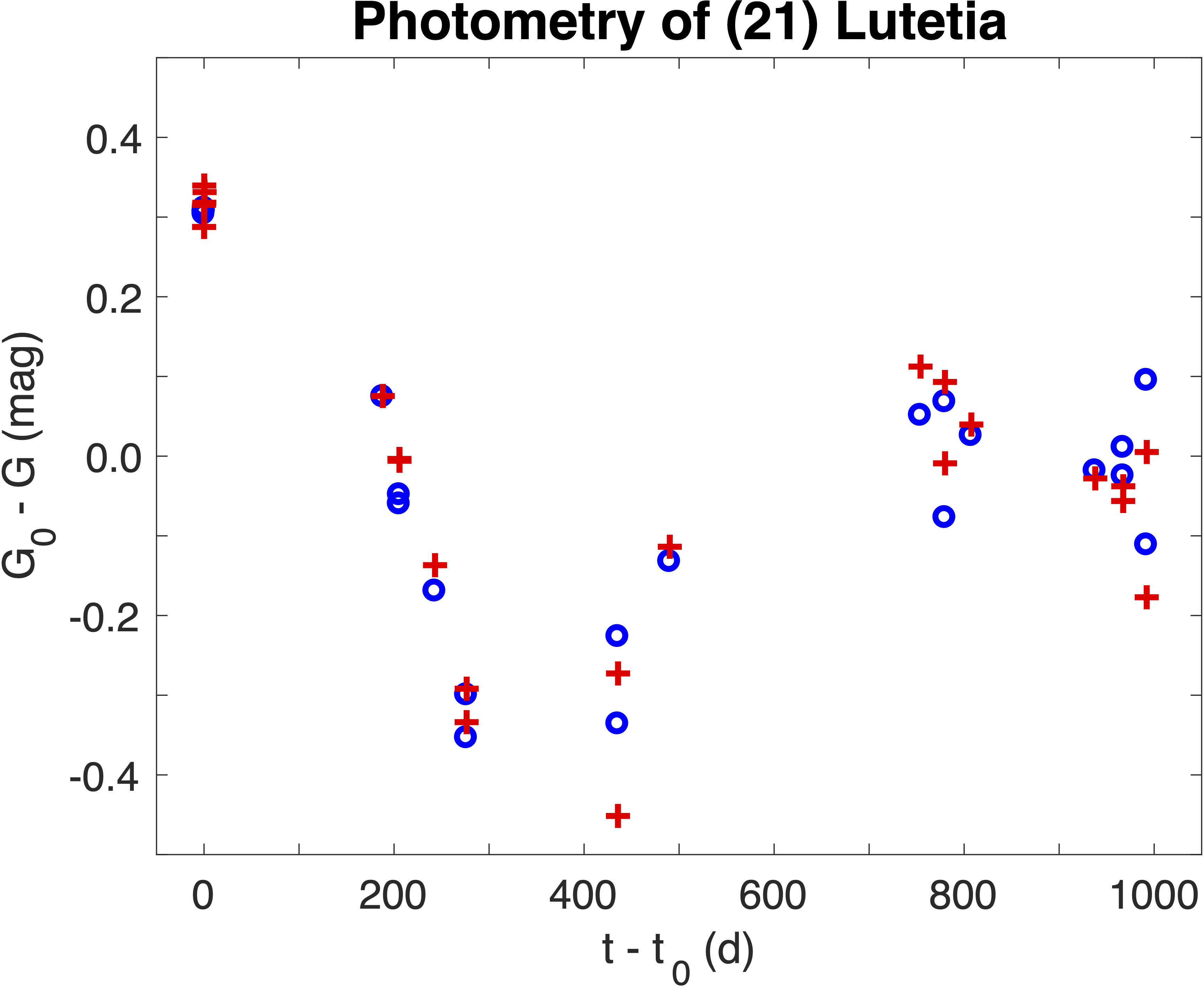}
        \hspace{1cm}
    \includegraphics[width=75mm]{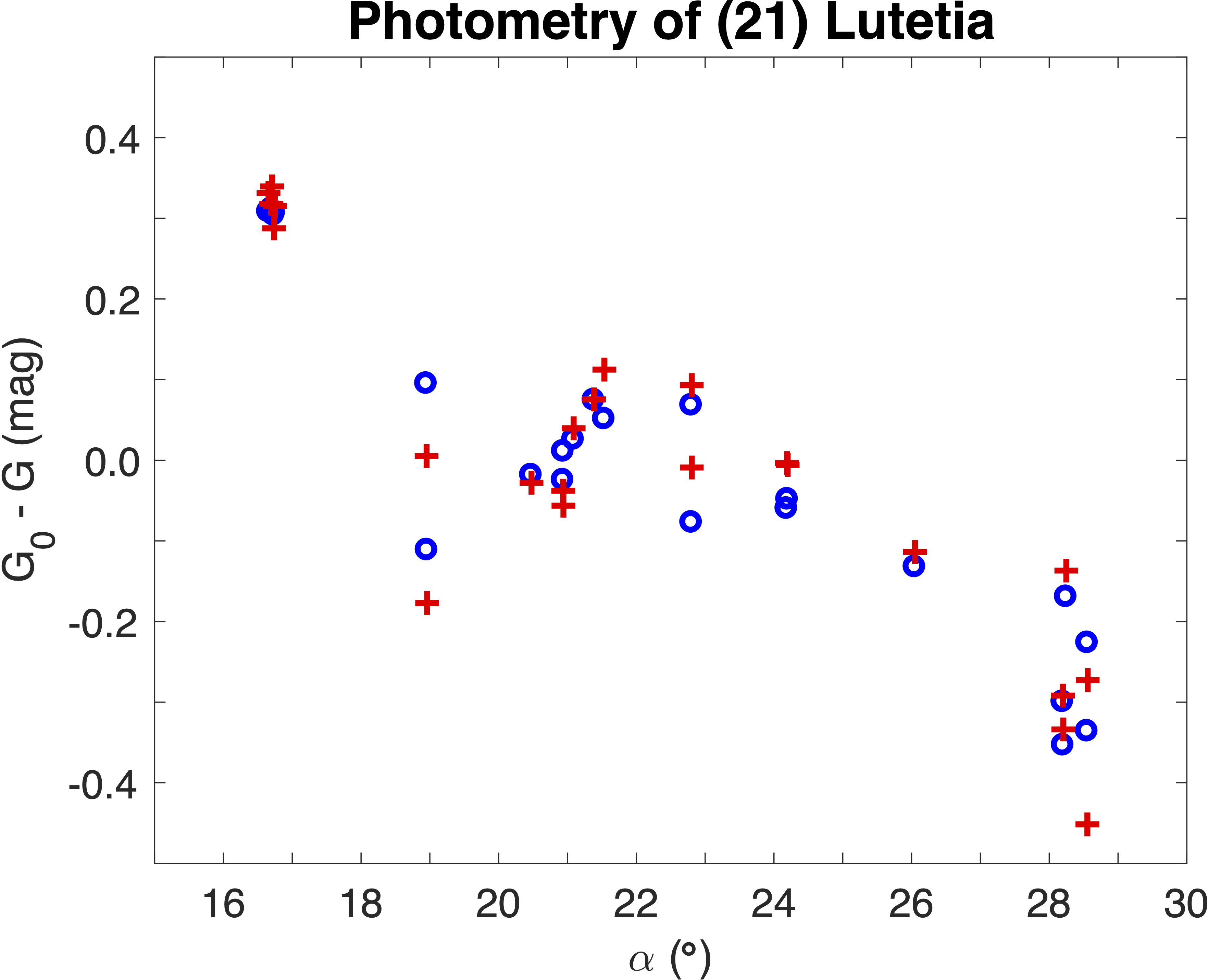}}
 \vspace{0.5cm}
\centerline{
  \includegraphics[width=75mm]{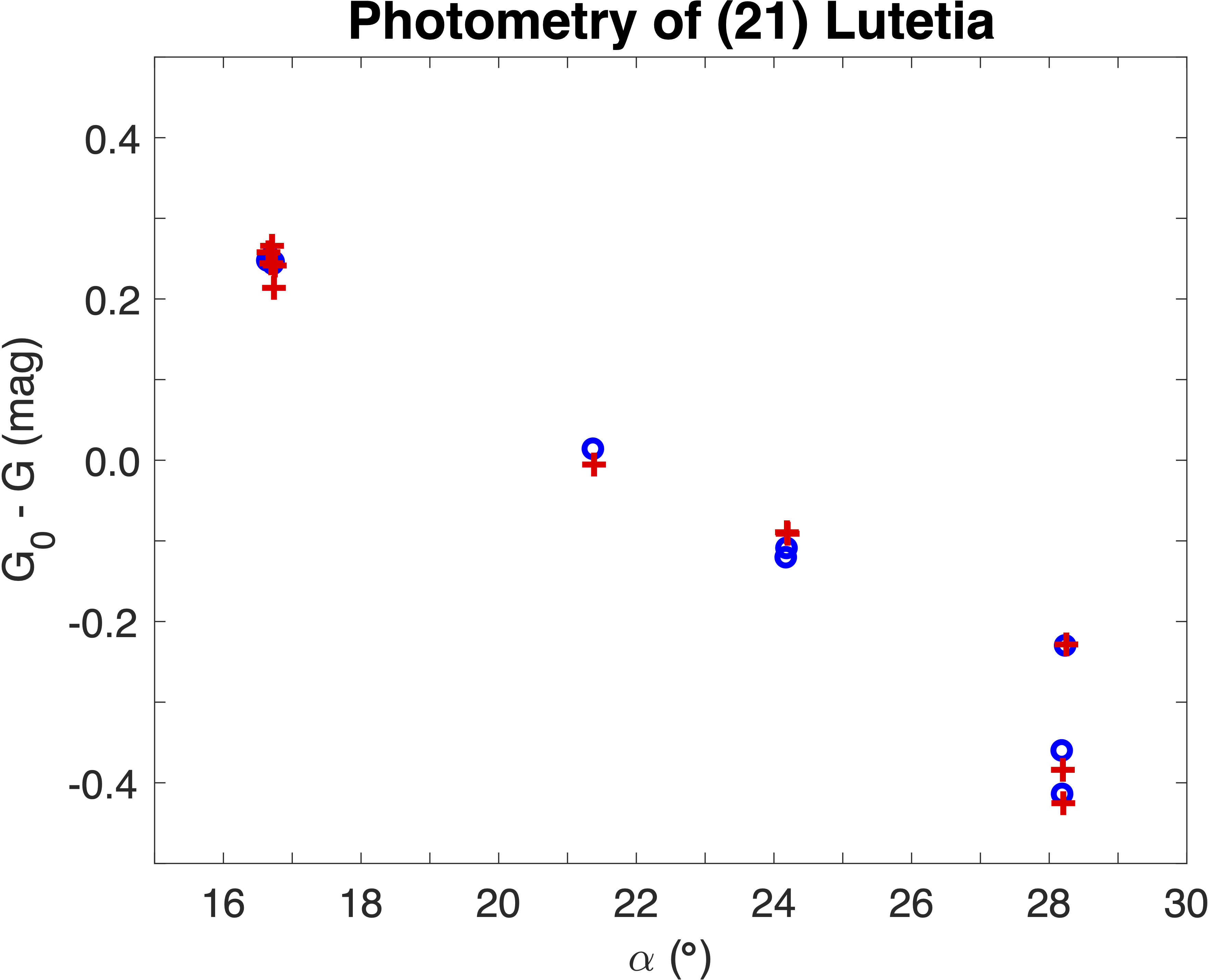}
      \hspace{1cm}
  \includegraphics[width=75mm]{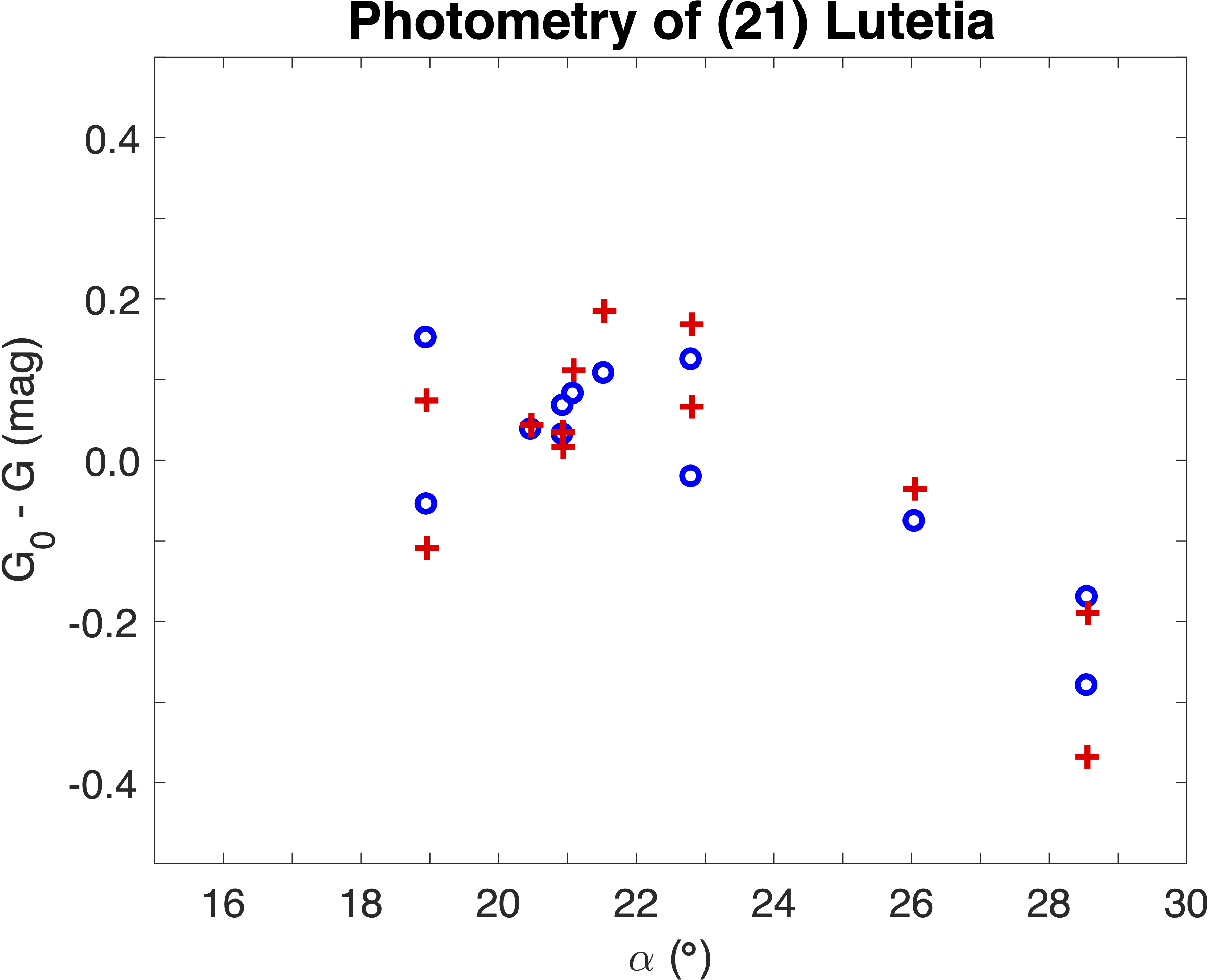}}
\caption[Modelled magnitudes of (21)~Lutetia] {Observed $G$
  magnitudes of asteroid (21)~Lutetia included in \gdrthree as a
  function of days after the first observation ($t_0$, top left) and against the phase
  angle ($\alpha$, top right) together with the modelled magnitudes. The bottom left panel 
  shows the same as in top right panel, but only for the observations concerning the Lutetia hemisphere
    observed by Rosetta.  The bottom right panel shows the same as in the top right panel, but only for the
    observations of the Lutetia hemisphere that was not observed by
    Rosetta.}
\label{fig:cu4sso_LutetiaM}
\end{figure*}

\begin{figure*}[t!] 
\centerline{
    \includegraphics[width=75mm]{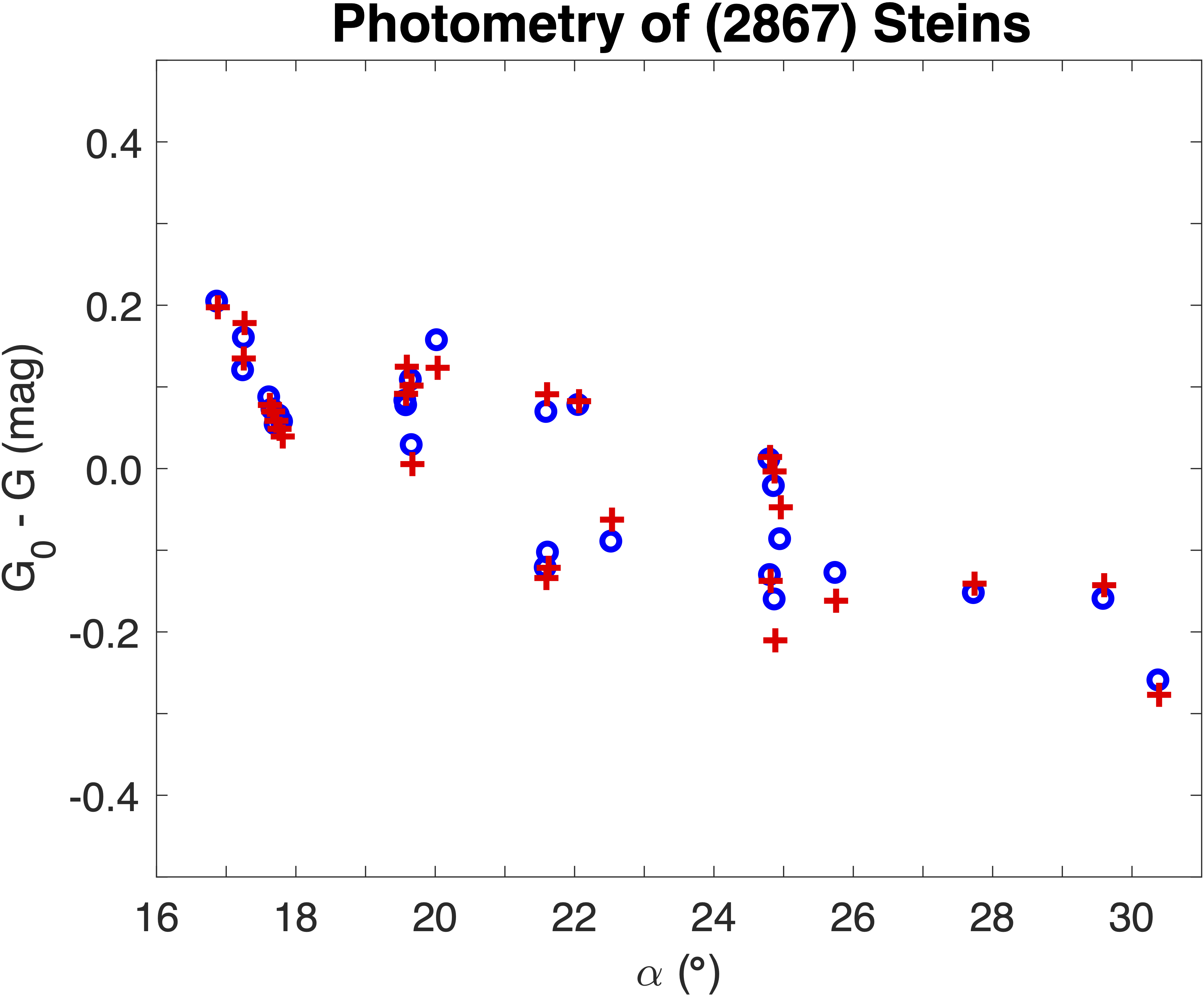}
    \hspace{1cm}
    \includegraphics[width=75mm]{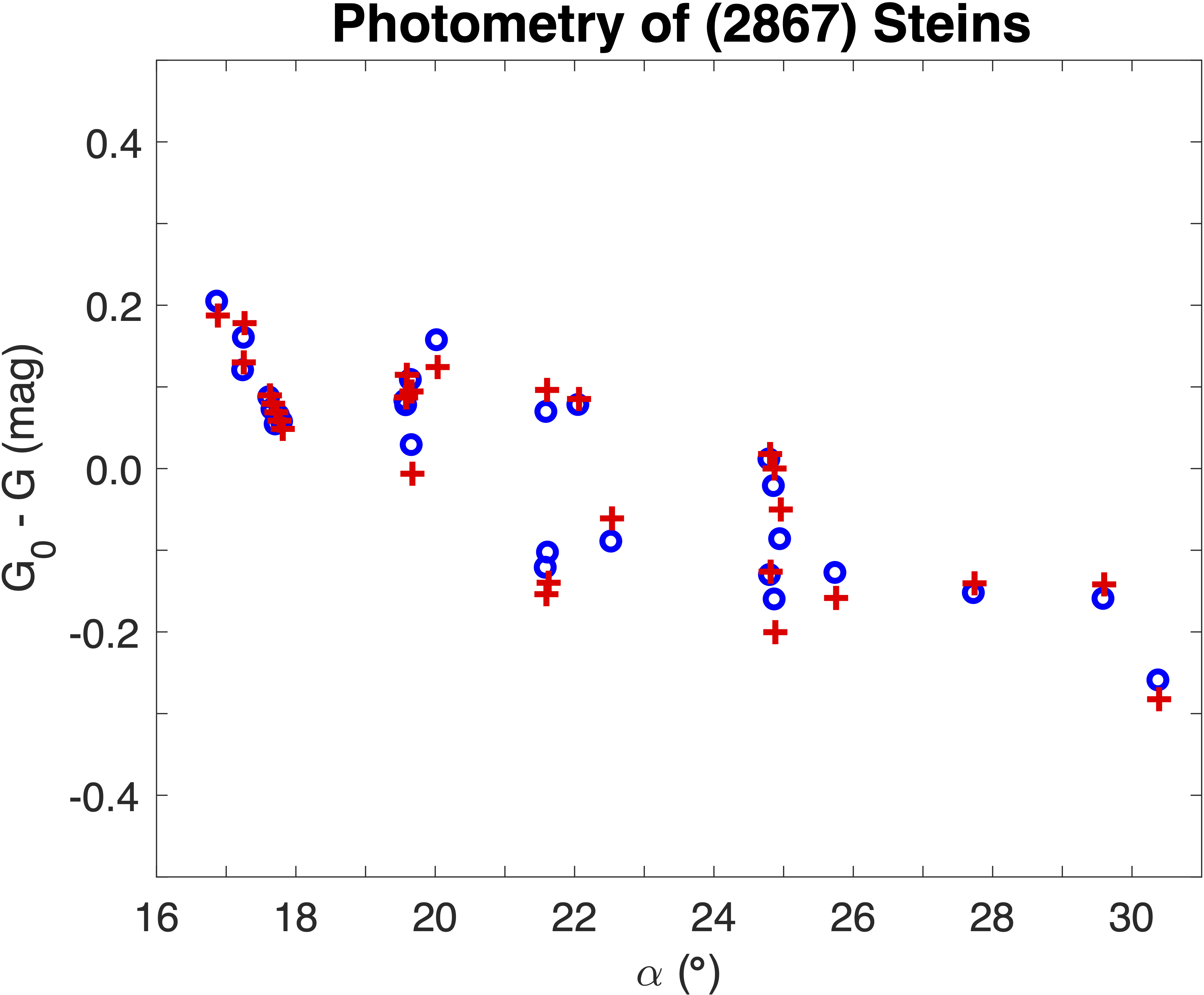} }
  \caption[Modelled magnitudes of (2867)~\v{S}teins]
  {Observed $G$ magnitudes of asteroid (2867)~\v{S}teins included
    in \gdrthree against the phase angle $\alpha$ 
    together with the modelled magnitudes. Two models are
    shown, corresponding to the pole orientations I (left) and II
    (right) given in Table~\ref{tab:cu4sso_LutetiaSteins}.}
\label{fig:cu4sso_SteinsM}
\end{figure*}

We studied (21)~Lutetia with the 23 transit magnitudes obtained by
\textit{Gaia}  (case~I) using a shape model derived from combined
ground-based relative photometry of 50 light curves with 4012 
photometric points in total \citep{Durechetal2010} and \textit{Gaia} photometry with convex 
inversion methods \citep{Muinonenetal2020,Kaasalainenetal2001,KaasalainenTorppa2001}. 
The results showed that the \textit{Gaia} measurements contributed to the shape
modelling with reasonable residuals, with a low RMS value of 0.016~mag. 
The RMS value for the entire data set of 51 light curves was 0.015~mag.
Next (case~II), it was seen that using the best available shape model 
\citep{Farnham2013}, constructed by combining a high-resolution 
shape model based on disk-resolved imaging by Rosetta and a lower-resolution 
model based on ground-based relative photometry and silhouette observations 
with adaptive optics, it was not possible to produce a straightforward fit 
to the \textit{Gaia} observations (see Fig.~\ref{fig:cu4sso_LutetiaM}). By optimising 
the rotational phase of Lutetia and the slope of the phase curve implied by 
the Lommel-Seeliger scattering model, the high-resolution shape model 
resampled at a 5-degree resolution reproduced the \textit{Gaia} photometry with a 
high RMS value of 0.047~mag, whereas for the entire data set, the RMS value 
was 0.017~mag. We interpreted the RMS values for cases I and II as giving a 
strong indication that the \textit{Gaia} data offer new information on the 
shape or surface properties of Lutetia.

A closer inspection of the time distribution of the 23 \gdrthree transit
magnitudes shows that for 12 of them, the Lutetia hemisphere
that predominantly visible to \textit{Gaia }was the hemisphere that was not{\it } mapped by Rosetta.
It is therefore highly probable that the {\gaia} data provide additional
information about this portion of the asteroid surface. In particular,
the \textit{Gaia} photometry, with its absolute phase angle dependence,
relates the {size} and geometric albedo characteristics of the two
hemispheres.

We used the high-resolution shape resampled in 5-degree 
resolution to model the photometry of the remaining 11 observations that were obtained with the
hemisphere mapped by Rosetta in view directly (case~III). Allowing for optimisation in
only the rotational phase and the slope of the phase curve implied by
the Lommel-Seeliger scattering model, the high-resolution shape model
reproduced the Gaia photometry with an excellent RMS value of 0.019~mag (Fig.~\ref{fig:cu4sso_LutetiaM}). We then applied an analogous approach to 
the 12 observations obtained with the hemisphere that was not mapped by Rosetta 
(case~IV). With constant geometric albedo characteristics, the RMS value was 
high at 0.057~mag.

The photometric slopes implied by the Lommel-Seeliger scattering model
were then compared in detail. In the cases of convex optimisation
using all Gaia measurements (case~I), optimisation of the rotational
phase and slope using the high-resolution shape model and all Gaia
measurements (II), measurements corresponding to the hemisphere
observed by Rosetta (III), and measurements corresponding to the
hemisphere that was not observed by Rosetta (IV), the slopes were 1.89~mag~rad$^{-1}$, 1.56~mag~rad$^{-1}$,
1.65~mag~rad$^{-1}$, and 1.47~mag~rad$^{-1}$, respectively. First, case I stands out 
as showing the steepest photometric slope for a convex shape solution. 
A plausible explanation is that global self-shadowing effects due to 
a non-convex shape steepen the photometric slope. Second, case IV stands 
out as showing the shallowest photometric slope for the high-resolution 
non-convex shape model. Third, case III shows the most realistic photometric 
slope for the high-resolution shape model: the least-squares fit to the \textit{Gaia} 
data is successful with a small RMS-value. Finally, the case II photometric slope 
is a compromise between cases III and IV. Whereas it is beyond the scope of 
the present study to improve the Lutetia model, the present analysis indicates varying scattering properties for the hemisphere that was not imaged 
by Rosetta. This is also supported by the fact that all the ground-based 
light curves, consisting of extensive observations of both hemispheres and 
treated in relative sense in the validation, were well fitted by the 
high-resolution shape model.

The final photometric slope analysis for Lutetia was carried out in case III
using {Markov chain Monte Carlo (MCMC)} sampling. The photometric slope for the hemisphere observed
by Rosetta was estimated to be (1.677$\pm$0.075)~mag~rad$^{-1}$  using
the high-resolution shape model resampled at a 5-degree resolution. Based
on \citet{Penttilaetal2016}, \citet{Muinonenetal2020}, and
\citet{Martikainenetal2021}, these photometric slopes agree with 
those for Tholen S- and M-class asteroids. 

For Lutetia, the apparent $G$ magnitudes range from 11.1 to 13.1~mag, 
with minute nominal observational errors of $0.00033-0.00070$~mag. 
Our RMS values above are significantly higher. We conclude that Gaia photometry 
has an accuracy better than 0.01-0.02~mag with the limitations on
first, the shape model accuracy, which does not allow us to push the analysis
further. Second, we assume that the same Lommel-Seeliger scattering
model is valid everywhere on the surface of Lutetia. In particular,
we assume that a single combination of geometric albedo and
scattering phase function is representative of the entire surface of
Lutetia. Third, it is not possible to exclude that some anomalous
brightness values are present. 

The results obtained for (2867)~\v{S}teins, whose high-resolution
shape model is not limited by hemisphere scales similar
to those of (21)~Lutetia, support the idea that \textit{Gaia} photometry is
indeed accurate. In the case of this object, there are two pole
solutions (\v{S}teins cases I-II, see
Table~\ref{tab:cu4sso_LutetiaSteins}) that essentially only differ in
terms of the zero-point of the rotational phase. By directly using the
shape model to reproduce the \textit{Gaia} data, resampled at a 5-degree
resolution, the RMS values of the observed-computed magnitudes are 0.022~mag and 0.023~mag for the two pole solutions. These
fits are based on 30 photometric points, 3 of which had to be omitted
as outliers. Again, the fits were obtained by optimising only the
rotational phase and the photometric slope implied by the
Lommel-Seeliger scattering model. 

We continued the analysis of \v{S}teins photometry by resampling the
shape model at a higher 3-degree resolution and
simultaneously fitting five parameters, that is, the rotation period, pole
orientation, rotational phase, and photometric slope. The RMS values
for the two pole solutions were lower at 0.018~mag and
0.019~mag, respectively. To compute these fits, the
rotational parameters were regularised to lie within their domains of
uncertainty (Table~\ref{tab:cu4sso_LutetiaSteins}). The photometric slope 
obtained values of 1.515~mag~rad$^{-1}$ (case I) and 1.522~mag~rad$^{-1}$ 
(II), that is, the values were essentially equal.

Finally, {MCMC} sampling was carried out for the 
rotational phase {and} photometric slope in cases I-II using the 
high-resolution shape model re-sampled at 5-degree resolution. 
For 5000 samples, the means and standard deviations of the photometric 
slope obtained the values of (1.497$\pm$0.064)~mag~rad$^{-1}$ (case I) 
and (1.458$\pm$0.062)~mag~rad$^{-1}$ (II). The values agree mutually, 
and they are realistic for high-albedo E-class asteroids \citep{Penttilaetal2016,Muinonenetal2020,Martikainenetal2021}. 
The mean values agreed within the given uncertainties with the least-squares 
values for the resampled 5-degree and 3-degree shape models.

For \v{S}teins, the apparent $G$ magnitudes ranged over $16.3-18.4$~mag, 
with nominal observational errors within $0.0028-0.014$~mag. 
Considering the simplicity of the surface-scattering modelling we used, 
our results can be considered completely satisfactory (see Fig.~\ref{fig:cu4sso_SteinsM}). 
The remaining limitations in the case of (2867)~\v{S}teins are related to 
details of the shape model and surface-scattering characteristics. 
In particular, certain assumptions were made on the scattering properties 
when the high-resolution shape model was derived from the Rosetta images. 
It remains possible that the three observations that we omitted as outliers were
omitted due to modelling issues rather than low observational
accuracy.

In conclusion, we cannot rule out the possibility that the sample published in \gdrthree
could still include a non-negligible fraction of anomalous data. We
recommend detailed analyses and careful verifications when the
\gdrthree photometry of asteroids is applied. However, especially for objects as bright as Lutetia,
our current approach to magnitude prediction cannot investigate $\gaia$ photometry of SSOs at the level that would be required by the very high accuracy. 

\section{Conclusions}

The Solar System data processing, developed and trained over several years, reaches maturity with \gdrthree by providing the expected large survey for asteroids and planetary satellites. The different data types (astrometry, photometry, and low-resolution reflectance spectra) and their accuracy and homogeneity are an impressive achievement of the \gaia mission. We have summarized the approach that we followed for the processing pipeline and in particular explained the procedures that clean the data set and provide a reliable outcome.

The highly accurate epoch astrometry obtained by \textit{Gaia} is confirmed by the orbit adjustment, resulting in residuals at submilliarcsecond level for {$G<18$}. The astrometric performance is clearly improved with respect to \gdrtwo.

This accuracy {for} a large number of asteroids brings new capabilities of investigation, revealing effects related to the partially resolved shape of asteroids. The amplitude of the photocentre wobble that we measured on (21) Lutetia shows that this effect probably affects the {orbit computation} even for asteroids that are three to four times smaller. The capability of detecting the wobbling, associated with the orbital motion of a satellite, discloses an impressive domain of investigation for the search of asteroid binaries with the astrometric method. This approach was simply not possible before. A comprehensive exploration of \gdrthree should reveal a variety of binary systems.

A search for the best orbital modelling for {near-Earth objects} that can exhibit the Yarkovsky {drift shows that in some circumstances,} \gaia can detect this effect, even in {the absence of the radar ranging data} that have been essential in the recent past. A careful coupling to ground-based data will fully disclose this potential.

The G-band photometric data are the other valuable source of information that will be exploited to obtain new constraints on the rotation and the shape of a very large number of objects. For them, it remains difficult to assess the quality at the level of their expected uncertainty, simply because there is no direct comparison to other data sets of comparable accuracy. The simulated photometry from accurate shape models brings, however, positive results, despite the remaining uncertainties. It also shows that some outliers remain probably present among the released measurements. 

In conclusion, our review shows that the data processing of Solar System data in \gdrthree has produced an extremely rich data set standing out for its unique properties with respect to other existing surveys in many aspects. \gdrthree will certainly be exploited in many ways by the community of planetary scientists. New properties and features beyond those illustrated in this article will probably be found, and will be a major driver for improvements of the data quality in future data releases. 


\begin{acknowledgements}
We dedicate this work to the memory of Dimitri Pourbaix, who managed the Coordination Unit 4 of DPAC with immense dedication, enthusiasm and intellectual honesty. His legacy is also present in this work.
This work presents results from the European Space Agency (ESA) space mission \gaia. \gaia\ data are being processed by the \gaia\ Data Processing and Analysis Consortium (DPAC). Funding for the DPAC is provided by national institutions, in particular the institutions participating in the \gaia\ MultiLateral Agreement (MLA). The \gaia\ mission website is \url{https://www.cosmos.esa.int/gaia}. The \gaia\ archive website is \url{https://archives.esac.esa.int/gaia}.
Acknowledgements are given in Appendix~\ref{ssec:appendixA}

\end{acknowledgements}

\bibliographystyle{aa} 
\bibliography{allref} 
\begin{appendix}
\section{}\label{ssec:appendixA}

The \gaia\ mission and data processing have financially been supported by, in alphabetical order by country:
\begin{itemize}
\item the Algerian Centre de Recherche en Astronomie, Astrophysique et G\'{e}ophysique of Bouzareah Observatory;
\item the Austrian Fonds zur F\"{o}rderung der wissenschaftlichen Forschung (FWF) Hertha Firnberg Programme through grants T359, P20046, and P23737;
\item the BELgian federal Science Policy Office (BELSPO) through various PROgramme de D\'{e}veloppement d'Exp\'{e}riences scientifiques (PRODEX) grants and the Polish Academy of Sciences - Fonds Wetenschappelijk Onderzoek through grant VS.091.16N, and the Fonds de la Recherche Scientifique (FNRS), and the Research Council of Katholieke Universiteit (KU) Leuven through grant C16/18/005 (Pushing AsteRoseismology to the next level with TESS, GaiA, and the Sloan DIgital Sky SurvEy -- PARADISE);  
\item the Brazil-France exchange programmes Funda\c{c}\~{a}o de Amparo \`{a} Pesquisa do Estado de S\~{a}o Paulo (FAPESP) and Coordena\c{c}\~{a}o de Aperfeicoamento de Pessoal de N\'{\i}vel Superior (CAPES) - Comit\'{e} Fran\c{c}ais d'Evaluation de la Coop\'{e}ration Universitaire et Scientifique avec le Br\'{e}sil (COFECUB);
\item the Chilean Agencia Nacional de Investigaci\'{o}n y Desarrollo (ANID) through Fondo Nacional de Desarrollo Cient\'{\i}fico y Tecnol\'{o}gico (FONDECYT) Regular Project 1210992 (L.~Chemin);
\item the National Natural Science Foundation of China (NSFC) through grants 11573054, 11703065, and 12173069, the China Scholarship Council through grant 201806040200, and the Natural Science Foundation of Shanghai through grant 21ZR1474100;  
\item the Tenure Track Pilot Programme of the Croatian Science Foundation and the \'{E}cole Polytechnique F\'{e}d\'{e}rale de Lausanne and the project TTP-2018-07-1171 `Mining the Variable Sky', with the funds of the Croatian-Swiss Research Programme;
\item the Czech-Republic Ministry of Education, Youth, and Sports through grant LG 15010 and INTER-EXCELLENCE grant LTAUSA18093, and the Czech Space Office through ESA PECS contract 98058;
\item the Danish Ministry of Science;
\item the Estonian Ministry of Education and Research through grant IUT40-1;
\item the European Commission’s Sixth Framework Programme through the European Leadership in Space Astrometry (\href{https://www.cosmos.esa.int/web/gaia/elsa-rtn-programme}{ELSA}) Marie Curie Research Training Network (MRTN-CT-2006-033481), through Marie Curie project PIOF-GA-2009-255267 (Space AsteroSeismology \& RR Lyrae stars, SAS-RRL), and through a Marie Curie Transfer-of-Knowledge (ToK) fellowship (MTKD-CT-2004-014188); the European Commission's Seventh Framework Programme through grant FP7-606740 (FP7-SPACE-2013-1) for the \gaia\ European Network for Improved data User Services (\href{https://gaia.ub.edu/twiki/do/view/GENIUS/}{GENIUS}) and through grant 264895 for the \gaia\ Research for European Astronomy Training (\href{https://www.cosmos.esa.int/web/gaia/great-programme}{GREAT-ITN}) network;
\item the European Cooperation in Science and Technology (COST) through COST Action CA18104 `Revealing the Milky Way with \gaia (MW-Gaia)';
\item the European Research Council (ERC) through grants 320360, 647208, and 834148 and through the European Union’s Horizon 2020 research and innovation and excellent science programmes through Marie Sk{\l}odowska-Curie grant 745617 (Our Galaxy at full HD -- Gal-HD) and 895174 (The build-up and fate of self-gravitating systems in the Universe) as well as grants 687378 (Small Bodies: Near and Far), 682115 (Using the Magellanic Clouds to Understand the Interaction of Galaxies), 695099 (A sub-percent distance scale from binaries and Cepheids -- CepBin), 716155 (Structured ACCREtion Disks -- SACCRED), 951549 (Sub-percent calibration of the extragalactic distance scale in the era of big surveys -- UniverScale), and 101004214 (Innovative Scientific Data Exploration and Exploitation Applications for Space Sciences -- EXPLORE);
\item the European Science Foundation (ESF), in the framework of the \gaia\ Research for European Astronomy Training Research Network Programme (\href{https://www.cosmos.esa.int/web/gaia/great-programme}{GREAT-ESF});
\item the European Space Agency (ESA) in the framework of the \gaia\ project, through the Plan for European Cooperating States (PECS) programme through contracts C98090 and 4000106398/12/NL/KML for Hungary, through contract 4000115263/15/NL/IB for Germany, and through PROgramme de D\'{e}veloppement d'Exp\'{e}riences scientifiques (PRODEX) grant 4000127986 for Slovenia;  
\item the Academy of Finland through grants 299543, 307157, 325805, 328654, 336546, and 345115 and the Magnus Ehrnrooth Foundation;
\item the French Centre National d’\'{E}tudes Spatiales (CNES), the Agence Nationale de la Recherche (ANR) through grant ANR-10-IDEX-0001-02 for the `Investissements d'avenir' programme, through grant ANR-15-CE31-0007 for project `Modelling the Milky Way in the \gaia era’ (MOD4Gaia), through grant ANR-14-CE33-0014-01 for project `The Milky Way disc formation in the \gaia era’ (ARCHEOGAL), through grant ANR-15-CE31-0012-01 for project `Unlocking the potential of Cepheids as primary distance calibrators’ (UnlockCepheids), through grant ANR-19-CE31-0017 for project `Secular evolution of galxies' (SEGAL), and through grant ANR-18-CE31-0006 for project `Galactic Dark Matter' (GaDaMa), the Centre National de la Recherche Scientifique (CNRS) and its SNO \gaia of the Institut des Sciences de l’Univers (INSU), its Programmes Nationaux: Cosmologie et Galaxies (PNCG), Gravitation R\'{e}f\'{e}rences Astronomie M\'{e}trologie (PNGRAM), Plan\'{e}tologie (PNP), Physique et Chimie du Milieu Interstellaire (PCMI), and Physique Stellaire (PNPS), the `Action F\'{e}d\'{e}ratrice \gaia' of the Observatoire de Paris, the R\'{e}gion de Franche-Comt\'{e}, the Institut National Polytechnique (INP) and the Institut National de Physique nucl\'{e}aire et de Physique des Particules (IN2P3) co-funded by CNES;
\item the German Aerospace Agency (Deutsches Zentrum f\"{u}r Luft- und Raumfahrt e.V., DLR) through grants 50QG0501, 50QG0601, 50QG0602, 50QG0701, 50QG0901, 50QG1001, 50QG1101, 50\-QG1401, 50QG1402, 50QG1403, 50QG1404, 50QG1904, 50QG2101, 50QG2102, and 50QG2202, and the Centre for Information Services and High Performance Computing (ZIH) at the Technische Universit\"{a}t Dresden for generous allocations of computer time;
\item the Hungarian Academy of Sciences through the Lend\"{u}let Programme grants LP2014-17 and LP2018-7 and the Hungarian National Research, Development, and Innovation Office (NKFIH) through grant KKP-137523 (`SeismoLab');
\item the Science Foundation Ireland (SFI) through a Royal Society - SFI University Research Fellowship (M.~Fraser);
\item the Israel Ministry of Science and Technology through grant 3-18143 and the Tel Aviv University Center for Artificial Intelligence and Data Science (TAD) through a grant;
\item the Agenzia Spaziale Italiana (ASI) through contracts I/037/08/0, I/058/10/0, 2014-025-R.0, 2014-025-R.1.2015, and 2018-24-HH.0 to the Italian Istituto Nazionale di Astrofisica (INAF), contract 2014-049-R.0/1/2 to INAF for the Space Science Data Centre (SSDC, formerly known as the ASI Science Data Center, ASDC), contracts I/008/10/0, 2013/030/I.0, 2013-030-I.0.1-2015, and 2016-17-I.0 to the Aerospace Logistics Technology Engineering Company (ALTEC S.p.A.), INAF, and the Italian Ministry of Education, University, and Research (Ministero dell'Istruzione, dell'Universit\`{a} e della Ricerca) through the Premiale project `MIning The Cosmos Big Data and Innovative Italian Technology for Frontier Astrophysics and Cosmology' (MITiC);
\item the Netherlands Organisation for Scientific Research (NWO) through grant NWO-M-614.061.414, through a VICI grant (A.~Helmi), and through a Spinoza prize (A.~Helmi), and the Netherlands Research School for Astronomy (NOVA);
\item the Polish National Science Centre through HARMONIA grant 2018/30/M/ST9/00311 and DAINA grant 2017/27/L/ST9/03221 and the Ministry of Science and Higher Education (MNiSW) through grant DIR/WK/2018/12;
\item the Portuguese Funda\c{c}\~{a}o para a Ci\^{e}ncia e a Tecnologia (FCT) through national funds, grants SFRH/\-BD/128840/2017 and PTDC/FIS-AST/30389/2017, and work contract DL 57/2016/CP1364/CT0006, the Fundo Europeu de Desenvolvimento Regional (FEDER) through grant POCI-01-0145-FEDER-030389 and its Programa Operacional Competitividade e Internacionaliza\c{c}\~{a}o (COMPETE2020) through grants UIDB/04434/2020 and UIDP/04434/2020, and the Strategic Programme UIDB/\-00099/2020 for the Centro de Astrof\'{\i}sica e Gravita\c{c}\~{a}o (CENTRA);  
\item the Slovenian Research Agency through grant P1-0188;
\item the Spanish Ministry of Economy (MINECO/FEDER, UE), the Spanish Ministry of Science and Innovation (MICIN), the Spanish Ministry of Education, Culture, and Sports, and the Spanish Government through grants BES-2016-078499, BES-2017-083126, BES-C-2017-0085, ESP2016-80079-C2-1-R, ESP2016-80079-C2-2-R, FPU16/03827, PDC2021-121059-C22, RTI2018-095076-B-C22, and TIN2015-65316-P (`Computaci\'{o}n de Altas Prestaciones VII'), the Juan de la Cierva Incorporaci\'{o}n Programme (FJCI-2015-2671 and IJC2019-04862-I for F.~Anders), the Severo Ochoa Centre of Excellence Programme (SEV2015-0493), and MICIN/AEI/10.13039/501100011033 (and the European Union through European Regional Development Fund `A way of making Europe') through grant RTI2018-095076-B-C21, the Institute of Cosmos Sciences University of Barcelona (ICCUB, Unidad de Excelencia `Mar\'{\i}a de Maeztu’) through grant CEX2019-000918-M, the University of Barcelona's official doctoral programme for the development of an R+D+i project through an Ajuts de Personal Investigador en Formaci\'{o} (APIF) grant, the Spanish Virtual Observatory through project AyA2017-84089, the Galician Regional Government, Xunta de Galicia, through grants ED431B-2021/36, ED481A-2019/155, and ED481A-2021/296, the Centro de Investigaci\'{o}n en Tecnolog\'{\i}as de la Informaci\'{o}n y las Comunicaciones (CITIC), funded by the Xunta de Galicia and the European Union (European Regional Development Fund -- Galicia 2014-2020 Programme), through grant ED431G-2019/01, the Red Espa\~{n}ola de Supercomputaci\'{o}n (RES) computer resources at MareNostrum, the Barcelona Supercomputing Centre - Centro Nacional de Supercomputaci\'{o}n (BSC-CNS) through activities AECT-2017-2-0002, AECT-2017-3-0006, AECT-2018-1-0017, AECT-2018-2-0013, AECT-2018-3-0011, AECT-2019-1-0010, AECT-2019-2-0014, AECT-2019-3-0003, AECT-2020-1-0004, and DATA-2020-1-0010, the Departament d'Innovaci\'{o}, Universitats i Empresa de la Generalitat de Catalunya through grant 2014-SGR-1051 for project `Models de Programaci\'{o} i Entorns d'Execuci\'{o} Parallels' (MPEXPAR), and Ramon y Cajal Fellowship RYC2018-025968-I funded by MICIN/AEI/10.13039/501100011033 and the European Science Foundation (`Investing in your future');
\item the Swedish National Space Agency (SNSA/Rymdstyrelsen);
\item the Swiss State Secretariat for Education, Research, and Innovation through the Swiss Activit\'{e}s Nationales Compl\'{e}mentaires and the Swiss National Science Foundation through an Eccellenza Professorial Fellowship (award PCEFP2\_194638 for R.~Anderson);
\item the United Kingdom Particle Physics and Astronomy Research Council (PPARC), the United Kingdom Science and Technology Facilities Council (STFC), and the United Kingdom Space Agency (UKSA) through the following grants to the University of Bristol, the University of Cambridge, the University of Edinburgh, the University of Leicester, the Mullard Space Sciences Laboratory of University College London, and the United Kingdom Rutherford Appleton Laboratory (RAL): PP/D006511/1, PP/D006546/1, PP/D006570/1, ST/I000852/1, ST/J005045/1, ST/K00056X/1, ST/\-K000209/1, ST/K000756/1, ST/L006561/1, ST/N000595/1, ST/N000641/1, ST/N000978/1, ST/\-N001117/1, ST/S000089/1, ST/S000976/1, ST/S000984/1, ST/S001123/1, ST/S001948/1, ST/\-S001980/1, ST/S002103/1, ST/V000969/1, ST/W002469/1, ST/W002493/1, ST/W002671/1, ST/W002809/1, and EP/V520342/1.
\end{itemize}

The GBOT programme  uses observations collected at (i) the European Organisation for Astronomical Research in the Southern Hemisphere (ESO) with the VLT Survey Telescope (VST), under ESO programmes
092.B-0165,
093.B-0236,
094.B-0181,
095.B-0046,
096.B-0162,
097.B-0304,
098.B-0030,
099.B-0034,
0100.B-0131,
0101.B-0156,
0102.B-0174, and
0103.B-0165;
%
%
and (ii) the Liverpool Telescope, which is operated on the island of La Palma by Liverpool John Moores University in the Spanish Observatorio del Roque de los Muchachos of the Instituto de Astrof\'{\i}sica de Canarias with financial support from the United Kingdom Science and Technology Facilities Council, and (iii) telescopes of the Las Cumbres Observatory Global Telescope Network.

L. Liberato acknowledges support by the Coordena\,c\~ao de Aperfei\,coamento de Pessoal de N\'ivel Superior – Brasil (CAPES) – Finance Code 001, also by CAPES-PRINT Process 88887.570251/2020-00.

The authors want to acknowledge Valéry Lainey (IMCCE, Paris obseervatory) for providing the ephemerides of planetary satellites, and Josselin Desmars (IMCCE, Paris observatory) for providing extensive external and quality checks on the orbit computations from the NIMA software.\\
We made use of the software products \href{http://www.starlink.ac.uk/topcat/}{TOPCAT}, \citep{taylorTOPCATSTILStarlink2005}; Matplotlib \citep{Hunter:2007};
Astropy, a community-developed core Python package for Astronomy \citep{astropy:2013, astropy:2018}.

\end{appendix}
\end{document}